\documentclass[12pt,sort&compress]{elsarticle}        % final
\usepackage[a4paper,left =2.25cm, bottom=2.25cm, top=2.25cm, right=2.25cm]{geometry}     
\usepackage{setspace}
\usepackage{lineno}

\usepackage{pstricks, pst-plot}	% pstricks
\usepackage{graphicx} % enables import of various formats
\usepackage{wrapfig}  % enables floating wrapped figures
\usepackage[figuresright]{rotating}

\usepackage{amsmath,bm}
\usepackage{amsthm}
\usepackage{amssymb}
\usepackage{dsfont}
\usepackage{ragged2e}
\usepackage{xcolor}
\usepackage[english]{babel}
\usepackage{blindtext}
\usepackage{caption}
\usepackage{subcaption}
\usepackage{tabularx}
\usepackage[ruled,vlined]{algorithm2e}
\usepackage{algpseudocode}
\usepackage{float}

\usepackage{booktabs} % for professional tables
\usepackage{multirow}
\usepackage{pifont}
\usepackage{csquotes}

\usepackage{enumitem}
\setlist{leftmargin=15pt,labelindent=15pt}
\setlist[enumerate]{align=left}
\usepackage{ulem}
\newcommand*\dif{\mathop{}\!\mathrm{d}}

\DeclareMathAlphabet{\mathpzc}{OT1}{pzc}{m}{it}

\usepackage{bm}
\usepackage{mathtools}

\usepackage{placeins}

\newcolumntype{Y}{>{\centering\arraybackslash}X}

\usepackage[obeyFinal]{easy-todo}

\begin{document}
    \setlength\abovedisplayskip{4pt}
    \setlength\abovedisplayshortskip{4pt}
    \setlength\belowdisplayskip{4pt}
    \setlength\belowdisplayshortskip{4pt}
\begin{frontmatter}%

\title{Machine learning for viscoelastic constitutive model identification and parameterisation using Large Amplitude Oscillatory Shear} 

\author[a]{TP John\corref{cor1}}
\ead{thomas.john@manchester.ac.uk}
\author[a]{M Mowbray\corref{equal_cont}}
\ead{max.mowbray@manchester.ac.uk}
\author[a]{A Alalwyat}
\author[a]{M Vousvoukis}
\author[a]{P Martin}
\author[b]{A Kowalski}
\author[a]{CP Fonte}
\cortext[cor1]{Corresponding author}
\cortext[equal_cont]{Equal contribution}

\address[a]{Department of Chemical Engineering, The University of Manchester, Manchester, M13 9PL, United Kingdom}
\address[b]{Unilever R\&D, Port Sunlight Laboratory, Quarry Road East, Bebington, Wirral, CH63 3JW, United Kingdom}


%\listoftodos

\begin{abstract} 

Identification and parameterisation of constitutive models can be a challenging task in rheology. We investigate the use of Random Forest (RF) regression to estimate viscoelastic constitutive model parameters using Large Amplitude Oscillatory Shear (LAOS) data. Specifically, we deploy the RF to predict constitutive model parameters using the spectra of Chebyshev coefficients pertaining to the stress-strain-strain rate Lissajous curves. As far as we know, this is the first time Machine Learning (ML) has been utilised for this predictive task. We test three constitutive models: the linear and exponential Phan-Thien-Tanner (PTT) models, and the RoLiE-Poly model. For both PTT models, the RF estimator predicts the model parameters from the Chebyshev spectra with high accuracy. For the RoLiE-Poly model, the RF estimator demonstrates lower accuracy in model validation, however the predicted parameters still accurately capture the rheological behaviour in both oscillatory shear and oscillatory extensional flow. This is because multiple sets of RoLiE-Poly model parameters can yield very similar rheological responses and indicates practical identifiability issues under the investigated conditions. This does not represent an inadequacy of the framework, but rather a fundamental challenge in estimating model parameters using rheometric data. Overall, the study highlights the strong potential ML has to offer for selecting and parametrising constitutive models based on rheometry data. Our results can help in allowing for widespread robust, modelling of viscoelastic fluid flows. 

\end{abstract}
\end{frontmatter}

%\linenumbers

%\clearpage

\section{Introduction}

\subsection{Identifying constitutive viscoelastic models}

Viscoelastic fluid flows are ubiquitous across a wide range of industries and sectors, ranging from Fast Moving Consumer Goods (FMCG) \cite{Mackley1994, Datta2020} to health-care \cite{Bilgi2020,Giannokostas2021,Oliveira2011,Zhang2023}. With respect to the the former example, constitutive models are particularly important for understanding how product design and formulation affects rheological behavoiour in process equipment (mixing devices \cite{John2023b,Michael2022}, pipe bends \cite{Sutton2022}, contractions \cite{Sousa2011} etc). This is a key step in transitioning from product development and accelerating towards manufacture. Moreover, given the current climate emergency and the need to make our societies more sustainable, a particular challenge within FMCG and many other sectors at present is the rapid incorporation of novel and sustainable raw ingredients. Since the introduction of new materials into products can influence their rheology \cite{Beltrao2020}, it is vitally important that we develop tools to accurately and rapidly identify suitable constitutive models, to allow for robust digital modelling. Without access to robust digital modelling tools, industries still have to rely on trial-and-error experiments to scale-up and roll-out new products, which wastes time and increases carbon emissions. Additionally, although it gives rise to complex fluid behaviour, viscoelasticity can be a desirable material property in many cases. For example, this is the case for hydrogels used within the context of regenerative medicine. Viscoelasticity is a key property of biological soft-tissues, as it helps create the mechanical environments required to direct cell morphology appropriately \cite{charrier2018control,senior2019fabrication}. Therefore, the effective design and manufacture of viscoelastic hydrogels via digital modelling is crucial for regenerative technology.

A vast number of constitutive equations (or models) have been proposed to enable continuum scale modelling of viscoelastic fluid flows. To model these flows with, for example, advanced numerical tools such as Computational Fluid Dynamics (CFD), one must accurately select and parameterise (or fit) a particular constitutive model using some experimental data. Usually, this data is obtained uing shear rheometery. Provided that an analytical solution exists for the chosen constitutive model under the chosen rheometric test (or kinematics), model parameter estimation is relatively straightforward as solution to a nonlinear program \cite{costa2012generalized}. However, if no analytical solution exists for the particular choice of constitutive model and flow, parameter estimation becomes more challenging. For such cases, the constitutive model can be solved numerically using an appropriate scheme \cite{alves2021numerical}. The numerical solution can then be incorporated into an optimisation scheme for estimation of the parameters \cite{Jeyaseelan1993, Zhou2014}. This is, however, more complex and time-consuming than the case where the model can be solved analytically, and in practice model nonlinearity poses a considerable challenge. %Optimisation algorthims can include methods such as particle swarm optimisation \cite{Zheng2010}, which in general are not guaranteed to identify local optima.

Some of the most physically advanced and recently developed constitutive models tend to include large numbers of fitting parameters. This can pose a challenge to parameter estimation both from the perspective of practical identifiability and computation. In the former case, this might be resolved through the collection of more rheometric data. Note that in the previous discussion regarding model fitting, it is already assumed that an appropriate constitutive model has already been selected. If one is not certain as to which constitutive model is best to use for a given material, this parameterisation process must be repeated for a range of models in order to assess which one best fits the experimental data. Moreover, if spatial heterogeneity arises in the flow due to complex phenomena such as shear banding \cite{Divoux2016,Olmsted2008} (which is commonly observed for many personal care products such as shampoos and shower gels \cite{Zhou2012}), this must also be captured in the numerical modelling in order to properly fit the model response to the rheometric data. This means the numerical simulations must resolve the flow spatially, which dramatically increases the number of discretised equations to be solved and makes the parameterisation process even more complex and time-consuming. Considering the disadvantages of the traditional fitting methods discussed previously, we therefore propose, in this proof-of-concept study, an alternative approach for constitutive model identification and parameterisation based on Machine Learning (ML) \cite{Mahesh2020}. 

In the following sub-sections, we will give brief reviews of the deployment of Machine Learning within the fields of chemical engineering and then, more specifically, rheology and constitutive modelling. We will also introduce to the reader constitutive modelling for viscoelastic materials, and detail the particular constitutive models employed in this investigation.

\subsection{Machine Learning in Chemical Engineering}

ML has been widely investigated in Chemical Engineering applications as far back as the early 1990s. The process systems engineering community, which develops computational tools for design and operation of multi-scale systems, has demonstrated particular interest in these algorithms \cite{lee2018machine, mowbray2022industrial}. What unites this interest is the use of flexible, black-box functions to approximate relationships for which one has observations, but few processes to reason about mechanistically. For example, in operational decision problems such as control there has been interest in the development of Reinforcement Learning (RL) \cite{shin2019reinforcement}. RL can be conceptualized as a data-driven approach to identifying an approximately optimal policy function approximation. Depending on the problem, one may be agnostic of the optimal policy function structure and parameters. In these cases, RL provides an attractive solution method. As a result of its generality, it has observed applications in production scheduling \cite{hubbs2020deep}, supply chain optimisation \cite{perez2021algorithmic} and integrated design and control \cite{sachio2022integrating}.

Another domain where model structure may evade reasoning processes include complex reaction systems and their process dynamics. For example, Gaussian process have been used to model the discrete-time state transitions of biochemical reaction systems \cite{bradford2018dynamic}. However, purely black-box approaches have limited extrapolation ability. For this reason, hybridization of ML and first principles models has had academic and industrial impact. This is because the resultant models perform in both extrapolative and interpolative predictive tests \cite{von2014hybrid, narayanan2023hybrid, mowbray2023reinforcement}. 

Highly dimensional data sources have historically provided challenge to model identification. However, many ML models are able to identify structure within the data, learn lower dimensional representations of it, and then make robust predictions. For example, \cite{zhuang2022ensemble} proposed the use of deep learning and data assimilation to forecast drop dynamics in a microfluidic device from real-time video recordings. This is also exploited in application areas such as soft-sensing \cite{mowbray2022probabilistic, ching2021advances}, which provides means to estimate process variables that are conventionally difficult to measure by utilising other available process sensor measurements.

ML has also demonstrated application, beyond operational decision-making, in design and analysis. For example, surrogate modelling of CFD simulations presents as attractive means of model-based design \cite{humbert2022combined, savage2022deep}. In general, these approaches aim to identify a nonlinear relationship between the design decisions and the design objective for exploitation by model-based optimisation. Finally, it is worth noting that although ML describes a set of black-box models, once the model has been identified, it can be interrogated for the purpose of analysis. For example, \citet{kim2022supervised} demonstrated the use of ML for analysing the relationship between features of batch production scheduling mixed-integer linear programs and respective solution times.

Having provided wider context and perspectives on the motivation for use of ML across chemical engineering, we now review its application within rheology.

\subsection{Machine Learning in rheology}

Recently a number of authors have employed ML for data-driven constitutive modelling of complex fluids \cite{Mahmoudabadbozchelou2021, Lennon2023, Jin2023, Saadat2022, Mahmoudabadbozchelou2022, Mahmoudabadbozchelou2022a}. In the large majority of these studies, the ML algorithm is trained to reproduce observed rheological behaviour by formulating a `generalised' or `meta' non-linear constitutive model. That is, a generic model which satisfies required physical principles such as material-frame indifference \cite{Edwards2023}, but is not absolutely restricted in terms of the forms or positions of the non-linear functions within the model. Therefore, these data-driven models could represent blends of pre-existing constitutive models or even new constitutive models entirely. \citet{Mahmoudabadbozchelou2021} presents a multi-fidelity neural network approach for modelling the rheology of complex fluids. In this method, high fidelity data (experimental or numerical) is used in combination with low fidelity data generated for simple Generalised Newtonian Fluid (GNF) models (in particular the power-law and Herschel-Bulkley models) for the neural network training. The results showed that the multi-fidelity approach was able to reproduce experimental rheometric data much more accurately than the more traditional approach, which is purely data-driven and is not trained based on constitutive models. \citet{Mahmoudabadbozchelou2022a} present a framework which the authors call `Rheoloy-informed Graph neural Netorks' or RhiGNets. In this study, the authors use a physics-informed approach to recover rheological behaviour based on a small set of experimental data for a given material. This is particularly useful for constitutive models containing very large numbers of fitting parameters, where the generation of large training data sets is computationally demanding due to the dimensionality of the problem. This is particularly the case for recently developed constitutive models \cite{Wei2018, Armstrong2016, Dimitrious2014} for Thixotropic Elasto-Visco-Plastic (TEVP) materials, which typically contain more than 10 fitting parameters \cite{Larson2019}.

For the majority of available CFD codes (at least those primarily of interest in industry settings), the user is still restricted to using a set of chosen constitutive models, rather than ML informed models. Also, even if ML informed constitutive models are implemented and become the norm in major CFD codes, it is still important to be able to quickly and accurately select and parameterise well-known constitutive models derived from micro-structural theories. This is particularly the case when one wishes to undertake fundamental studies of the fluid dynamics of these constitutive models, which is still a popular and impactful topic \cite{Wu2022, Ahmed2023, Abreu2022, De2022, Guo2022}.

In this proof-of-concept study, we explore a means by which to approximate the mapping between rheometric data and an optimal parameterisation of a constitutive model by leveraging the highly flexible, and nonlinear set of models encompassed by Machine Learning. Specifically, we explore the use of Random Forest (RF) regression. The major benefit of this is that we mitigate dependence on optimisation-based parameter estimation routines to quickly provide a model estimate. The manuscript is organized as follows: in Section \ref{sec:prelim}, viscoelastic constitutive models of interest to this work are introduced;  Section \ref{sec:method} presents the proposed framework to enable parameterization of a constitutive model from Large Amplitude Oscillatory Data (LAOS) via the inference process of an ML model; in Section \ref{sec:R&D}, key results and discussions are presented before we subsequently present our conclusions from the study.

\section{Preliminary: Viscoelastic constitutive modelling}\label{sec:prelim}

Here, viscoelastic constitutive models of interest to this work are introduced. For a more thorough introduction to constitutive modelling and viscoelasticity, the reader is referred to the book by \citet{Bird1978}. We start by detailing the Phan-Thien-Tanner \cite{Thien1977} constitutive models. A generic polymer network-type constitutive model may be written as 

\begin{equation}
\overset{\kern0em\Box}{\boldsymbol{\mathrm{A}}} = -\frac{1}{\lambda}(D(A)\boldsymbol{\mathrm{A}}-C(A)\boldsymbol{\mathrm{I}}) ,\label{eq:networkmodel}
\end{equation}

\noindent where $\boldsymbol{\mathrm{A}}$ is the conformation, or configuration, tensor describing the stretching and orientation of the ensemble of microscopic entities (polymers or micelles etc), $\lambda$ is the viscoelastic relaxation time, and $D(A)$ and $C(A)$ are scalar functions describing the rates of microstructural destruction and creation, respectively. $A$ here is defined as $A\equiv \mathrm{tr}(\boldsymbol{\mathrm{A}}) \equiv A_{11}+A_{22}+A_{33}$. $\overset{\kern0em\Box}{\boldsymbol{\mathrm{A}}}$ denotes the objective Gordon-Schowalter time derivative  \cite{Gordon1972} defined as

\begin{equation}
    \overset{\kern0em\Box}{\boldsymbol{\mathrm{A}}} \equiv \frac{\partial \boldsymbol{\mathrm{A}}}{\partial t} + \boldsymbol{u}\boldsymbol{\cdot} \nabla \boldsymbol{\mathrm{A}} - (\boldsymbol{\mathrm{A}}\boldsymbol{\cdot} {\nabla \boldsymbol{u}} + {\nabla \boldsymbol{u}}^{\mathrm{T}}\boldsymbol{\cdot} \boldsymbol{\mathrm{A}}) + \zeta(\boldsymbol{\mathrm{A}}\boldsymbol{\cdot} \boldsymbol{\mathrm{D}} + \boldsymbol{\mathrm{D}} \boldsymbol{\cdot} \boldsymbol{\mathrm{A}}), \label{eq:gordonschowalter}
\end{equation}

\noindent where $\boldsymbol{\mathrm{D}}$ is the rate-of-strain tensor defined as $\boldsymbol{\mathrm{D}}\equiv (\nabla \boldsymbol{u} + {\nabla \boldsymbol{u}}^{\mathrm{T}})/2$. The parameter $\zeta$ (where $0 \leq \zeta \leq 2$) physically describes "slip" between the polymeric material and the solvent. In the case that $\zeta = 0$, the upper-convected (or Oldroyd \cite{Oldroyd1950}) derivative is obtained, denoted by $\overset{\kern0em\nabla}{\boldsymbol{\mathrm{A}}}$, where the reference frame translates, rotates, and deforms affinely with the material. In the case that $\zeta = 1$, the co-rotational (or Jaumann) derivative is obtained, where the reference frame only translates and rotates with the material. Finally, in the case that $\zeta = 2$, the (less frequently employed) lower-convected derivative is obtained. The extra-stress tensor $\boldsymbol{\tau}$ (the total or Cauchy stress tensor minus the isotropic pressure) is recovered from $\boldsymbol{\mathrm{A}}$ by

\begin{equation}
    \boldsymbol{\tau} = \frac{\eta_p}{(1-\zeta)\lambda}(\boldsymbol{\mathrm{A}}-\boldsymbol{\mathrm{I}}) + 2\eta_s\boldsymbol{\mathrm{D}}.\label{eq:kramersequation}
\end{equation}
 
In the Phan-Thien-Tanner (PTT) family of viscoelastic models, which are popular choices of constitutive models for many viscoelastic materials \cite{Omowunmi2010, Yu2014, Crawford2010}, the assumption is made that $D(A) = C(A)$. The original PTT model \cite{Thien1977}, which we denote as the linear PTT (lPTT) model, uses a linear function for $D(A)$, which is given by 

\begin{equation}
    D(A) \equiv 1+\varepsilon(\mathrm{tr}(\boldsymbol{\mathrm{A}})-3),\label{eq:pttlinearfunction}
\end{equation}

\noindent where $\varepsilon$ represents the rate of destruction of the junction points in the Lodge-Yamamoto network theory, from which the PTT model is derived. The PTT model was, shortly after, modified to include an exponential term for the junction breakage \cite{Thien1978}. In this model, which we denote as the exponential PTT (ePTT) model, $D(A)$ is defined as 

\begin{equation}
    D(A) \equiv e^{(\mathrm{tr}(\boldsymbol{\mathrm{A}})-3)}.\label{eq:pttexponentialfunction}
\end{equation}

\noindent Therefore, if one wishes to use either the lPTT or ePTT model to simulate the flow of a real-world viscoelastic material, the parameters $\eta_p$, $\eta_s$, $\lambda$, $\zeta$, and $\varepsilon$ must be obtained via fitting with experimental data. The analytical or numerical solution for the constitutive model is obtained, most often, by substituting into Equations \eqref{eq:networkmodel} and \eqref{eq:kramersequation} known kinematics (i.e. a prescribed velocity gradient tensor $\nabla \boldsymbol{u}$), which may be time-dependent. In the case of simple (or homogeneous) shear flow, where the 1,2, and 3 components of the orthonormal coordinate system are in the velocity, velocity gradient, and vorticity directions, respectively, the only nonzero component of $\nabla \boldsymbol{u}$ is $(\nabla u)_{21} = \dot{\gamma}$. The resulting solution, with initially guessed model parameters, is then to be compared with the experimental data for the fitting of the parameters.

We also test in this study the ROuse LInear Entangled POLYmers (RoLiE-Poly) model, which was proposed by \citet{Likhtman2003}. The RoLiE-Poly model is derived from a refined version of the Doi-Edwards tube theory for entangled polymers \cite{Graham2003,Doi1988} and accounts for the physical (microscopic) processes of reptation, convective constraint release (CCR), chain stretch and retraction \cite{Reis2013}. The model is given by

\begin{equation}
    \lambda \overset{\kern0em\nabla}{\boldsymbol{\mathrm{A}}} = -(\boldsymbol{\mathrm{A}} - \boldsymbol{\mathrm{I}}) - 2 z\bigg(1-\sqrt{3/\mathrm{tr}(\boldsymbol{\mathrm{A}})}\bigg)\bigg[ \boldsymbol{\mathrm{A}} + \beta_{\mathrm{CCR}} \bigg( \frac{\mathrm{tr}(\boldsymbol{\mathrm{A}})}{3} \bigg)^{\delta} (\boldsymbol{\mathrm{A}} - \boldsymbol{\mathrm{I}}) \bigg],
    \label{eq:roliepolymodelintro}
\end{equation}

\noindent and $\boldsymbol{\tau}$ is then recovered using Equation \eqref{eq:kramersequation} with $\zeta = 0$, since the RoLiE-Poly model uses the upper-convected derivative. The parameter $z$ defines the ratio between the reptation relaxation time $\lambda$ and the Rouse relaxation time $\lambda_R$. Note that the entanglement number $Z$ is often used in the literature and is defined as $Z=z/3$. For clarity, we do not refer to lowercase $z$ in Equation \eqref{eq:roliepolymodelintro} as the entanglement number, it is simply, for our purpose, a model parameter that we desire to fit. $\beta_{\mathrm{CCR}}$ is the convective constraint release parameter, where $0\leq \beta_{\mathrm{CCR}}\leq 1$. $\delta$ is a fitting parameter, whose optimum value has been found to be $-0.5$ \cite{Graham2003}. Accordingly, we use $\delta = -0.5$ in this study, and so the non-linear model parameters (those other than $\lambda$, $\eta_p$, and $\eta_s$) we seek are $z$ and $\beta_{\mathrm{CCR}}$.

\section{Methodology}\label{sec:method}

As this is primarily a proof-of-concept study, we use only the three aforementioned constitutive models to highlight the methodology we are proposing. The methodology is designed such that it can be easily repeated for further sets of constitutive models. This is particularly useful since new and improved constitutive models are frequently being proposed. In this section, we will introduce the reader to the rheometric flows we use for model fitting and validation, and then we will detail the RF algorithm and its implementation in the proposed workflow for this study.

\subsection{Rheometric flows}

\subsubsection{Large Amplitude Oscillatory Shear (LAOS)}

In this study, we focus solely on Large Amplitude Oscillatory Shear (LAOS) as the rheometric test to be employed for the fitting of the model parameters. During an ideal oscillatory shear flow, the strain $\gamma (t)$ follows a sine wave as ${\gamma = \gamma_0 \mathrm{sin}(\omega t)}$ where $\gamma_0$ and $\omega$ represent the amplitude and the frequency of the oscillation respectively. The strain rate $\dot{\gamma}(t) \equiv \dif \gamma /\dif t$, is then given as ${\dot{\gamma} = \dot{\gamma}_0\mathrm{cos}(\omega t)}$ where $\dot{\gamma}_0 \equiv \gamma_0 \omega$ is the strain rate amplitude. The velocity gradient tensor is therefore expressed as

\begin{equation}
\nabla \boldsymbol{u} \, (t) = 
\begin{bmatrix}
0 & 0 & 0\\
\dot{\gamma}(t) & 0 & 0 \\
0 & 0 & 0
\end{bmatrix}_{123} = \begin{bmatrix}
0 & 0 & 0\\
\dot{\gamma}_0 \mathrm{cos}(\omega t) & 0 & 0 \\
0 & 0 & 0
\end{bmatrix}_{123}.
\label{eq:velocitygradienttensor}
\end{equation}

\noindent Due to the time-periodicity of the applied flow,  the response of a viscoelastic material or constitutive model to oscillatory shear flow can be decomposed, in an ideal case, as

\begin{equation}
\tau_{12} = \gamma_0 \sum_{n, \mathrm{odd}} (G^{'}_n(\gamma_0,\omega)\mathrm{sin}(n\omega t)+G^{''}_n(\gamma_0,\omega)\mathrm{cos}(n\omega t)), \label{eq:stressresponsefourier}   
\end{equation}

\noindent where $G^{'}$ and $G^{''}$ are, respectively, the storage and elastic moduli, which quantify elasticity and viscosity, respectively \cite{Hyun2011}. For Small Amplitude Oscillatory Shear (SAOS), the stress response of the material or model is linear such that a single mode of $G^{'}$ and $G^{''}$ is sufficient to reconstruct the stress (i.e. the response is linear). During a LAOS experiment or simulation, the non-linearity in the stress response can be interpreted as higher-order harmonics in the stress response. Experimentally, non-linearity arises due to intra-cycle microstructural changes which are induced by the large deformations (or rates of deformation) in the flow. 

% \begin{figure}[h!]
%     \centering
%     \includegraphics[trim={0cm 0.1cm 0cm 0cm},clip,width=1\textwidth]{Figures/Workflow/ShearFlowSchematic.png}
%     \caption{Schematic of homogeneous oscillatory shear flow. Left shows $\tilde{\nabla \boldsymbol{u}}_{21}$ \textit{versus} $\tilde{t}$. Right shows shows velocity vectors (black arrows) and streamlines (grey lines) at various times in the oscillation.}
%     \label{fig:ShearFLowSchematic}
% \end{figure}

LAOS is considered to be especially useful for fitting constitutive models to experimental data due to the fact it captures both transient and non-linear viscoelastic behaviour \cite{Jeyaseelan1993}.  Responses of viscoelastic materials and constitutive models under LAOS flow can be rich with physical information. Moreover, even in the non-linear viscoelastic regime, the responses of some viscoelastic models can only be distinguished from one another in transient flows such as LAOS \cite{Davoodi2022, John2023}. However, whilst analytical solutions exist for various constitutive models in SSSF, it is not always possible to obtain analytical solutions in LAOS for even relatively simple constitutive models. Various authors have obtained analytical solutions, utilising symbolic computational tools, in a regime denoted as Medium Amplitude Oscillatory Shear (MAOS) \cite{Hyun2007}, where the response is only weakly non-linear. In this case, the fitting of these viscoelastic constitutive models to experimental MAOS data does not have to involve numerical simulations. In the MAOS regime, however, the stress response might be largely insensitive to the values of the model parameters, due to the response only being weakly non-linear. Currently, for LAOS, the only viable option for fitting constitutive models involves minimising the error between a numerical solution and experimental data through optimisation. This can often be challenging and expensive due to high model nonlinearity. ML presents therefore as a potential tool for quickly fitting constitutive models to experimental LAOS data, by parameterising the mapping between LAOS data and optimal model parameters.

We now define the following dimensionless variables

\begin{equation}
    \tilde{t}=t\omega, \quad \tilde{(\nabla \boldsymbol{u})} = \frac{\nabla \boldsymbol{u}}{\dot{\gamma}_0}, \quad \tilde{\boldsymbol{\mathrm{D}}}=\frac{\boldsymbol{\mathrm{D}}}{\dot{\gamma}_0}, \quad \tilde{\boldsymbol{\tau}}=\frac{\boldsymbol{\tau}}{\dot{\gamma}_0(\eta_p+\eta_s)}.\label{eq:dimensionlessvariables}
\end{equation}

\noindent Therefore, according to Equations \eqref{eq:networkmodel} through \eqref{eq:roliepolymodelintro}, and Equation \eqref{eq:dimensionlessvariables}, the lPTT and ePTT models can be written for homogeneous oscillatory shear flow in dimensionless form as

\begin{equation}
    De \frac{\dif \boldsymbol{\mathrm{A}}}{\dif \tilde{t}} - Wi (\boldsymbol{\mathrm{A}}\boldsymbol{\cdot}\tilde{(\nabla \boldsymbol{u}}) + \tilde{(\nabla \boldsymbol{u}})^{\mathrm{T}} \boldsymbol{\cdot} \boldsymbol{\mathrm{A}}) + Wi \, \zeta(\boldsymbol{\mathrm{A}}\boldsymbol{\cdot} \tilde{\boldsymbol{\mathrm{D}}} + \tilde{\boldsymbol{\mathrm{D}}}\boldsymbol{\cdot} \boldsymbol{\mathrm{A}}) = -D(A)(\boldsymbol{\mathrm{A}}-\boldsymbol{\mathrm{I}}),\label{eq:pttdimensionless}
\end{equation}

\noindent where

\begin{equation}
  D(A) \equiv \begin{cases}
      1+\varepsilon(\mathrm{tr}(\boldsymbol{\mathrm{A}})-3) & \text{for lPTT}, \\[10pt]
      e^{(\mathrm{tr}(\boldsymbol{\mathrm{A}})-3)} & \text{for ePTT},
  \end{cases}
  \label{eq:destructionrates}
\end{equation}

\noindent The RoLiE-Poly model can be written as 

\begin{multline}
    De \frac{\dif \boldsymbol{\mathrm{A}}}{\dif \tilde{t}} - Wi (\boldsymbol{\mathrm{A}} \boldsymbol{\cdot} \tilde{\nabla \boldsymbol{u}} + \tilde{\nabla \boldsymbol{u}}^{\mathrm{T}} \boldsymbol{\cdot} \boldsymbol{\mathrm{A}}) = -(\boldsymbol{\mathrm{A}} - \boldsymbol{\mathrm{I}}) \\ - 2 z\bigg(1-\sqrt{3/\mathrm{tr}(\boldsymbol{\mathrm{A}})}\bigg)\bigg[ \boldsymbol{\mathrm{A}} + \beta_{\mathrm{CCR}} \bigg( \frac{\mathrm{tr}(\boldsymbol{\mathrm{A}})}{3} \bigg)^{\delta} (\boldsymbol{\mathrm{A}} - \boldsymbol{\mathrm{I}}) \bigg].
    \label{eq:roliepolymodel}
\end{multline}

\noindent We then recover the dimensionless extra-stress (reaffirming that $\zeta=0$ for the RoLiE-Poly model) with

\begin{equation}
    \tilde{\boldsymbol{\tau}} = \frac{(1-\beta)}{(1-\zeta)Wi}(\boldsymbol{\mathrm{A}}-\boldsymbol{\mathrm{I}}) + 2\beta\tilde{\boldsymbol{\mathrm{D}}}.\label{eq:kramersequationdimensionless}
\end{equation}

\noindent The dimensionless groups which arise from the non-dimensionalisation are the Deborah number $De = \lambda\omega$, representing dimensionless frequency, the Weissenberg number $Wi = \lambda\dot{\gamma}_0$, representing dimensionless amplitude, and the viscosity ratio $\beta = \eta_s/(\eta_s+\eta_p)$. Note here that we assume the flow is spatially uniform (creeping flow and no shear-banding), and so $\nabla \boldsymbol{\mathrm{A}} = \nabla \boldsymbol{\tau} = \mathbf{0}$. The dimensionless velocity gradient tensor is given for the oscillatory shear flow as 

\begin{equation}
\tilde{(\nabla \boldsymbol{u})} \, (\tilde{t}) = 
\begin{bmatrix}
0 & 0 & 0\\
\mathrm{cos}(\tilde{t}) & 0 & 0 \\
0 & 0 & 0
\end{bmatrix}_{123}. 
\label{eq:dimensionlessvelocitygradienttensor}
\end{equation}

\noindent Equations \eqref{eq:dimensionlessvariables} through \eqref{eq:dimensionlessvelocitygradienttensor} form a set of closed Ordinary Differential Equations (ODEs), which we solve numerically using MATLAB's \textit{ode15s} solver for each set of model parameters. This solver uses built-in adaptive time-stepping. All simulations are initialised with $\boldsymbol{\mathrm{A}}=\boldsymbol{\mathrm{I}}$ (i.e. $\boldsymbol{\tau}=\boldsymbol{0}$) and are run until the response converges to the limit cycle (i.e. a steady-periodic state is reached).  The equations solved in this study are presented in full in \ref{eq:PTTODES} and \ref{eq:RPODES}.

The Deborah number quantifies the unsteadiness of the system (i.e. the ratio of a flow time-scale and a material time-scale), whilst the Weissenberg number represents the degree of anisotropy in the flow (i.e. the ratio of elastic and viscous forces). For oscillatory flows, the $De-Wi$ space is known as Pipkin space. LAOS corresponds to the region in Pipkin space where the value of $De$ is moderately large and the value of $Wi$ is large. In this study, for the lPTT and ePTT models, we use values of $De = 1$ and $Wi = 10$ for the ML training. We use $De = 1$ and $Wi = 50$ for the training for the RoLiE-Poly model (this will be discussed later). For the trained ML algorithm to be utilised for fitting of experimental data, this means $\lambda$ must be measured separately using a SAOS measurement. Only then can $\omega$ and $\dot{\gamma}_0$ be calculated such that the experimental LAOS data corresponds to the chosen values of $De$ and $Wi$ for the training. If we were to train the machine using a fixed values of $\omega$ and $\dot{\gamma}_0$, and were to include $\lambda$ as one of the fitting parameters, many of the responses in the training data would likely be linear, meaning the shear stress response would not be sensitive to the non-linear model parameters (hence they would not be able to be determined). Also, since we are using a non-dimensional stress based on $\dot{\gamma_0}\eta_0$ where $\eta_0 = \eta_p + \eta_s$ is the zero-shear viscosity, the zero-shear viscosity must also be known for a given material for the implementation of this proposed methodology. However, both $\lambda$ and $\eta_0$ are typically very easy to measure. There is no reason that the ML algorithm could not be trained on data for multiple values of $De$ and/or $Wi$ simultaneously (e.g. data for an amplitude sweep). However, as this is primarily a proof-of-concept study, and we wish to limit the amount of generated training data, we use one fixed point in Pipkin space. 

The assumption that the flow is homogeneous, at least in the direction of the velocity gradient, is not realistic since it is widely known that the inclusion of a nonzero slip parameter $\zeta$ in the constitutive model results in shear-banding for a wide range of model and system parameters \cite{Varchanis2022,Lu2000,Alves2001}. The RoLiE-Poly model is also widely studied specifically for its ability to predict shear banding \cite{Chung2011, Adams2011}. We could relax this assumption by solving both the constitutive model and the momentum equation in a 1D gap of fluid and allowing the flow to shear-band, as is done in the study of \citet{John2023}. However, since we wish to create a large training database containing thousands of LAOS simulations for varying sets of parameters, this would take considerably longer. In this study, we just wish to highlight the potential for ML to correlate LAOS results with constitutive model parameters, and so we assume spatially uniform flow. If this method were to be implemented to process real rheometry data, the 1D method would need to be implemented in the RF training to account for potential shear-banding. 

\subsubsection{Large Amplitude Oscillatory Extension (LAOE)}

We also check the accuracy of the parameter predictions of the RF algorithm in a homogeneous oscillatory planar (i.e. 2D) extensional flow. We test this flow to check if the rheological behaviour under this flow is more sensitive to the model parameters than for LAOS. To simulate this flow, we impose a velocity gradient tensor given by

\begin{equation}
\tilde{(\nabla \boldsymbol{u})} \, (\tilde{t}) = 
\begin{bmatrix}
-\mathrm{cos}(\tilde{t}) & 0 \\
0 & \mathrm{cos}(\tilde{t})
\end{bmatrix}_{12}, 
\label{eq:dimensionlessvelocitygradienttensorLAOE}
\end{equation}

\noindent and solve the resulting system of Equations after substituting this into the constitutive model (see \eqref{eq:RPODES}), as was done for the oscillatory shear flow in previous sections. We use the first normal stress difference $N_1 = \tilde{\tau}_{22}-\tilde{\tau}_{11}$ as the measure of the polymeric stress. Note that for the non-dimensionalisation of the equations, the extensional strain rate $\dot{\varepsilon}$ is used instead of the shear strain rate $\dot{\gamma}$. Therefore, $Wi = \lambda \dot{\varepsilon}_0$. 

% \begin{figure}[h!]
%     \centering
%     \includegraphics[trim={0.2cm 0.5cm 0.3cm 0cm},clip,width=1\textwidth]{Figures/Workflow/ExtensionalFlowSchematic.png}
%     \caption{Schematic of hypothetical homogeneous oscillatory extensional flow. Left shows $\tilde{\nabla \boldsymbol{u}}_{22}$ and $\tilde{\nabla \boldsymbol{u}}_{11}$ \textit{versus} $\tilde{t}$. Right shows shows velocity vectors (black arrows) and streamlines (grey lines) at various times in the oscillation.}
%     \label{fig:ExtensionalFLowSchematic}
% \end{figure}

Whilst oscillatory homogeneous shear flows can be easily created experimentally with modern shear rheometers, homogeneous extensional flows are significantly more challenging to realise. Thus, whilst the methodologies and results for this study for LAOS can be readily implemented, this section of the study is more hypothetical in nature. Many candidates for extensional rheometers (planar, uniaxial, and biaxial) have been proposed. These include, but are not limited to, four-roll mills \cite{Feng1997}, fibre spinning devices \cite{Sridhar1988}, counter-rotating drums \cite{Sentmanat2005}, Capillary Break-up Extensional Rheometers (CaBER) \cite{McKinley2000}, Filament Stretching Extensional Rheometers (FiSER) \cite{Mckinley2001}, and microfluidic devices \cite{Galindo2013}. As shown in Equation \eqref{eq:dimensionlessvelocitygradienttensorLAOE}, we will limit our attention to planar extension in a $1-2$ plane. Perhaps one of the most promising candidates for a planar extensional rheometer is the Optimised Shape Cross-slot Extensional Rheometer (OSCER) proposed by \citet{Haward2012}.  

\subsection{Stress Decomposition and Chebyshev Analysis}

As shown by Equation \eqref{eq:stressresponsefourier}, a particular elegance of LAOS (and LAOE) data is that the periodic shear-stress response (usually containing hundreds or thousands of data points) can be represented as a relatively small set of basis functions and coefficients by considering only a sufficient number of Fourier modes. In the context of utilising ML for model fitting, this is useful in two ways. Firstly, the dimensionality of the problem is drastically reduced (i.e. the machine need only be trained on the small set of coefficients rather than the individual data points in the stress waveform). Secondly, this technique removes requirements to match the sampling rates of the experimental data with that used for training data; enabling a uniform characterisation of stress waveforms gathered from different users.

We decompose the periodic stress response using the Stress Decomposition (SD) technique introduced by \citet{Cho2005}. Here the shear-stress is decomposed into the sum of an instantaneous elastic stress $\tau_{e}$ and an instantaneous viscous stress $\tau_v$ as

\begin{equation}
    \tau_{12} = \tau_e(\gamma) + \tau_v(\dot{\gamma}) .\label{eq:stressdecomposition}
\end{equation}

\noindent Following the methodology first detailed by \citet{Ewoldt2008}, we then approximate $\tau_{e}$ and $\tau_v$ using Chebyshev polynomials of the first kind, denoted by $T$, as 

\begin{subequations}
    \begin{gather}
        \tau_e = \sum_{n,\mathrm{odd}} e_n(De,Wi;\boldsymbol{\psi})\,T_n(\gamma/\gamma_0) \label{eq:elasticchebyshev},\\
        \tau_v = \sum_{n,\mathrm{odd}} v_n(De,Wi;\boldsymbol{\psi})\,T_n(\dot{\gamma}/\dot{\gamma}_0), \label{eq:viscouschebyshev}
    \end{gather}
    \label{eq:chebyshev}
\end{subequations}

\noindent where $\gamma/\gamma_0=[-1,1]$ and $\dot{\gamma}/\dot{\gamma}_0=[-1,1]$ are the normalised strain and normalised strain rate, respectively. $G^{'}_n$ and $G^{''}_n$ are recovered from $e_n$ and $v_n$, respectively, as 

\begin{subequations}
    \begin{gather}
            e_n = G^{'}_n (-1)^{(n-1)/2},\label{eq:enconversion} \\
            v_n = G^{''}_n.\label{eq:vnconversion}
    \end{gather}
\end{subequations}

\noindent In general, the quality of this approximation increases as the order $n$ of the polynomials increases; however, we use a finite $n$ that provides an adequate representation of the waveform. Once the data for the limit-cycle has been generated for all data points in the model parameter space, we fit a series of $n$ Chebyshev polynomials to all waveforms. We then reconstruct the waveform from the fitted coefficient spectra and compute the total error across all data points between the real and reconstructed waveforms (real here meaning numerical, not experimental). We choose a value of $n$ where the error does not change with increasing $n$. In essence, we ensure the series representation of the waveform is converged to the real waveform for all waveforms. For the LAOE flow, we decompose and interpolate the waveform for $N_1$ using Chebyshev polynomials in the same way as described previously for LAOS.

In Equation \eqref{eq:chebyshev}, $\boldsymbol{\psi}$ denotes any variables other that $De$ and $Wi$ on which there is a dependence of $e_n$ and $v_n$. In the context of constitutive models, this would represent the set of non-linear fitting parameters (such as $L^2$ in the FENE-P model \cite{Bird1980} or $\varepsilon$ and $\zeta$ in the PTT models). Experimentally, whilst it is likely not possible to explicitly define $\boldsymbol{\psi}$, this might reflect, for example, variables such as polymer concentration, or the degree of chain-branching etc. The relationship between $e_n$ and $\boldsymbol{\psi}$, and that between $v_n$ and $\boldsymbol{\psi}$, can be found analytically for some constitutive models using MAOS solutions, as discussed. In this study, however, we wish to use ML to correlate the viscoelastic model fitting parameters $\boldsymbol{\psi}$ with the coefficients $e_n$ and $v_n$ in the fully non-linear (LAOS) regime. To the best of our knowledge, there exists no simple optimisation-free method for generally parameterising non-linear viscoelastic constitutive models under LAOS.

\subsection{Data generation}

For a given constitutive model to be tested, we first define the upper and lower bounds for each model parameter, and then we fill the model parameter space $\boldsymbol{\psi}$ with a number $n_p$ of data points, i.e. each point in this space corresponds to a particular parameterisation of the constitutive model. For some model parameters (i.e. in certain dimensions in the parameter space), it is intuitive to use log-spacing rather than linear spacing for the data points. For example, we know that for $\beta \rightarrow 1$, the response will practically be Newtonian. We desire that the training and validation data contains more responses for lower values of $\beta$, where the model response will be more non-linear, than for higher values of $\beta$. Therefore, we use a lower bound for $\beta$ of $10^{-5}$ and an upper bound of $0.9$, and we space the data points logarithmically (with base 10) in this dimension. For the PTT models, we use lower and upper bounds, respectively, of $0$ and $0.2$ for both $\zeta$ and $\varepsilon$. We chose to limit $\zeta$ at $0.2$ due to the fact that larger values led to highly un-realistic LAOS responses, at least for the viscoelastic materials we are generally concerned with. In any case, we only wish here to assess if the RF algorithm can learn the complex relationship between model parameters and Chebyshev spectra. Using $0 \leq \zeta \leq 0.2$ provided us with a wide range of Lissajous curves (and hence Chebyshev spectra) for this purpose. For the RoLiE-Poly model, we use lower and upper bounds of 1.2 and 100, respectively, for $z$. And we use lower and upper bounds of 0 and 1, respectively, for $\beta_{\mathrm{CCR}}$. For both of these parameters, logarithmic (base 10) spacing was used.

We use Latin Hypercube Sampling (LHS) to sample the parameter space \cite{lin2022latin}. This is particularly appealing because it provides better coverage of the space than na{\"i}ve random sampling, especially when the size of the dataset is relatively small. This is important given the predictive performance of ML models on interpolation tasks is strong, but generally poor under extrapolation.

% \begin{figure}[h!]
%     \centering
%     \subfloat[$n_p=4^3$]{
%     \includegraphics[trim={1cm 0 3cm 0},clip,width=0.35\textwidth]{Figures/Workflow/n_4.png}}
%     \subfloat[$n_p=8^3$]{
%     \includegraphics[trim={1cm 0 3cm 0},clip,width=0.35\textwidth]{Figures/Workflow/n_8.png}}

%     \subfloat[$n_p=12^3$]{
%     \includegraphics[trim={1cm 0 3cm 0},clip,width=0.35\textwidth]{Figures/Workflow/n_12.png}}
%     \subfloat[$n_p=16^3$]{
%     \includegraphics[trim={1cm 0 3cm 0},clip,width=0.35\textwidth]{Figures/Workflow/n_16.png}}
%     \caption{Example of Latin Hypercube sampling for PTT model with varying values of $n_p$.}
%     \label{fig:LHSsampling}
% \end{figure}

\subsection{Random Forest Modelling}

\begin{figure}[b!]
     \centering
     \begin{subfigure}[b]{0.45\textwidth}
         \centering
         \includegraphics[width=\textwidth]{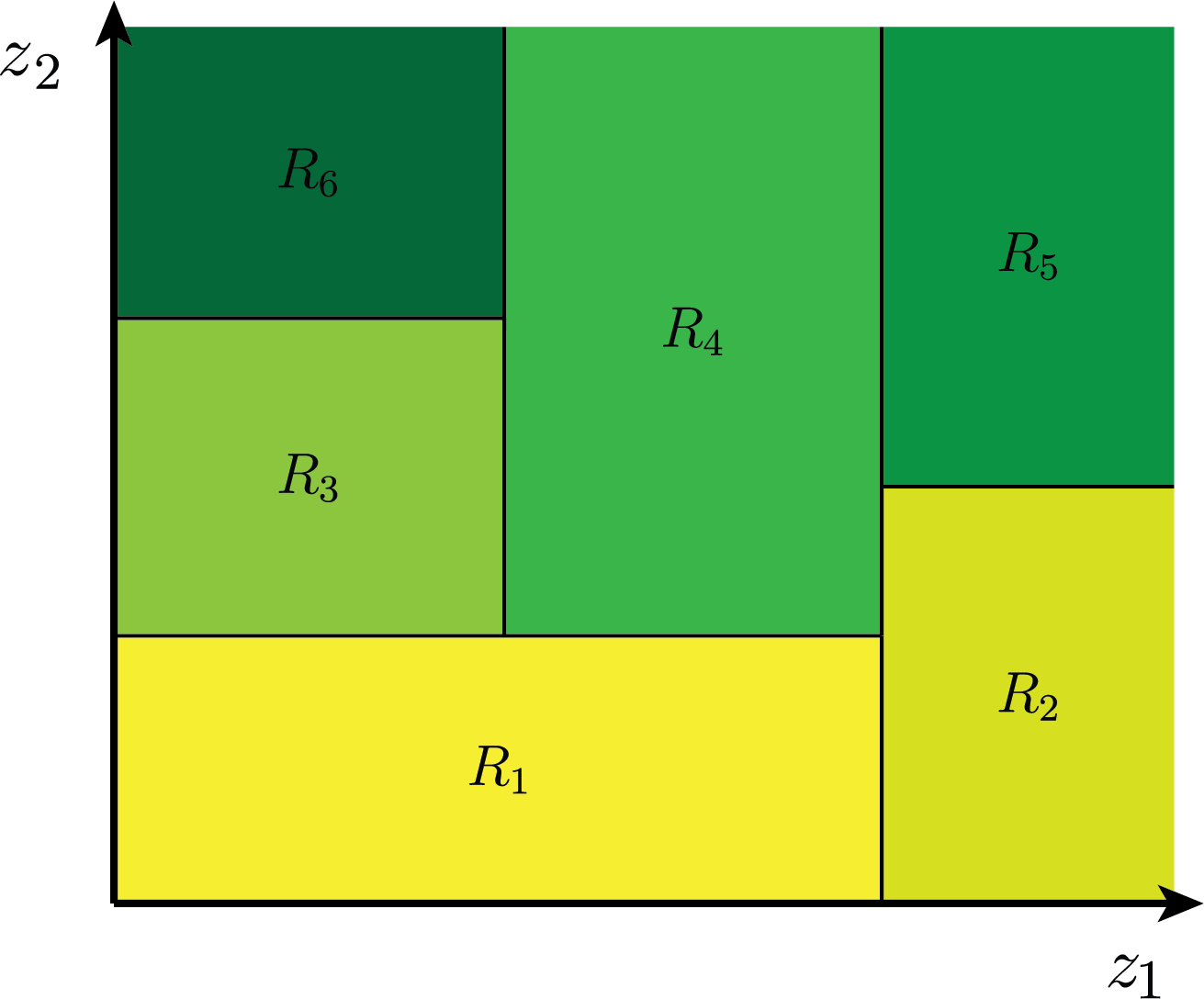}
         \caption{Domain partitioning via a decision tree}
         \label{fig:random_forests_a}
     \end{subfigure}
     \hfill
     \begin{subfigure}[b]{0.45\textwidth}
         \centering
         \includegraphics[width=\textwidth]{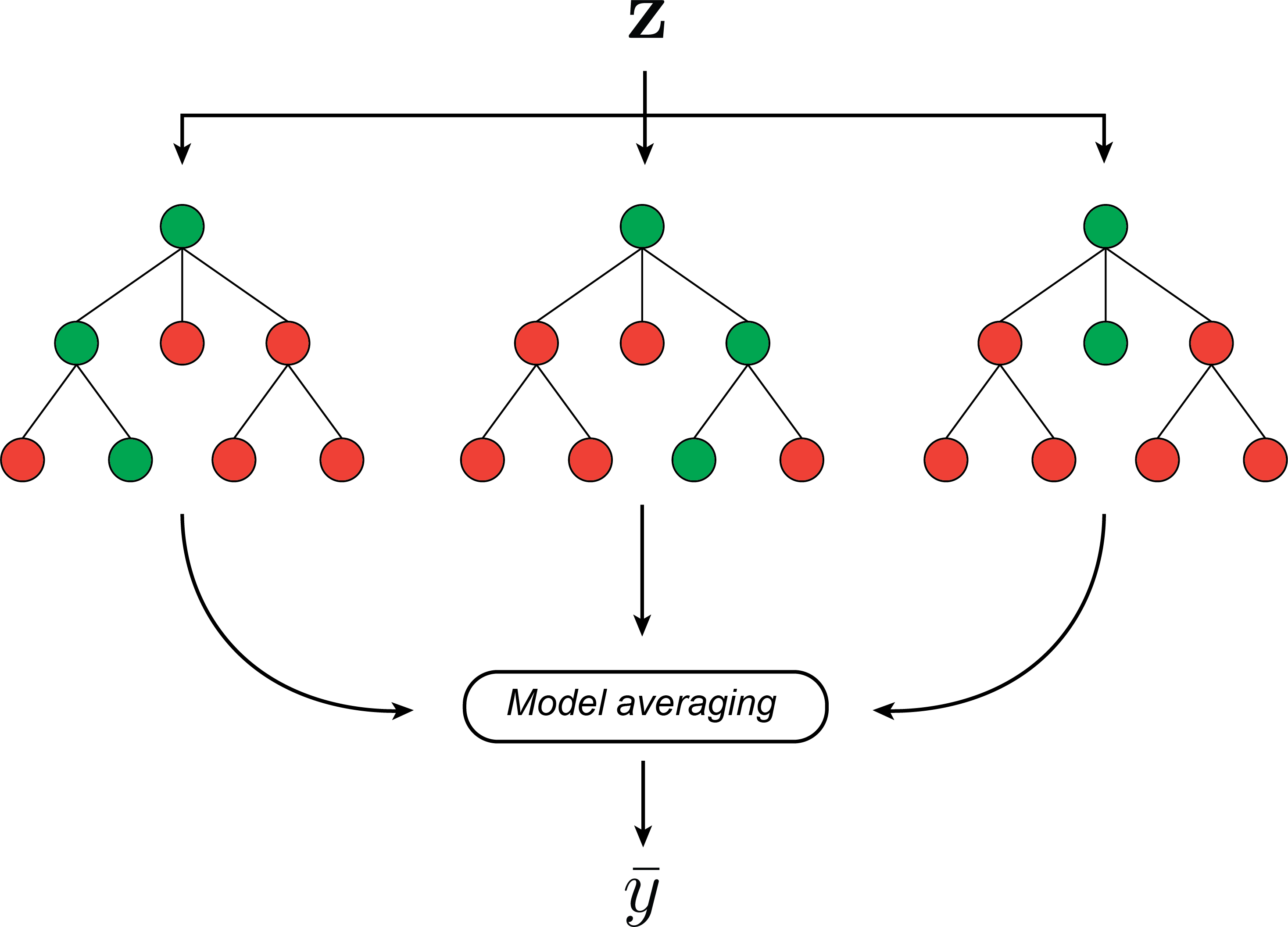}
         \caption{Decision tree model ensembling}
         \label{fig:random_forests_b}
     \end{subfigure}
    \caption{Figurative description of the key characteristics of decision tree models: a) partitioning of the domain through the identification of hyperplanes, b) ensembling of the weak estimator provided by each decision tree in the forest.}
    \label{fig:random_forests}
\end{figure}

The ML algorithm we utilise in this study is the Random Forest (RF) algorithm, which will now be detailed. RFs are bootstrap ensembles of Decision Trees (DTs), which are each identified on subsets of the training data available to the modeller. In the context of this work, DTs are mulitvariate, scalar-valued functions, $h: \mathbb{R}^{n_z}\rightarrow \mathbb{R}$. They have a logic structure, which partitions input data into a set of $K$ disjoint regions, $\{R_{k}\}_{k=1:K}$. Each disjoint region, $R_k\subset \mathbb{R}^{n_z}$, is associated with its own weak estimator, which generates a prediction given the model input \cite{hastie2009elements}. This is outlined by Fig. \ref{fig:random_forests}. Typically, the weak estimator is a constant, $y_{k}\in \mathbb{R}$, such that the prediction of the DT can be expressed as:
\begin{equation}
    h(\textbf{z}) = \sum_k y_k I_{k}(\textbf{z}),
\end{equation}
where $I_{k}(\textbf{z}): \mathbb{R}^{n_z} \rightarrow \mathbb{Z}_2$ is the indicator function that takes a value of 1 if the input, $\textbf{z}$, is partitioned into region, $R_k$, and a value of 0 otherwise. This means that given $M$ datapoints to train the DT, $\mathcal{D}_{DT} = \{\textbf{z}_i, v_i\}_{i=1:M}$, a region $R_k$ is associated with a subset of the input-output data, $\mathcal{D}_{DT}^{R_k} \subset \mathcal{D}_{DT}$. This data can be used to determine the  the weak estimator, $y_{k}$, and this is usually done in a non-parametric fashion as \cite{hastie2009elements}:
\begin{equation}
    y_k = \frac{1}{\lvert\mathcal{D}_{DT}^{R_k}\rvert}\sum_{(\textbf{z}_i, v_i)\in \mathcal{D}^{R_k}_{DT}} v_i
\end{equation}
When deployed within the bootstrap aggregation framework provided by the RF, predictions from each of $N$ DTs, are averaged to generate a prediction, $\bar{y}$:
\begin{equation}
    \bar{y} = \frac{1}{N}\sum_{j=1}^N h_j(\textbf{z}).
\end{equation}
where the index $j$ denotes a single DT estimator within the RF. Uncertainty estimates can also be gained by estimating the variance of the ensemble's predictions. Due to the interpretable structure of DTs one can also gain measures of the input features' importance to making predictions \cite{strobl2007bias}. This is quantified by metrics such as permutation importance \cite{breiman2001random}, and can provide information to the modeller about the calibration of the model provided one has intuition regarding those variables that are strongly correlated with the prediction.

RFs have a number of hyperparameters, which control the subsampling of data for identification of each DT, as well as the complexity of the model itself. In the case of the DT, complexity essentially refers to the definition and depth of the logic structure. This controls the number of disjoint regions and the order of interactions between input features captured in the prediction \cite{hastie2009elements}. When viewed in the context of a forest, complexity refers to the the selection of the number of DTs, $N$. We denote the depth of the trees in the forest (the number of splits that each tree can make) as $d_{\mathrm{max}}$. In the following, we approach the RF hyperparameter optimisation problem via grid search and 10-fold cross validation. 

It is worth noting, in this work we are interested in identifying a multivariate, vector-valued function to map from Chebyshev polynomial coefficients to the optimal parameterisation of a constitutive model. In order to do this, we construct independent scalar-valued RF models, one for each parameter in the constitutive model. We then combine the independent predictions of each RF model to provide a parameter estimate, given a setting of Chebyshev coefficients. In this way, we allow for a partitioning of the input space that accounts for the different relationships between Chebyshev polynomial coefficients and constitutive model parameters. 

\subsection{Workflow}

\begin{figure}[h!]
     \centering
     \begin{subfigure}[b]{0.32\textwidth}
         \centering
         \includegraphics[width=\textwidth]{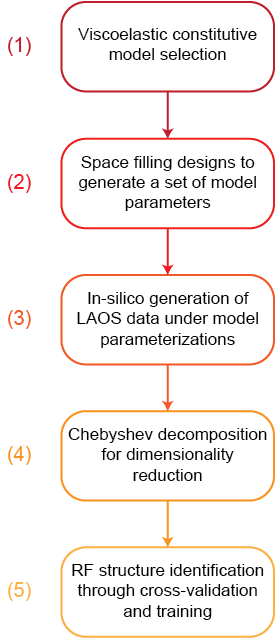}
         \caption{Workflow of model construction}
         \label{fig:workflow_a}
     \end{subfigure}
     \hfill
     \begin{subfigure}[b]{0.58\textwidth}
         \centering
         \includegraphics[width=\textwidth]{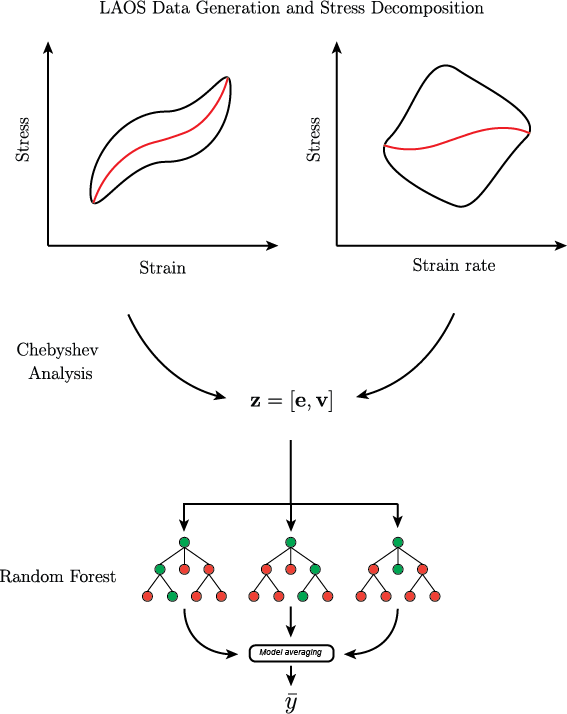}
         \caption{Model inference}
         \label{fig:workflow_b}
     \end{subfigure}
    \caption{Figurative description of a) the workflow for model construction, and b) model inference when presented with a new material.}
    \label{fig:workflow}
\end{figure}

Here, for clarity, we briefly outline the general workflow we are proposing to create a ML framework for constitutive model selection and fitting: \textbf{1} - Choose a viscoelastic constitutive model. \textbf{2} - Fill the space of constitutive model parameters with training and validation points through LHS sampling. \textbf{3} - Solve, for each point in the training and validation data, the system of ODEs for the constitutive model under LAOS for the limit-cycle (either solving directly for the limit-cycle \cite{Mittal2023} or solving in time until the response is converged to the limit-cycle). \textbf{4} - Obtain $e_n$ and $v_n$ for all data, ensuring that $n$ is large enough such that the most non-linear response in the data set is well-represented by the truncated series of Chebyshev polynomials $T_n$. \textbf{5} - Train the RF model to correlate the Chebyshev coefficients, $e_n$ and $v_n$, with constitutive model parameters. This is detailed by Fig. \ref{fig:workflow_a}.

Our method essentially solves an inverse problem for model parameterisation. Once steps 1 through to 5 have been completed, one then only needs to obtain some experimental LAOS data for a given material at $De = 1$ and $Wi = 10$ (or any other chosen values of $De$ and $Wi$ used for the training) and the trained RF algorithm will provide the values of the model parameters $\boldsymbol{\psi}$ which best fits the data. Steps 1 through to 5 only need to be completed once for a given constitutive model, whilst numerous sets of experimental data for different materials can be analysed almost instantly by the trained RF model, as depicted in Fig. \ref{fig:workflow_b}. For the task of constitutive model selection, one need only repeat steps 1 through 5 for a set of constitutive models. If experimental LAOS data is analysed by each trained machine (i.e. for each viscoelastic model), the user can then choose the best fitting model qualitatively, or some quantitative error can be used to select the most appropriate model. Note that we have chosen LAOS as rheometry protocol, but there is no reason this framework could not be extended for other rheometry protocols such as flow curves or CaBER tests etc.

\section{Results and Discussion}\label{sec:R&D}
\subsection{lPTT and ePTT models}

In this section, we present the results for the lPTT and ePTT models.

\subsubsection{lPTT model}

\begin{figure}[h!]
    \centering
    \includegraphics[width=0.9\textwidth]{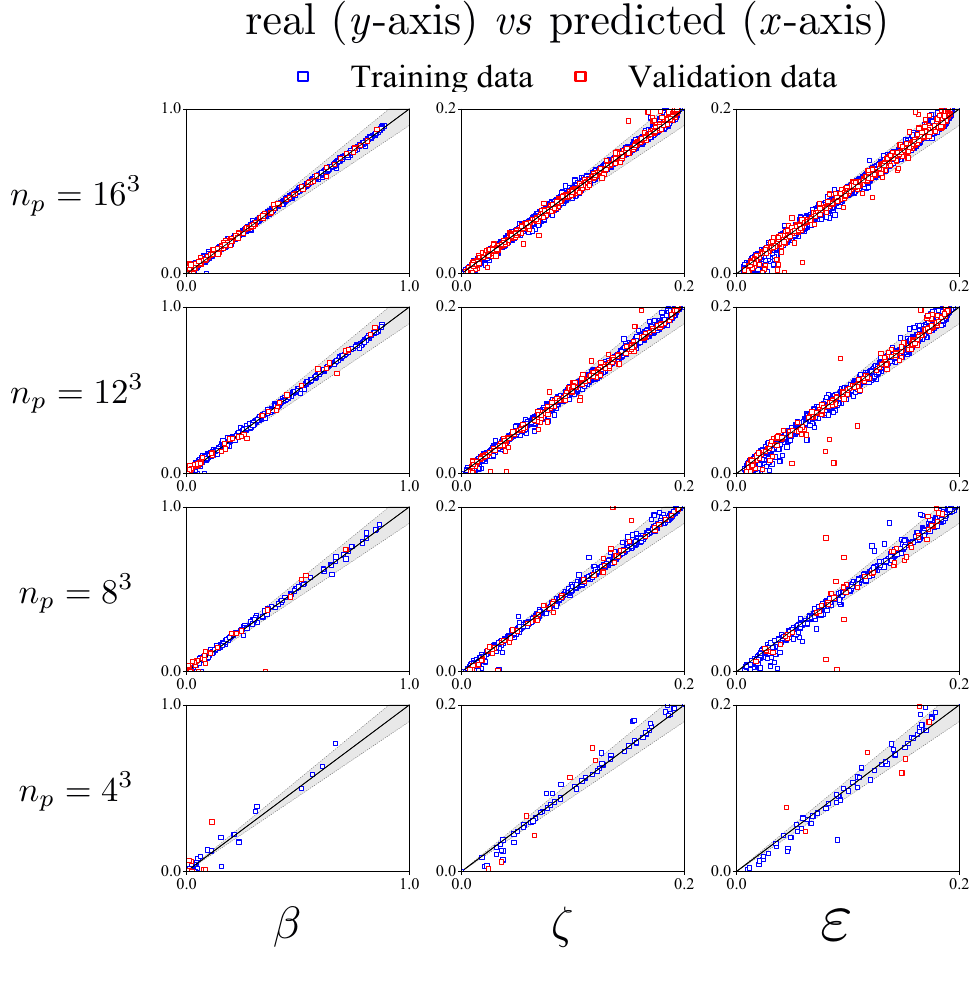}
    \caption{Parity plots (real \textit{versus} predicted) for lPTT model parameters. Hyperparameters: $d_{\mathrm{max}} = 10$ and $N=100$. Grey shaded area represents $\pm$10\% error. For each plot (training and validation) the value of $R^2$ is given in Table \ref{tab:datasizeRsquared}.}
    \label{fig:datasize}
\end{figure}

\begin{table}[]
\centering
\begin{tabularx}{\textwidth}{Y| Y Y Y| Y Y Y}
        \toprule
        \multirow{2}{*}{$n_p$} & \multicolumn{3}{c|}{Training } & \multicolumn{3}{c}{Validation} \\\cline{2-7} 
                & $\beta$ & $\zeta$ & $\varepsilon$ & $\beta$ & $\zeta$ & $\varepsilon$     \\\hline 
        $4^3$ & 0.9753 & 0.9944   & 0.9903  & 0.5048 & 0.9489 & 0.9676 \\ 
        $8^3$ & 0.9979        & 0.9989        & 0.9974      & 0.9215 & 0.9886 & 0.9679   \\
        $12^3$ & 0.9987        & 0.9993        & 0.9985        & 0.9959 & 0.9966 & 0.9904   \\
        $16^3$ & 0.9991 & 0.9994 & 0.9989 & 0.9977 & 0.9984 & 0.9969 \\\bottomrule
\end{tabularx}
\caption{\label{tab:datasizeRsquared}Values of $R^2$ for each of the parity plots in Figure \ref{fig:datasize}.}
\end{table}

For both PTT models, it was found from an initial grid search of hyperparameters that the optimal hyperparameters for the RF algorithm were $d_{\mathrm{max}} = 10$ with $N=100$.  Figure \ref{fig:datasize} shows, for the training data and validation data, parity plots for the RF predictions of the 3 viscoelastic model parameters ($\beta$, $\zeta$, and $\varepsilon$) with varying amounts of training data. Note that 90\% of the data is used for the regression (blue symbols) and the remaining 10\% (red symbols) is used to validate the predictive accuracy of the fit (i.e. it is not used in the regression). The training data is predicted well by the RF algorithm, even when relatively limited amounts of training data are used. The prediction of the validation data seems to be more sensitive to the amount of training data used, with the accuracy of interpolation increasing with dataset size. With $n_p = 16^3$ (i.e. an average of 16 data points in each dimension of the parameter space), the prediction of the validation data is highly accurate, with only a few outliers. The values of $R^2$ for each fit are displayed in Table \ref{tab:datasizeRsquared}. The values of $R^2$ increase with dataset size in all cases.

\begin{figure}[h!]
    \centering
    \includegraphics[width=0.325\textwidth]{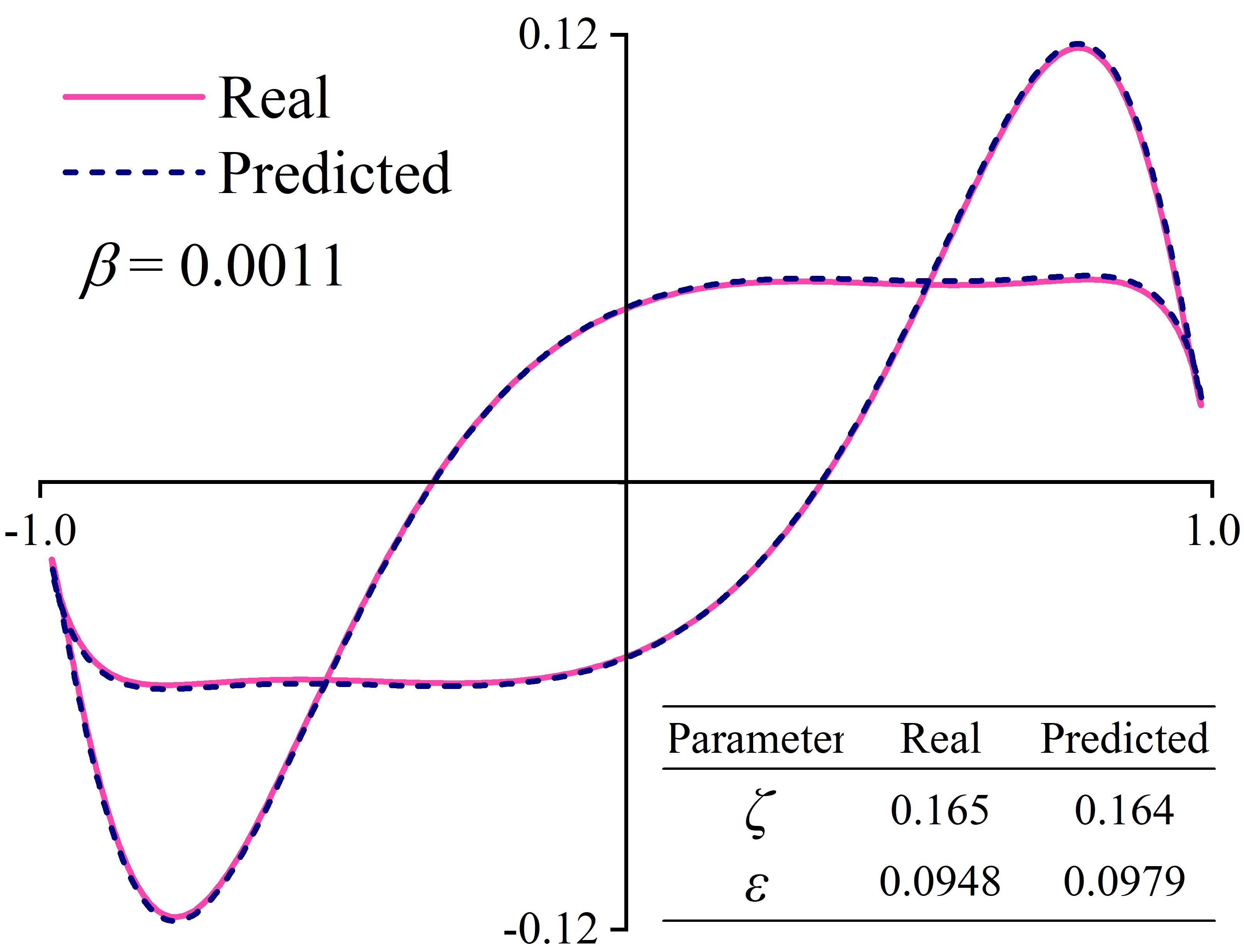}
    \includegraphics[width=0.325\textwidth]{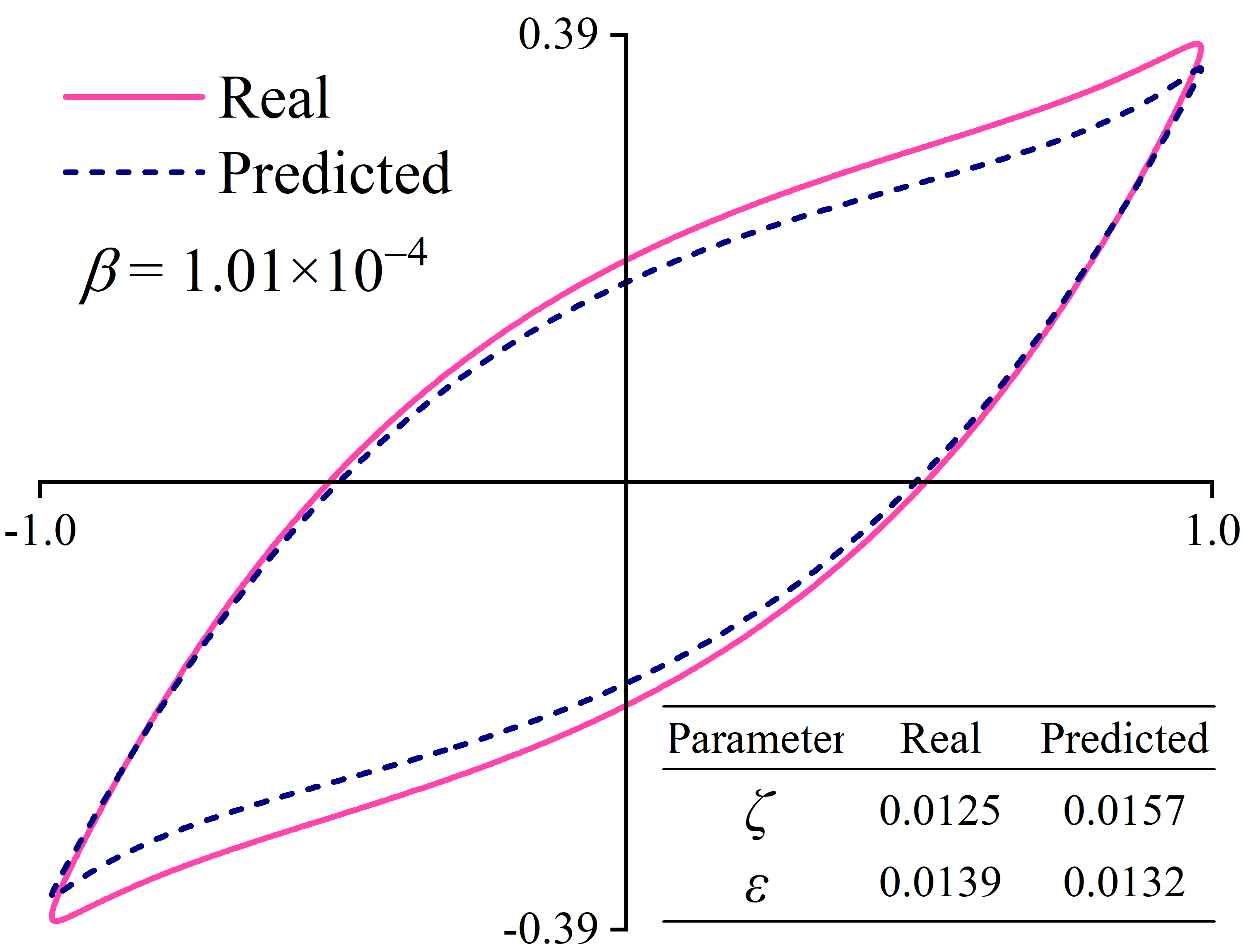}
    \includegraphics[width=0.325\textwidth]{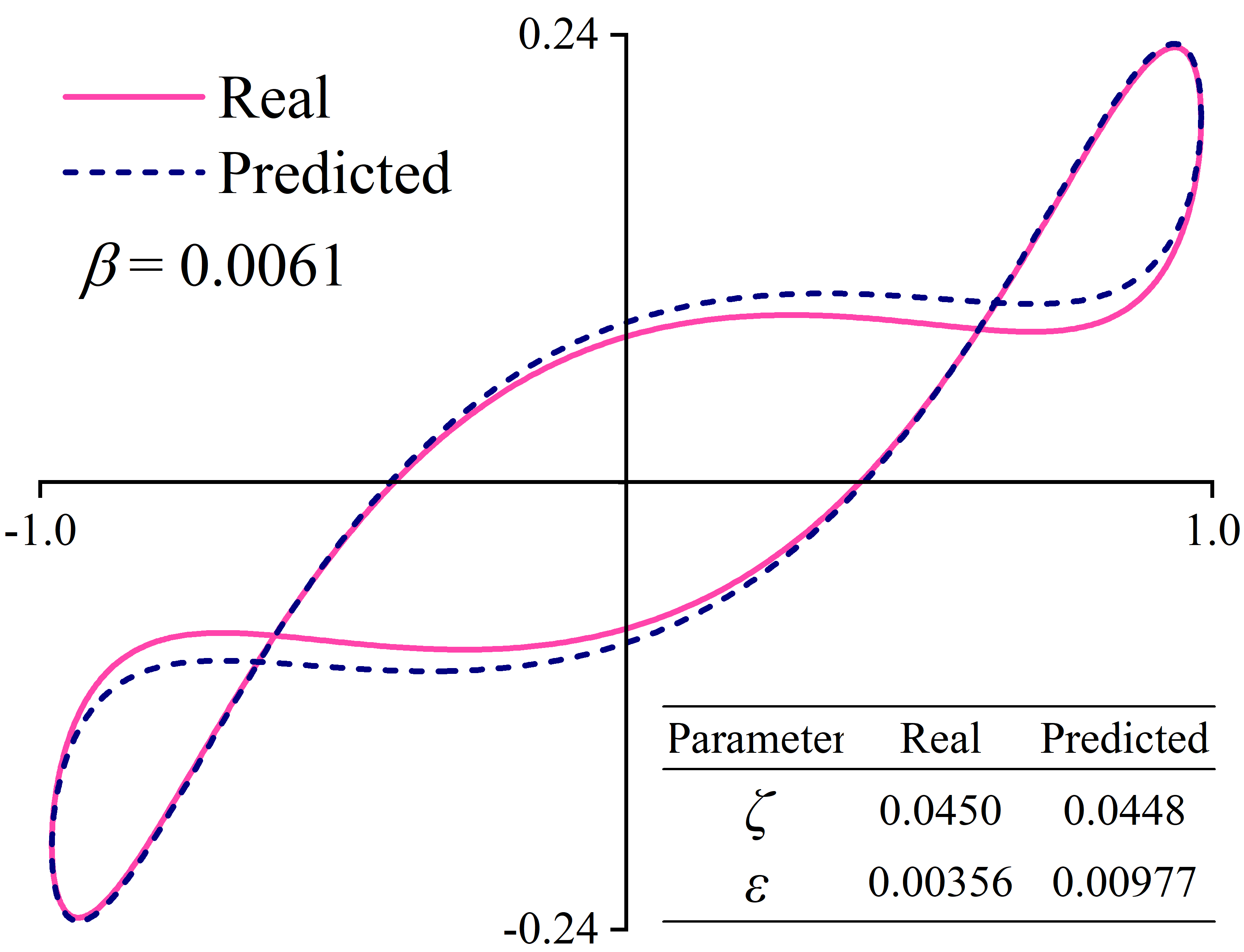}

    \includegraphics[width=0.325\textwidth]{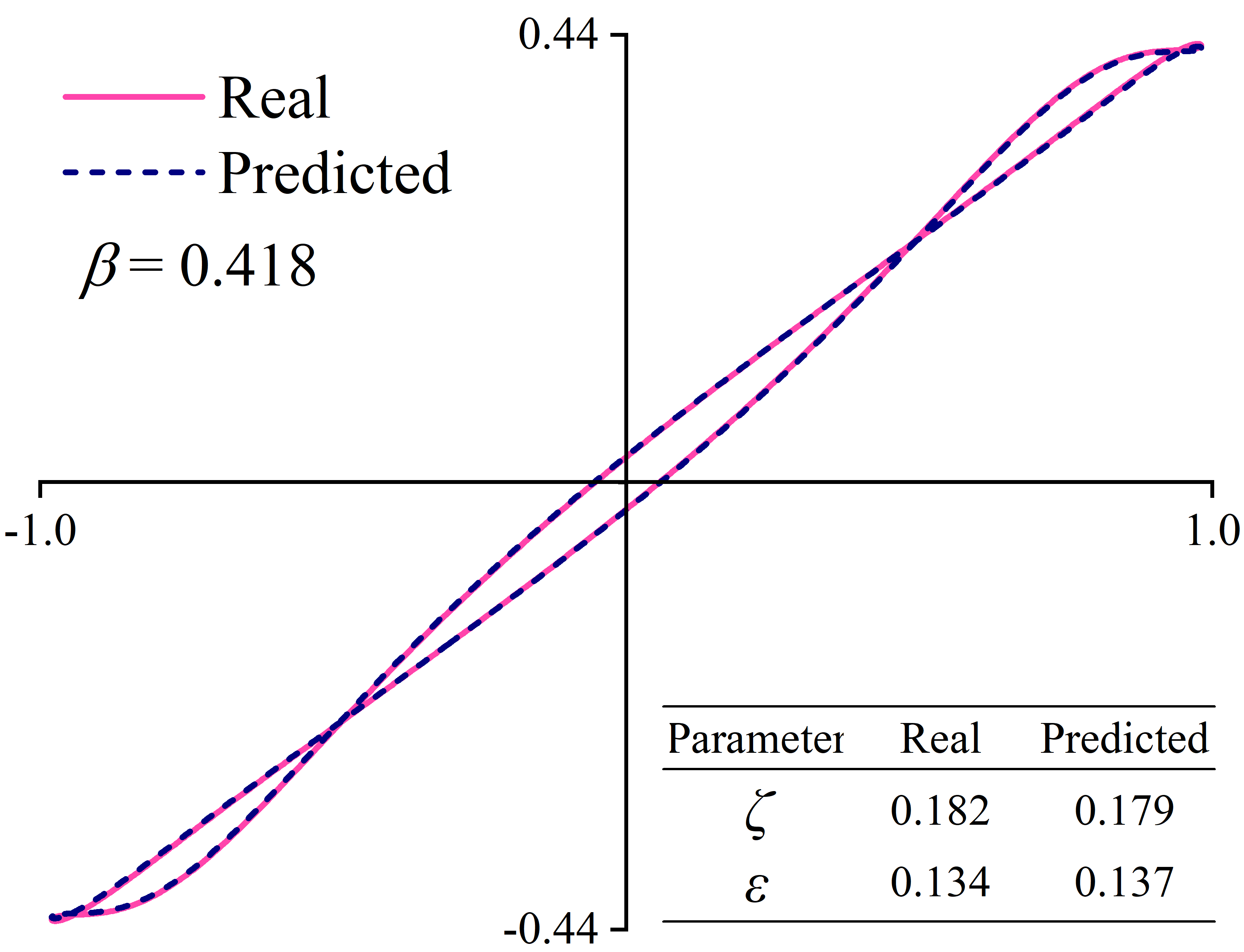}
    \includegraphics[width=0.325\textwidth]{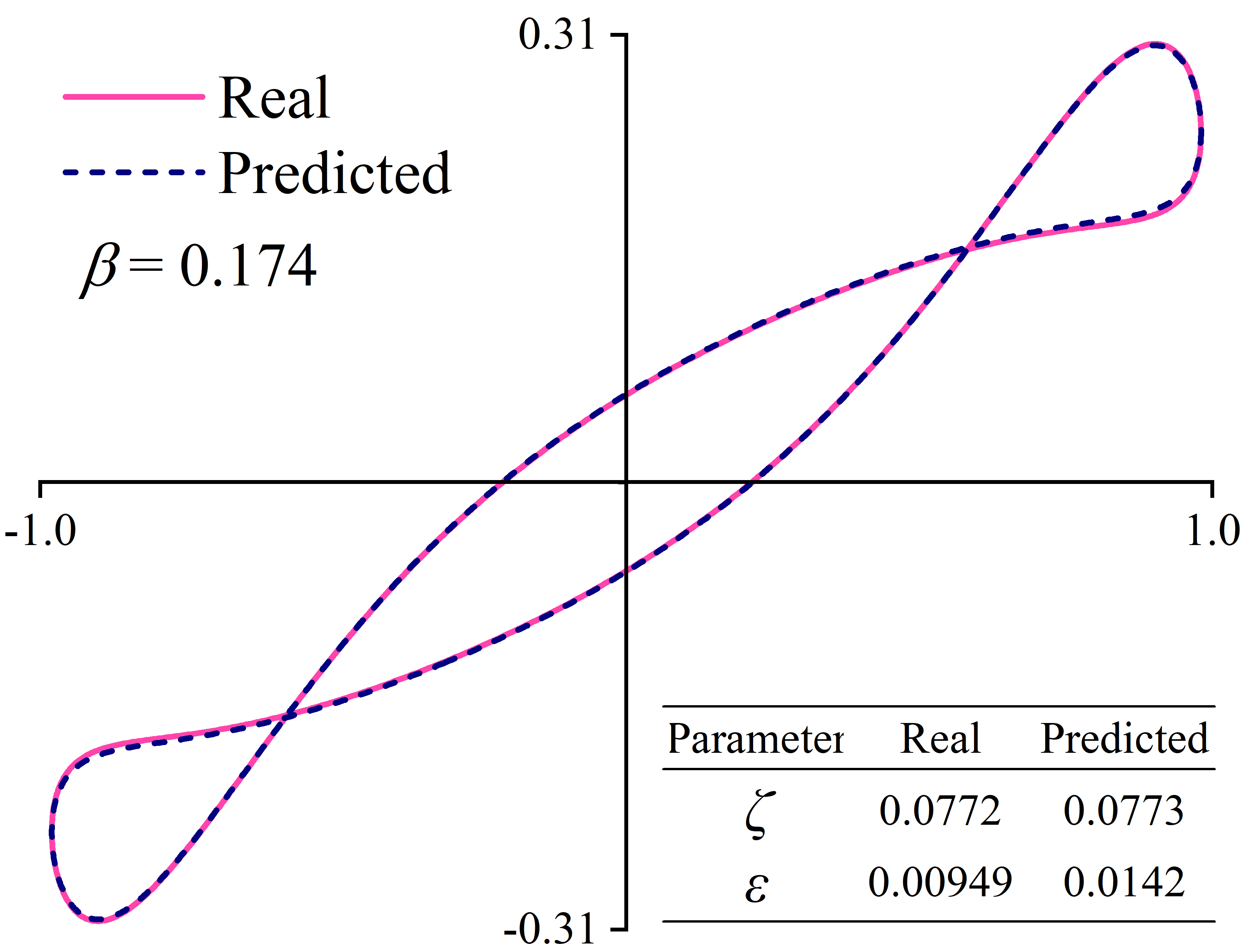}
    \includegraphics[width=0.325\textwidth]{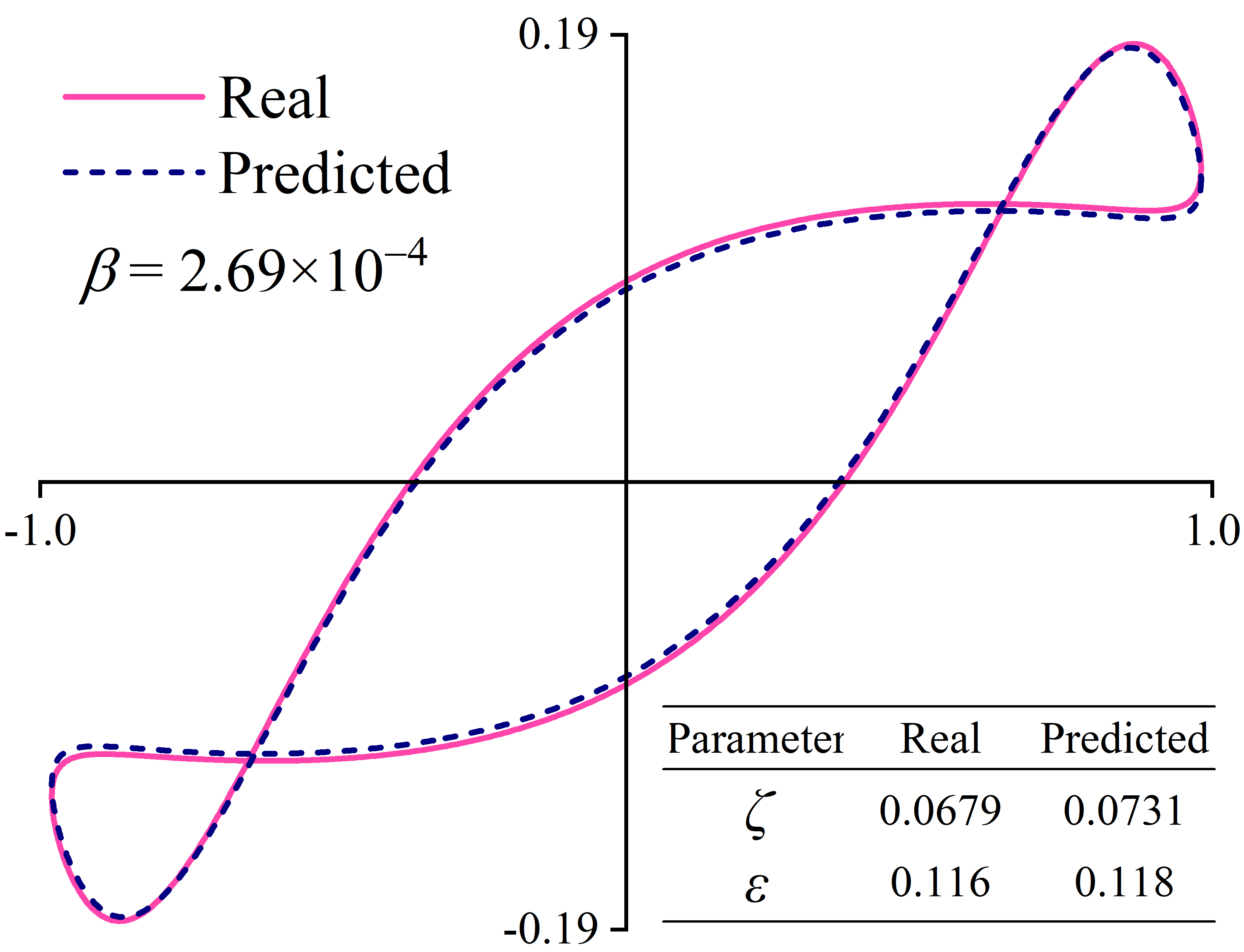}
    \caption{Examples of viscous Lissajous curve (stress \textit{vs} strain rate) predictions for lPTT model with $n_p=16^3$. Note that we have used the real value for $\beta$ rather than the RF predicted value. The value of $\beta$ is given in each plot. Each plot also contains a table with the real and RF predicted model parameters.}
    \label{fig:lissajousexamples}
\end{figure}

In order to test how well the RF predicted parameters predict the rheological behaviour under LAOS, we solve the ODEs for the lPTT model under LAOS (at the same conditions as those used for the training, $De = 1$ and $Wi = 10$) using both the real parameters and the RF predicted parameters. Of course, for perfect predictions of all the model parameters, there is no need to check the predictions of the rheological behaviour since the real and predicted parameterised constitutive models are identical. In Figure \ref{fig:lissajousexamples}, we highlight the accuracy of the predictions by showing the viscous Lissajous curves (stress $\textit{vs}$ strain rate) for 6 random data points in the validation data set. The pink solid lines are generated using the real model parameters, and the blue dashed lines are generated using the RF predicted parameters. Note that the RF predicted parameters are of course predicted from the real Lissajous curves (pink solid line). Here, we have used the real value for $\beta$ rather than the RF predicted value since this parameter can be measured directly using a flow curve measurement. However, the RF prediction for $\beta$ is still highly accurate, as shown in Figure \ref{fig:datasize}, and so the Lissajous curves would still be predicted exceptionally well even with the RF value of $\beta$. The results in Figure \ref{fig:lissajousexamples} show that the RF algorithm is able to predict the correct LAOS behaviour even for drastically different Lissajous curves.

\begin{figure}[h!]
    \centering
    \includegraphics[trim={0 0 2.5cm 0},clip,width=0.325\textwidth]{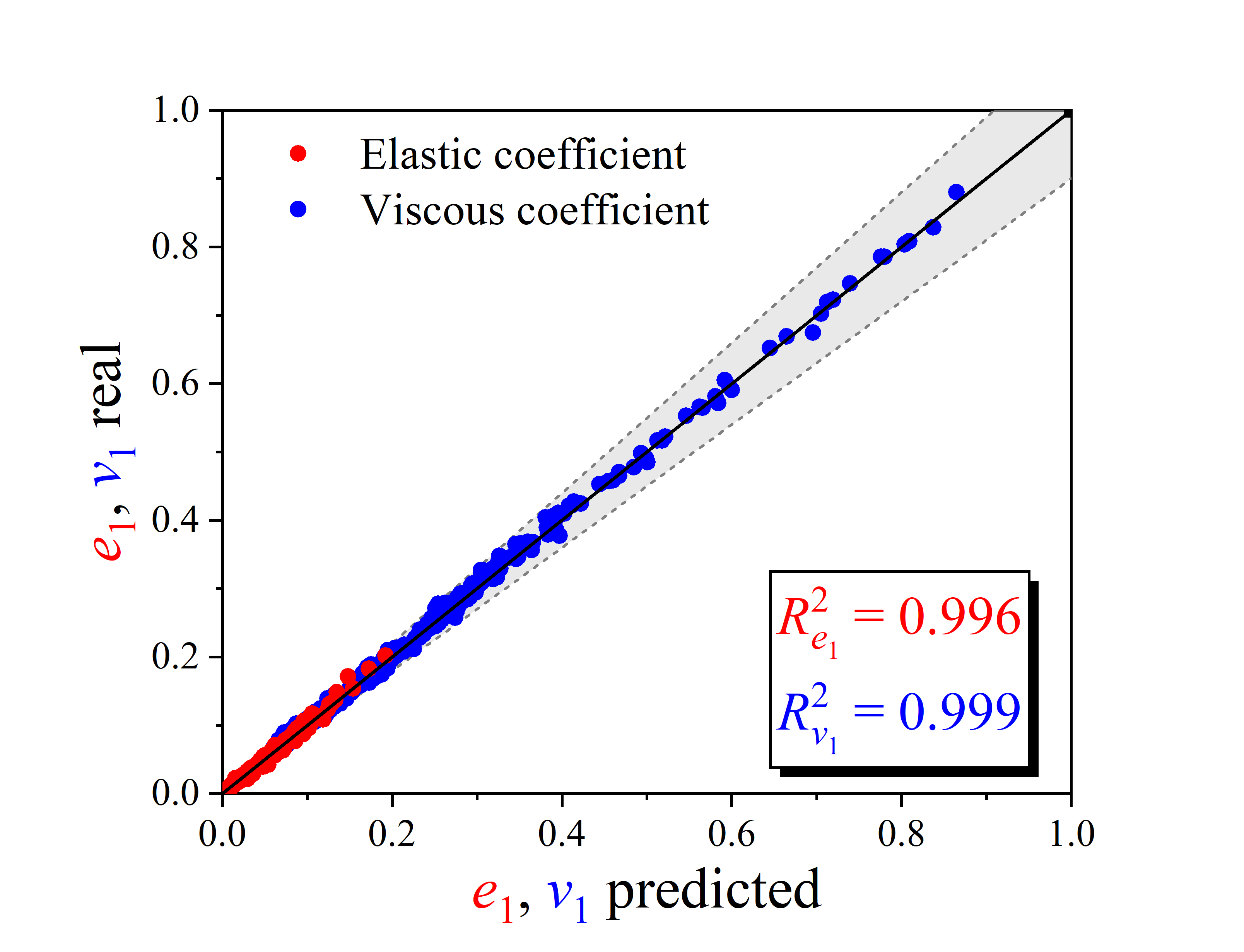}
    \includegraphics[trim={0 0 2.5cm 0},clip,width=0.325\textwidth]{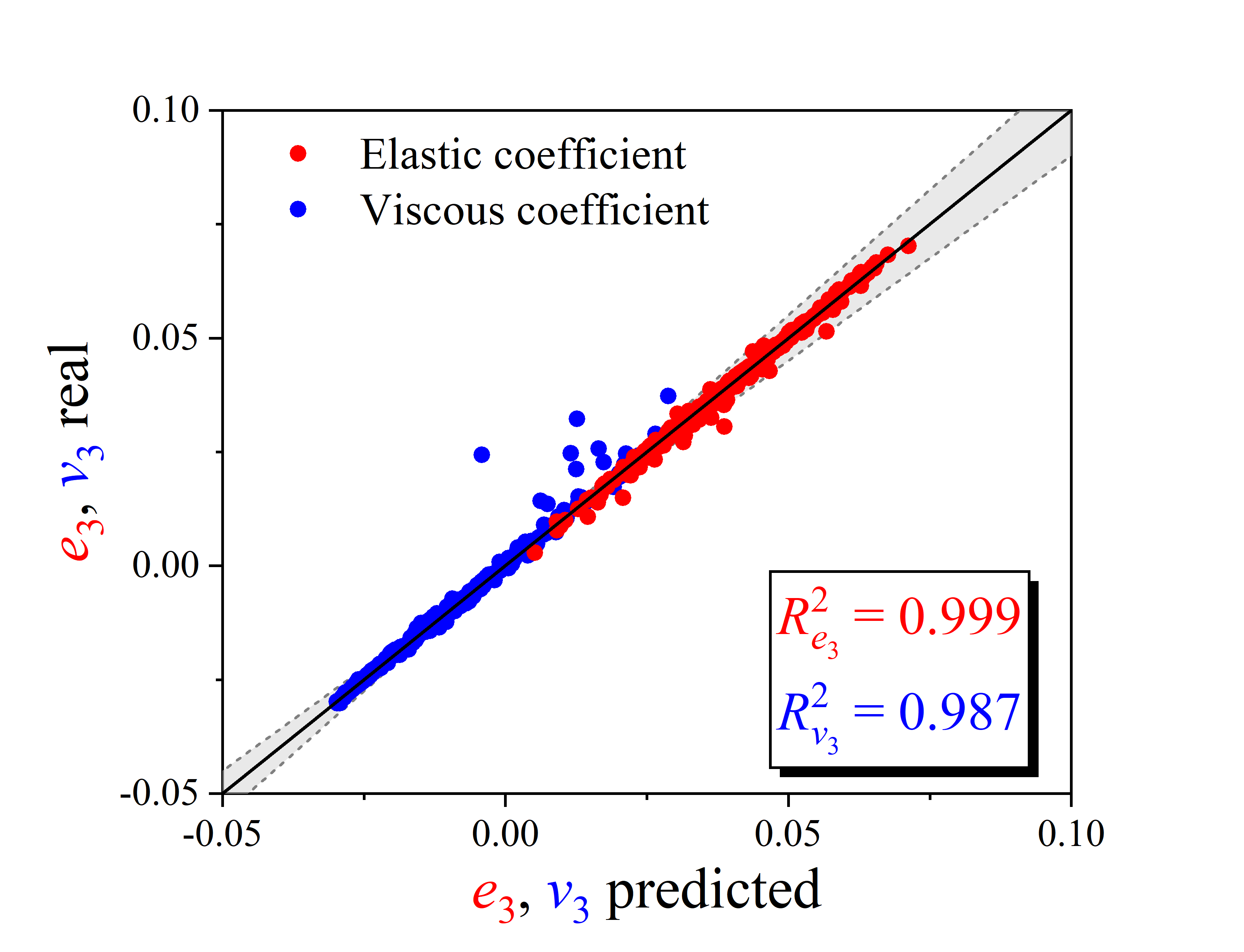}
    \includegraphics[trim={0 0 2.5cm 0},clip,width=0.325\textwidth]{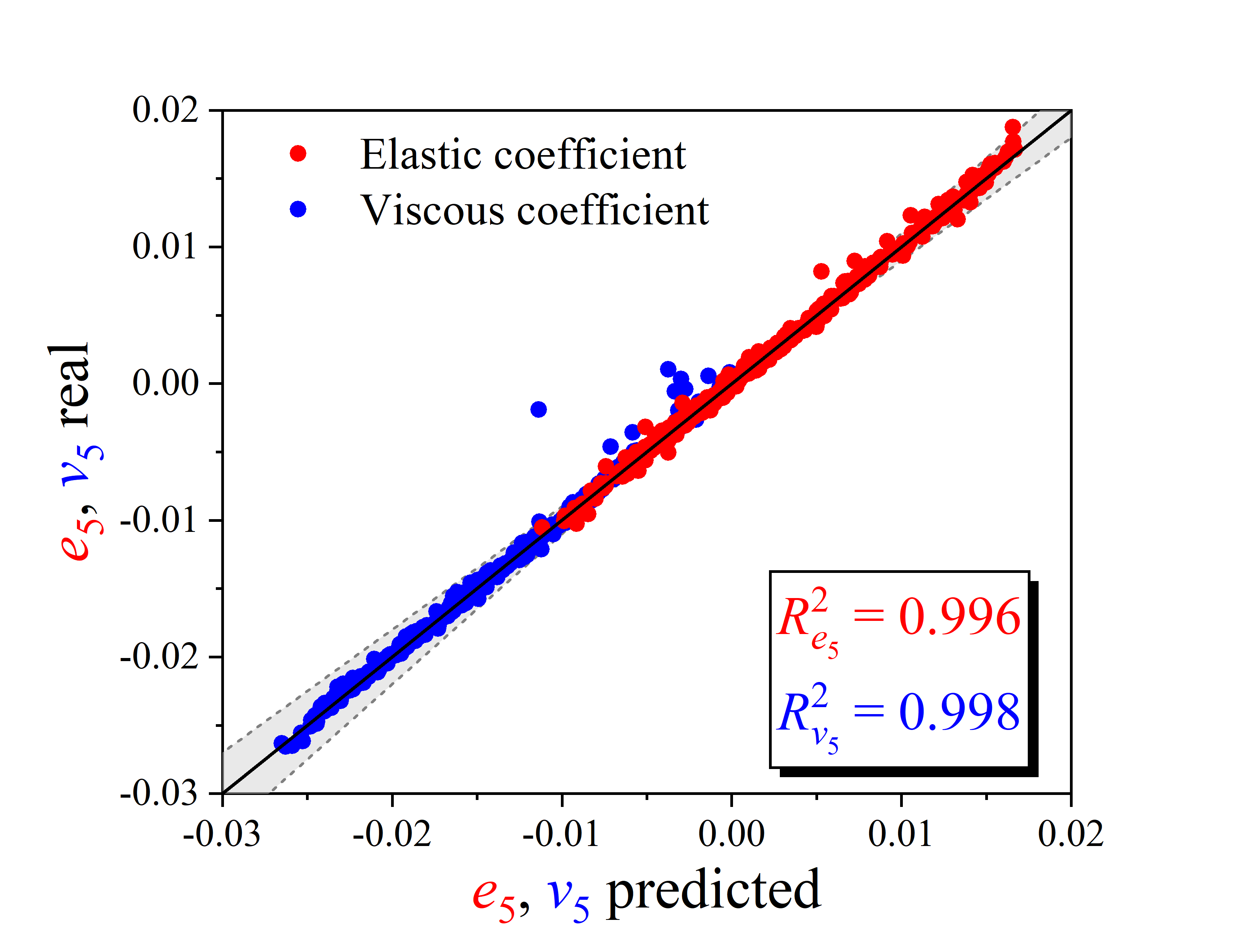}

    \includegraphics[trim={0 0 2.5cm 0},clip,width=0.325\textwidth]{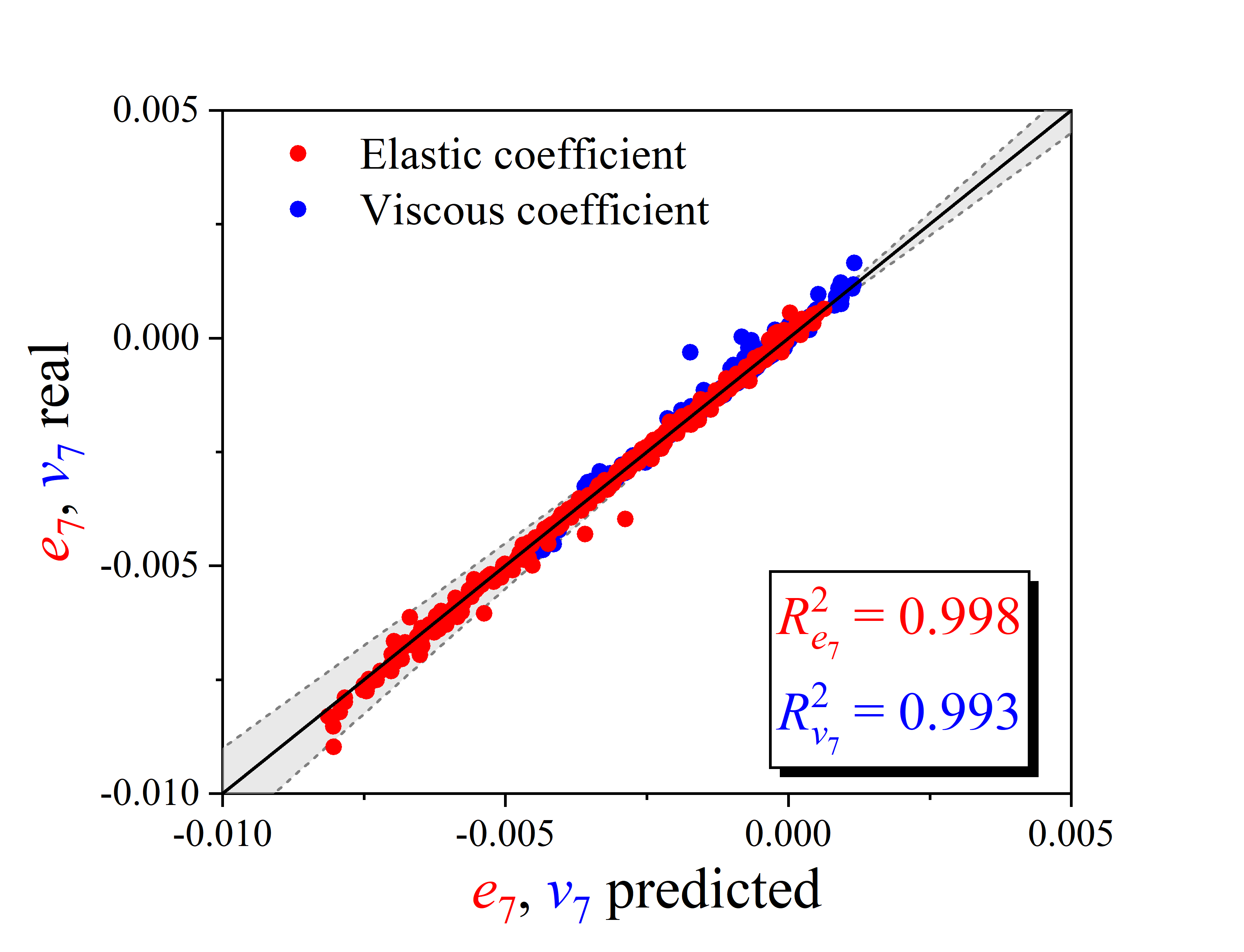}
    \includegraphics[trim={0 0 2.5cm 0},clip,width=0.325\textwidth]{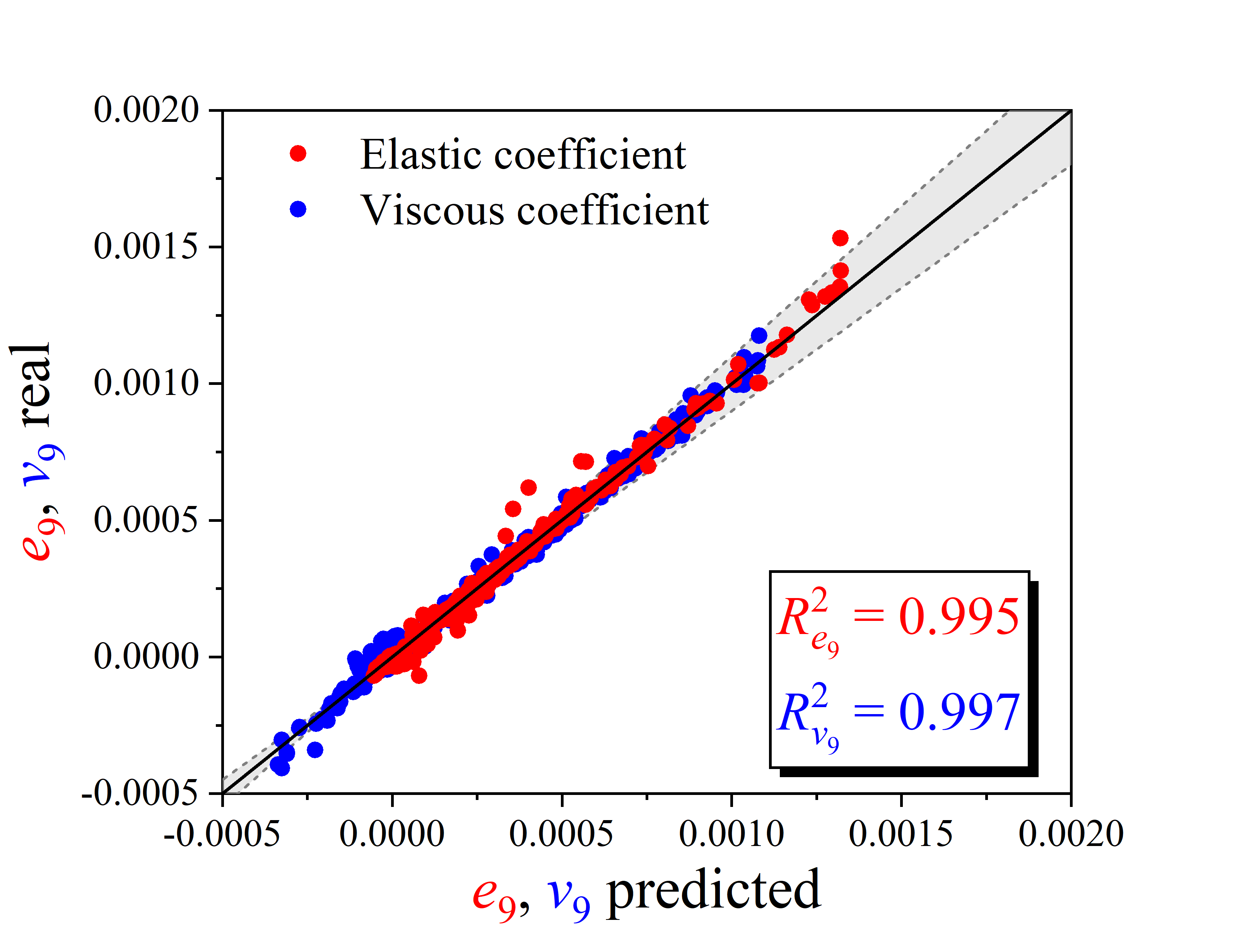}
    \caption{Parity plots (real \textit{vs} predicted) for Chebyshev coefficients for lPTT model with $n_p=16^3$. This data corresponds to the validation data shown in Figure \ref{fig:datasize}.}
    \label{fig:chebparity}
\end{figure}

To quantitatively assess the LAOS predictions for all validation data points, we compare the Chebyshev coefficients between the real and RF predicted Lissajous curves. In Figure \ref{fig:chebparity}, we show the parity plots for all viscous and elastic Chebyshev coefficients for the validation data in Figure \ref{fig:datasize} for $n_p = 16^3$. There is very good agreement between the predicted and real model parameters, as highlighted by the values of $R^2$ displayed in the plots. The RF algorithm is, thus, seemingly able to the map between Chebyshev coefficients and constitutive model parameters with high levels of accuracy. 

\subsubsection{ePTT model}

\begin{figure}[h!]
    \centering
    \includegraphics[trim={0.5cm 0 0.5cm 0},clip,width=\textwidth]{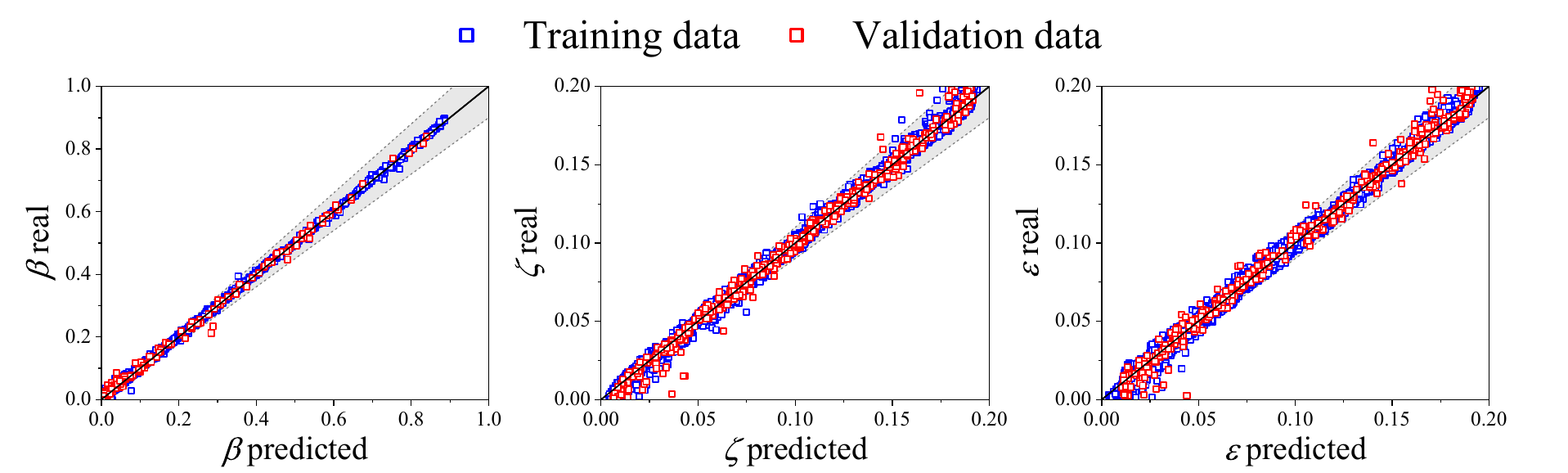}
    \caption{Parity plots for training and validation data for the ePTT model parameters. RF algorithm was trained at $De = 1$ and $Wi = 10$. RF hyperparameters: $d_{\mathrm{max}} = 10$ and $N=100$. $n_p = 16^3$.}
    \label{fig:ePTTParamFits}
\end{figure}

\begin{figure}[h!]
    \centering
    \includegraphics[trim={0 0 2.5cm 0},clip,width=0.325\textwidth]{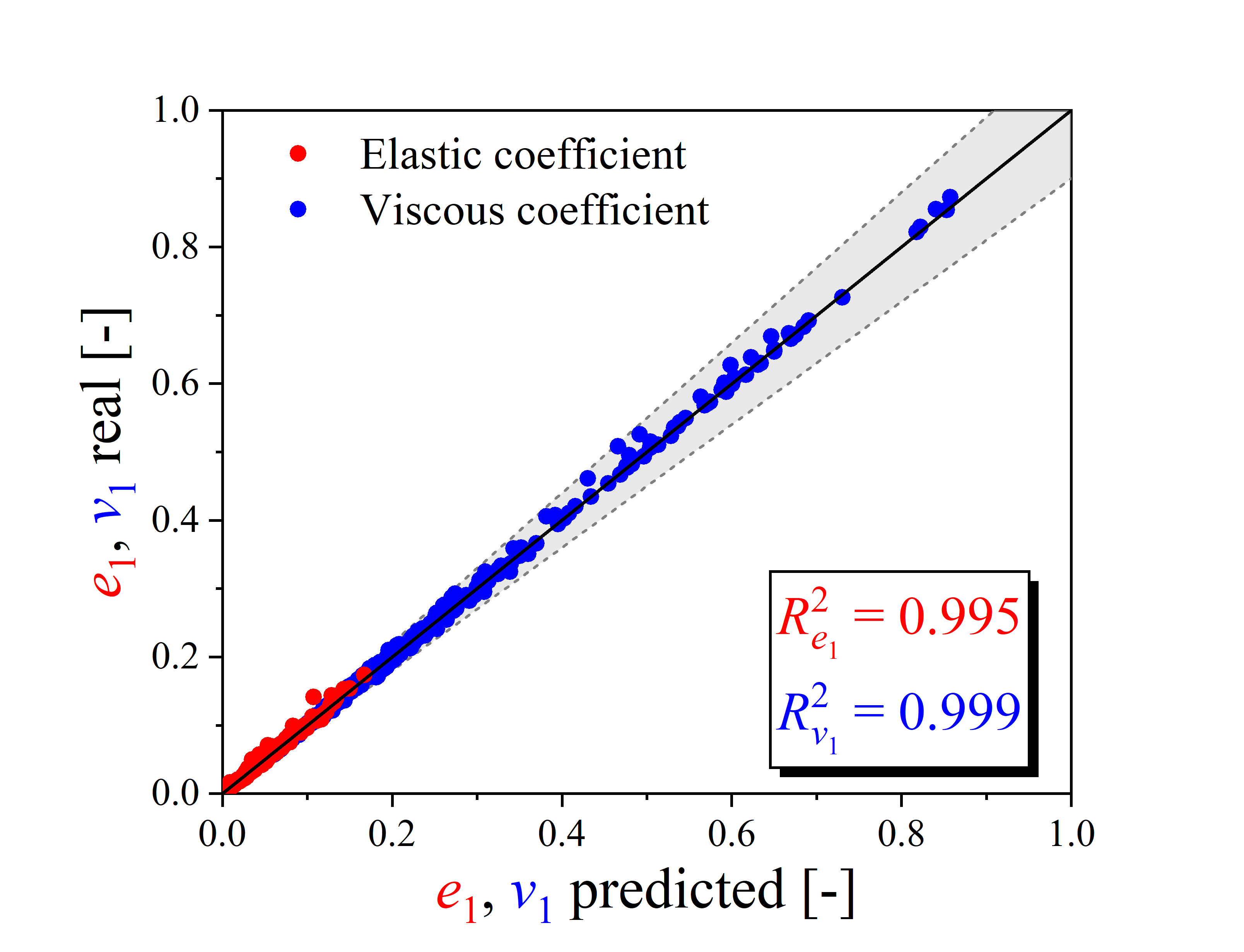}
    \includegraphics[trim={0 0 2.5cm 0},clip,width=0.325\textwidth]{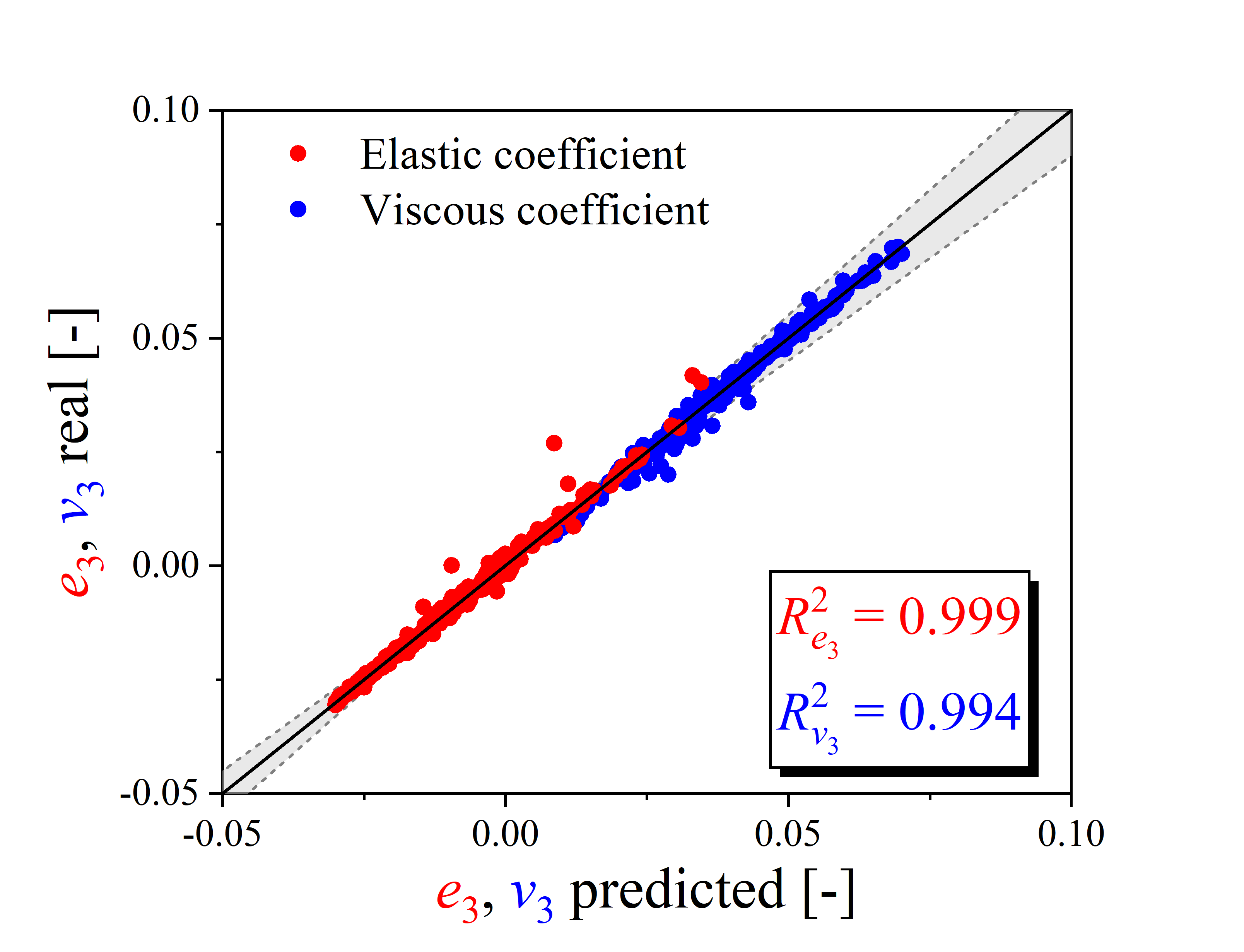}
    \includegraphics[trim={0 0 2.5cm 0},clip,width=0.325\textwidth]{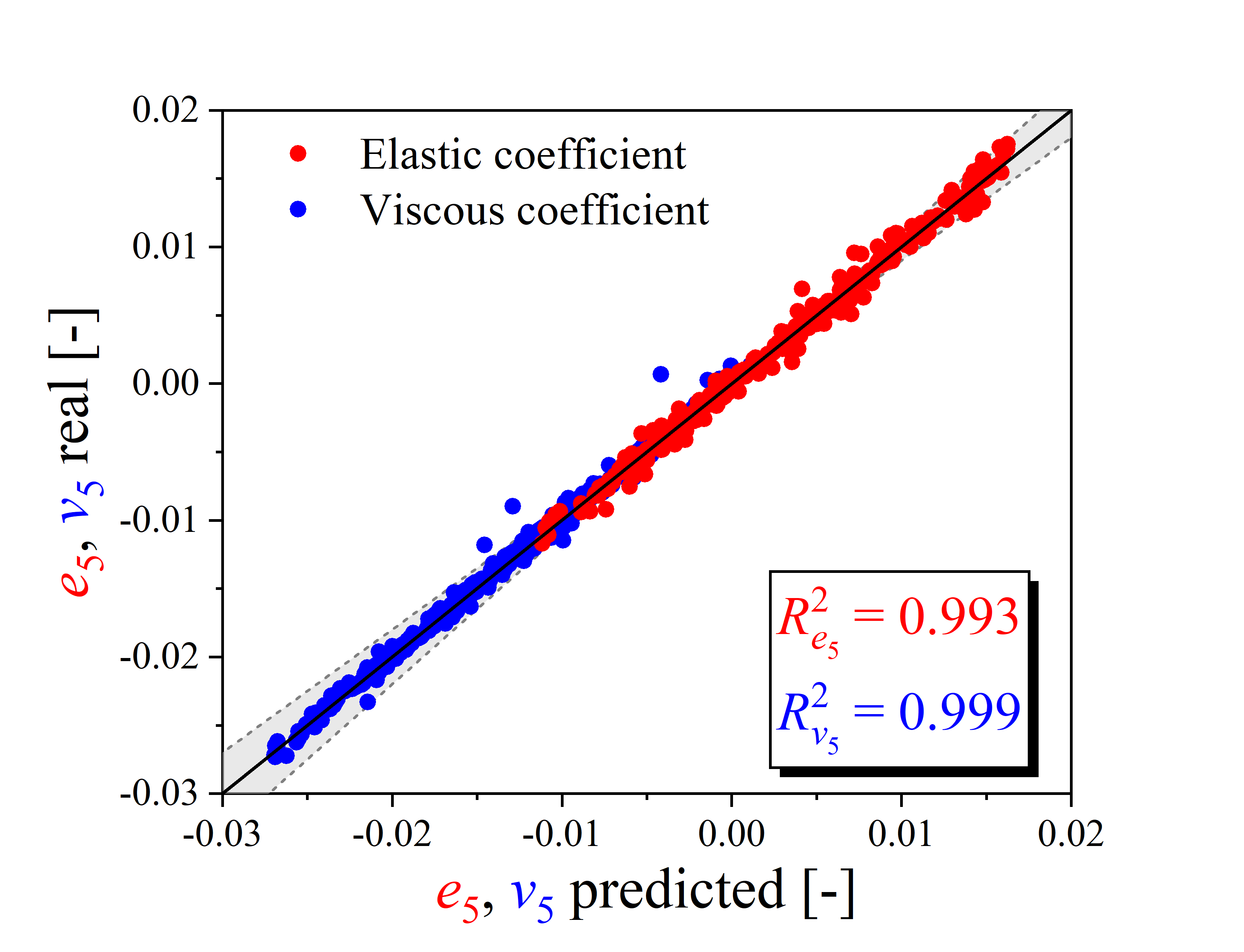}

    \includegraphics[trim={0 0 2.5cm 0},clip,width=0.325\textwidth]{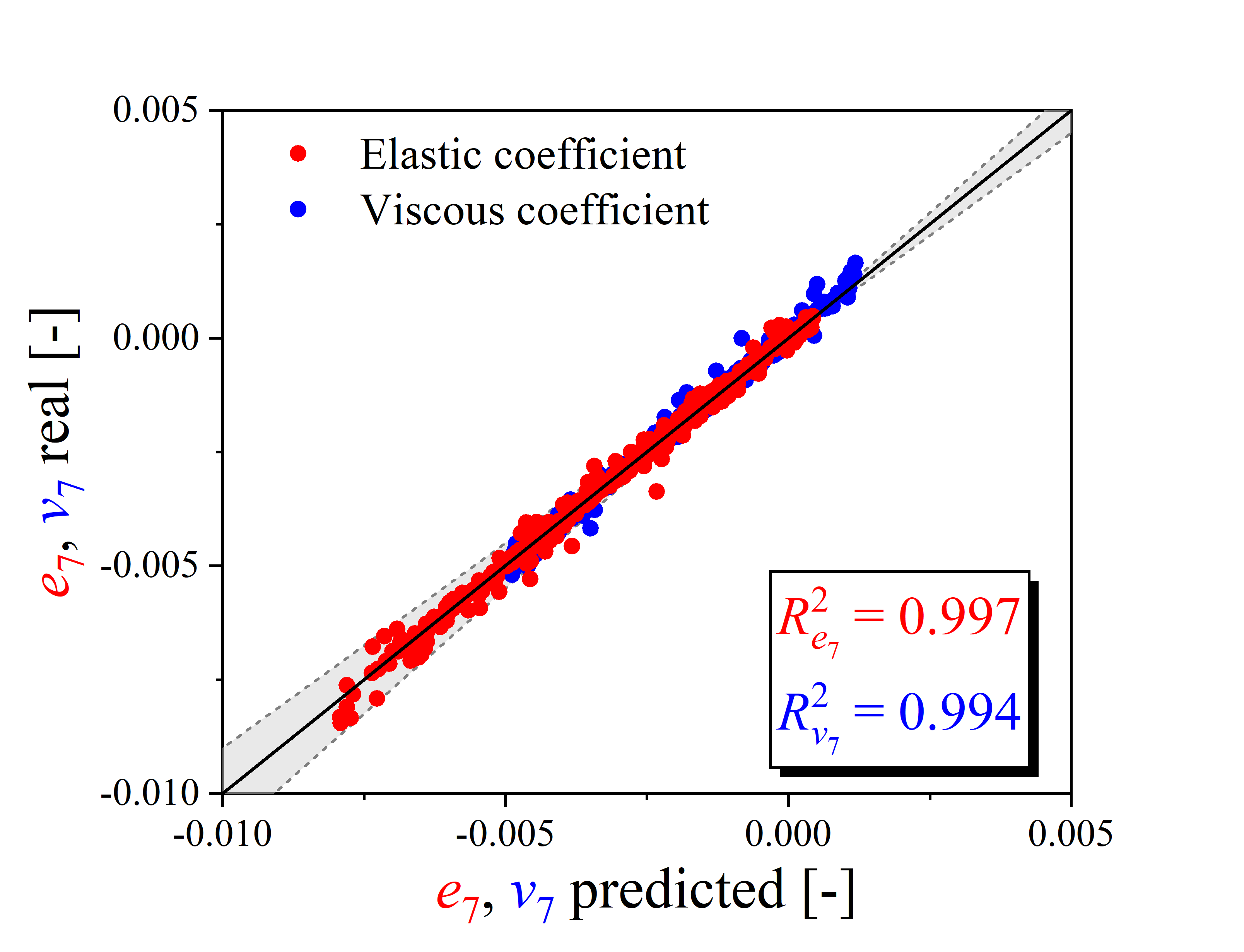}
    \includegraphics[trim={0 0 2.5cm 0},clip,width=0.325\textwidth]{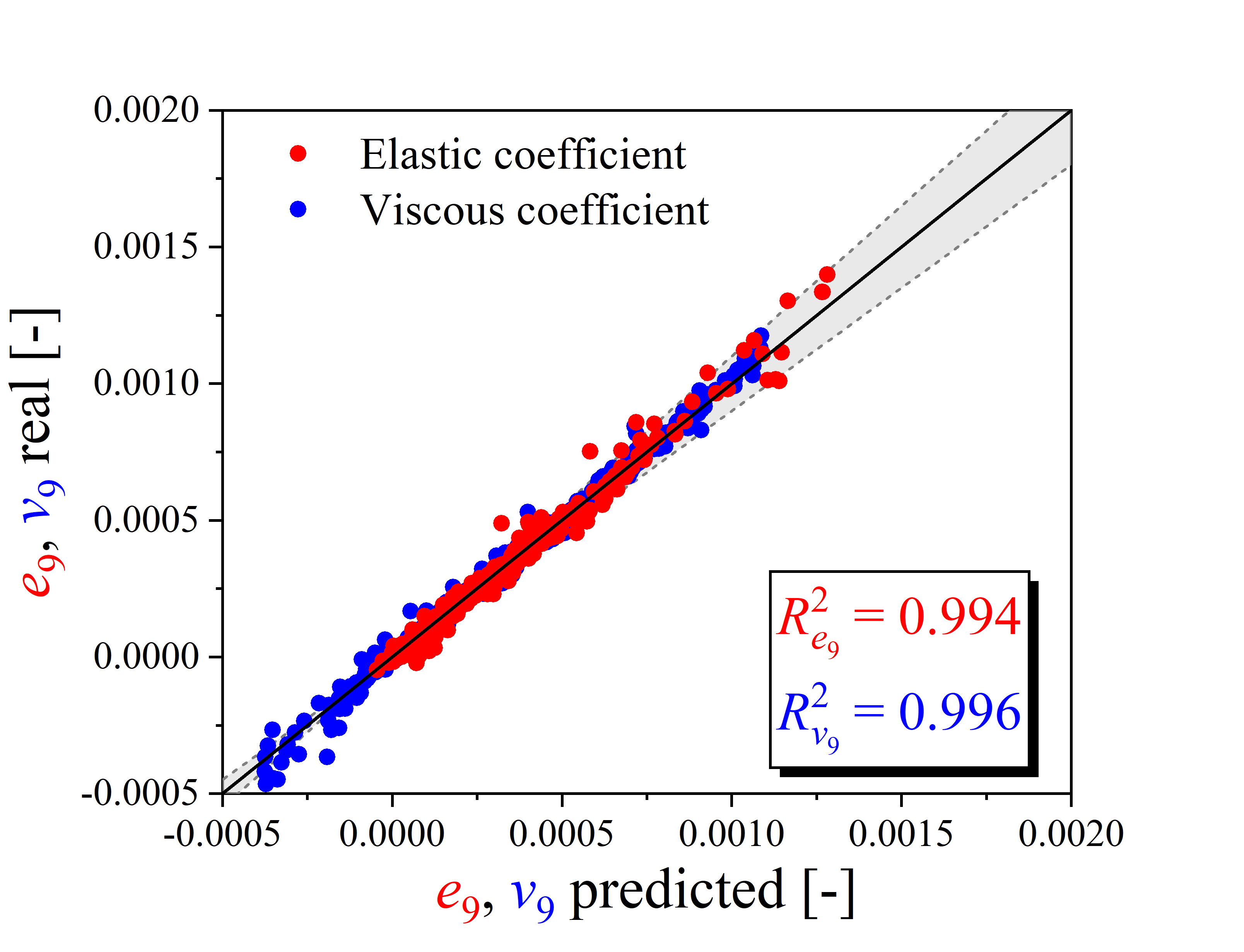}
    \caption{Parity plots (real \textit{vs} predicted) for Chebyshev coefficients for ePTT model with $n_p=16^3$. This data corresponds to the validation data shown in Figure \ref{fig:ePTTParamFits}.}
    \label{fig:ePTTchebparity}
\end{figure}

Next, we turn our attention to the ePTT model. It was found, again, that using hyperparameters of $d_{\mathrm{max}} = 10$ and $N=100$ provided optimal performance of the RF algorithm. Figure \ref{fig:ePTTParamFits} shows the model parameter parity plots for the training and validation data for the ePTT model using $n_p=16^3$. Again, the RF algorithm is able to make highly accurate predictions of the model parameters from the inputted LAOS data. This is reinforced by Figure \ref{fig:ePTTchebparity}, which visualizes the parity plot between the predicted and actual chebyshev coefficients in validation. 

\begin{figure}[h!]
    \centering
    \includegraphics[width=\textwidth]{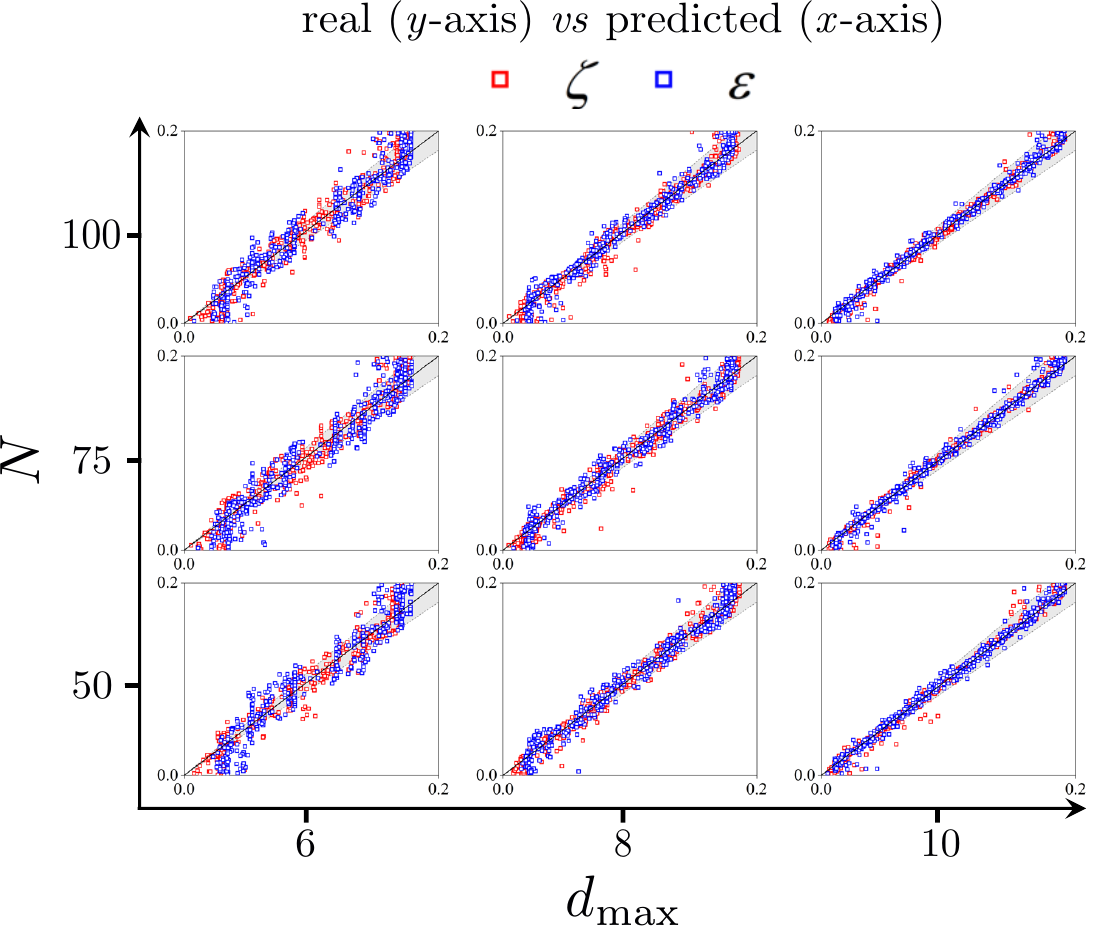}
    \caption{Parity plots for $\zeta$ and $\varepsilon$ predictions for ePTT model with various RF hyper-parameters. This data is validation data (i.e. not used for training). $n_p = 16^3$.}
    \label{fig:ePTTGridSearch}
\end{figure}

In Figure \ref{fig:ePTTGridSearch} we show the effect of the RF algorithm hyperparameters ($d_{\mathrm{max}}$ and $N$) on the predictions of $\zeta$ and $\varepsilon$. The data here is all validation data (i.e. not used for training). $n_p = 16^3$ in all cases. The predictive accuracy does seem to be somewhat sensitive to the hyperparameters used, which is completely expected and is frequently observed in other applications of RF. The accuracy of the predictions of both model parameters is increased as $d_{\mathrm{max}}$ is increased (increasing beyond 10 did not yield any further improvement in the predictions). The number of estimators $N$ does not seem to have much impact on the accuracy between 50 and 100, indicating $N>50$ is sufficient for this application of RF. It is often found for RF that the predictive error decreases initially as $N$ is increased and then converges to a plateau at a particular value of $N$. Evidently, $N=50$ is beyond this value. Overall, this highlights that the RF algorithm is a very robust regression method for this application of constitutive model parameterisation, and could potentially be useful in other areas of rheology.

\begin{table}[]
\centering
\begin{tabularx}{\textwidth}{Y Y Y| Y Y Y}
        \toprule
         \multicolumn{3}{c|}{Training} & \multicolumn{3}{c}{Validation} \\\hline
                 $\beta$ & $\zeta$ & $\varepsilon$ & $\beta$ & $\zeta$ & $\varepsilon$     \\\hline 
         0.9993 & 0.9992  & 0.9989  & 0.9982 & 0.9976 & 0.9974 \\\bottomrule
\end{tabularx}
\caption{\label{tab:ePTTRsquared}Values of $R^2$ for each of the parity plots in Figure \ref{fig:ePTTParamFits}.}
\end{table}

\begin{table}[]
\centering
\begin{tabularx}{\textwidth}{Y| Y Y Y| Y Y Y}
        \toprule
        \multirow{2}{*}{$d_{\mathrm{max}}$} & \multicolumn{3}{c|}{$\zeta$ } & \multicolumn{3}{c}{$\varepsilon$} \\\cline{2-7} 
                & $N = 50$ & $N = 75$ & $N = 100$ & $N = 50$ & $N = 75$ & $N = 100$     \\\hline 
        $6$ & 0.9877 & 0.9862   & 0.9844  & 0.9809 & 0.9828 & 0.9858 \\ 
        $8$ & 0.9949        & 0.9931        & 0.9929      & 0.9936 & 0.9930 & 0.9928   \\
        $10$ & 0.9972        & 0.9976        & 0.9976        & 0.9975 & 0.9970 & 0.9974   \\\bottomrule
\end{tabularx}
\caption{\label{tab:GridSeachRsquared}Values of $R^2$ for each of the parity plots in Figure \ref{fig:ePTTGridSearch}.}
\end{table}

\subsection{RoLiE-Poly model}

In this subsection, we present the results for the RoLiE-Poly model. We have observed already that the RF algorithm is very capable at predicting the lPTT and ePTT model parameters from the Chebyshev spectrum of LAOS responses. We now test the developed workflow/methodology for the more physically complex RoLiE-Poly model. Note that $Wi$ and $De$ in Equation \eqref{eq:roliepolymodel} are based on the reptation relaxation time, and not on the Rouse relaxation time. If instead we base $Wi$ and $De$ on the Rouse relaxation time, the value of $\lambda_R$ would need to be known for the material beforehand. Basing $Wi$ and $De$ on the reptation relaxation time $\lambda$ simplifies matters as $\lambda$ can be determined fairly easily from a SAOS measurement. We train the RF algorithm for the RoLiE-Poly model under LAOS at $De = 1$ and $Wi = 50$. A larger value of $Wi$ is used for this model due to the fact the effects of Rouse relaxation (and hence the sensitivity of the Chebyshev spectra on the parameter $z$) are only significant provided $Wi$ is large enough relative to $z$ \cite{Reis2013}. In the limit that $z\rightarrow 0$ (see Equation \eqref{eq:roliepolymodelintro}, the RoLiE-Poly model converges to the quasi-linear Oldroyd-B model.

\begin{figure}[h!]
    \centering
    \includegraphics[trim={0.2cm 0 2.8cm 0},clip,width=0.45\textwidth]{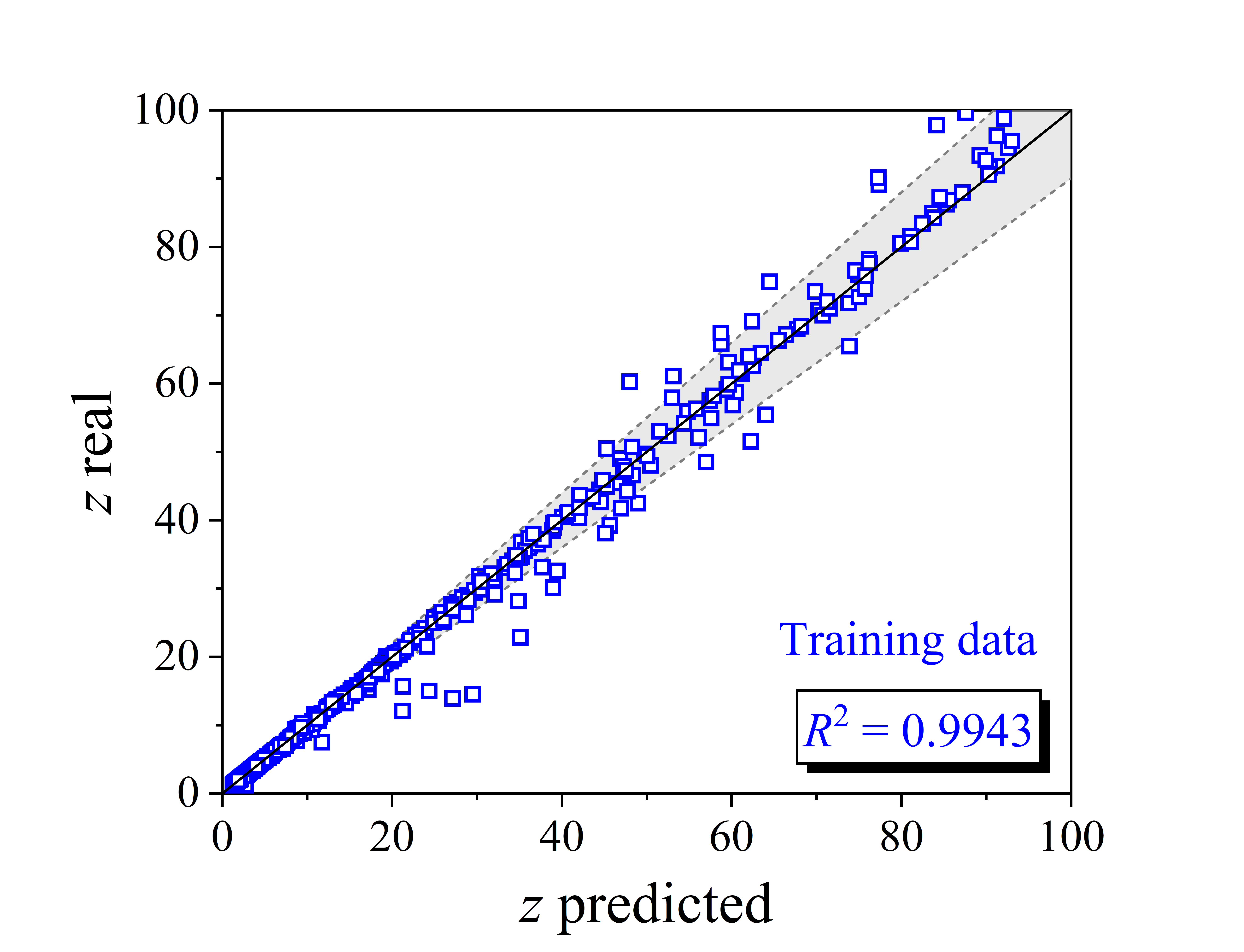}
    \includegraphics[trim={0.2cm 0 2.8cm 0},clip,width=0.45\textwidth]{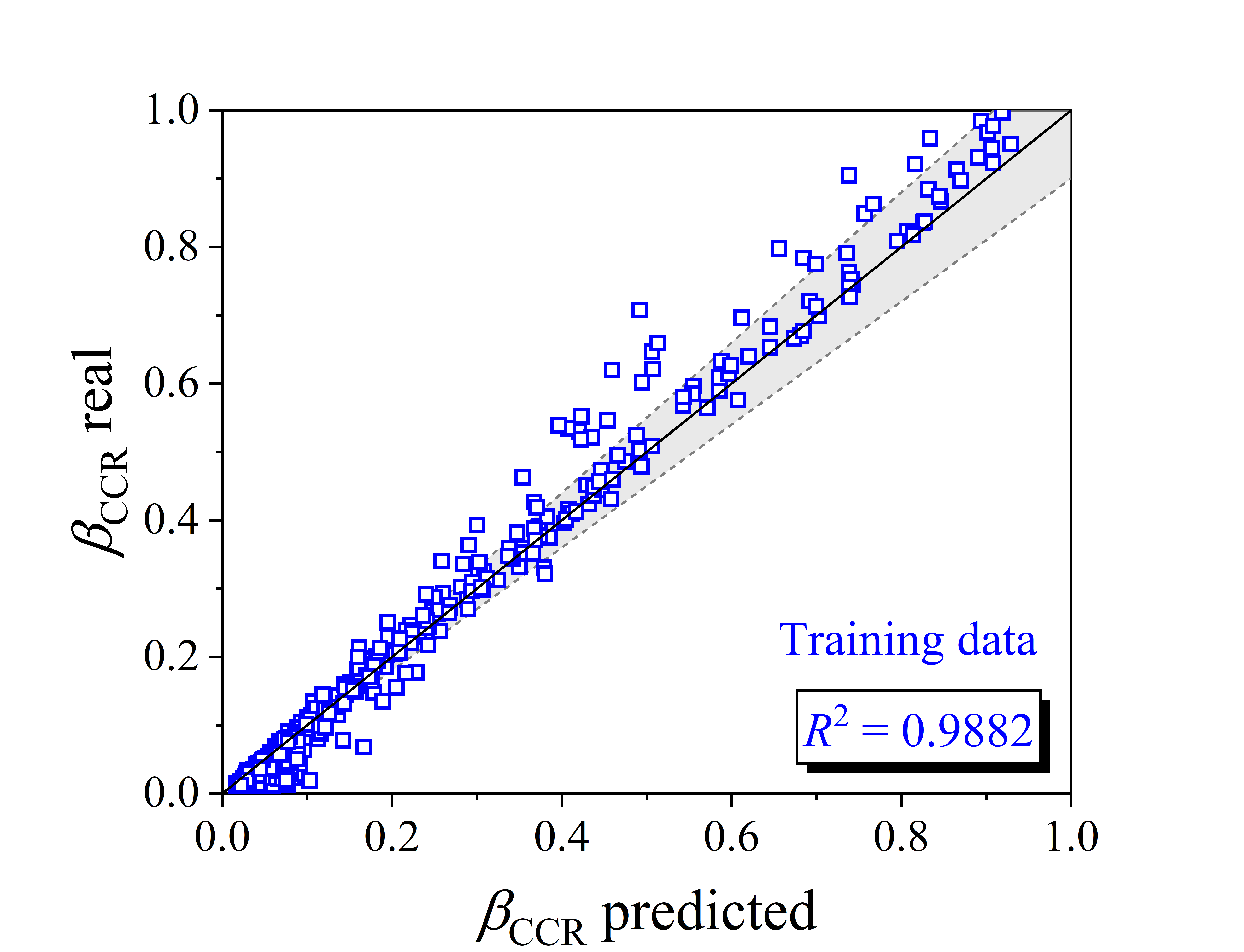}
    \caption{Parity plots for training data for the RoLiE-Poly model parameters. RF algorithm was trained at $De = 1$ and $Wi = 50$. RF hyperparameters: $d_{\mathrm{max}} = 20$ and $N=200$. $n_p = 8^3$.}
    \label{fig:RPTrainingFits}
\end{figure}

\begin{figure}[h!]
    \centering
    \includegraphics[trim={0.2cm 0 2.8cm 0},clip,width=0.45\textwidth]{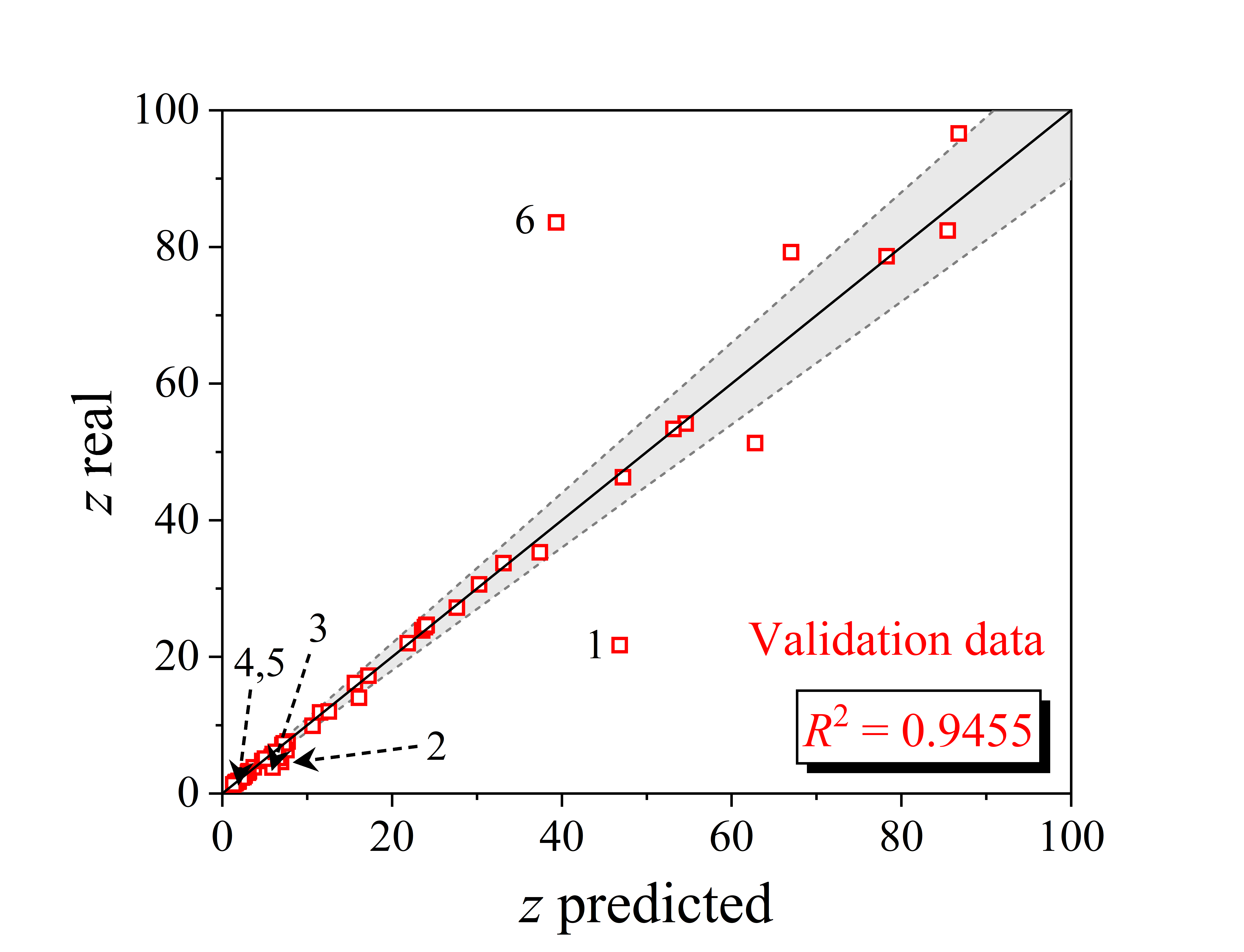}
    \includegraphics[trim={0.2cm 0 2.8cm 0},clip,width=0.45\textwidth]{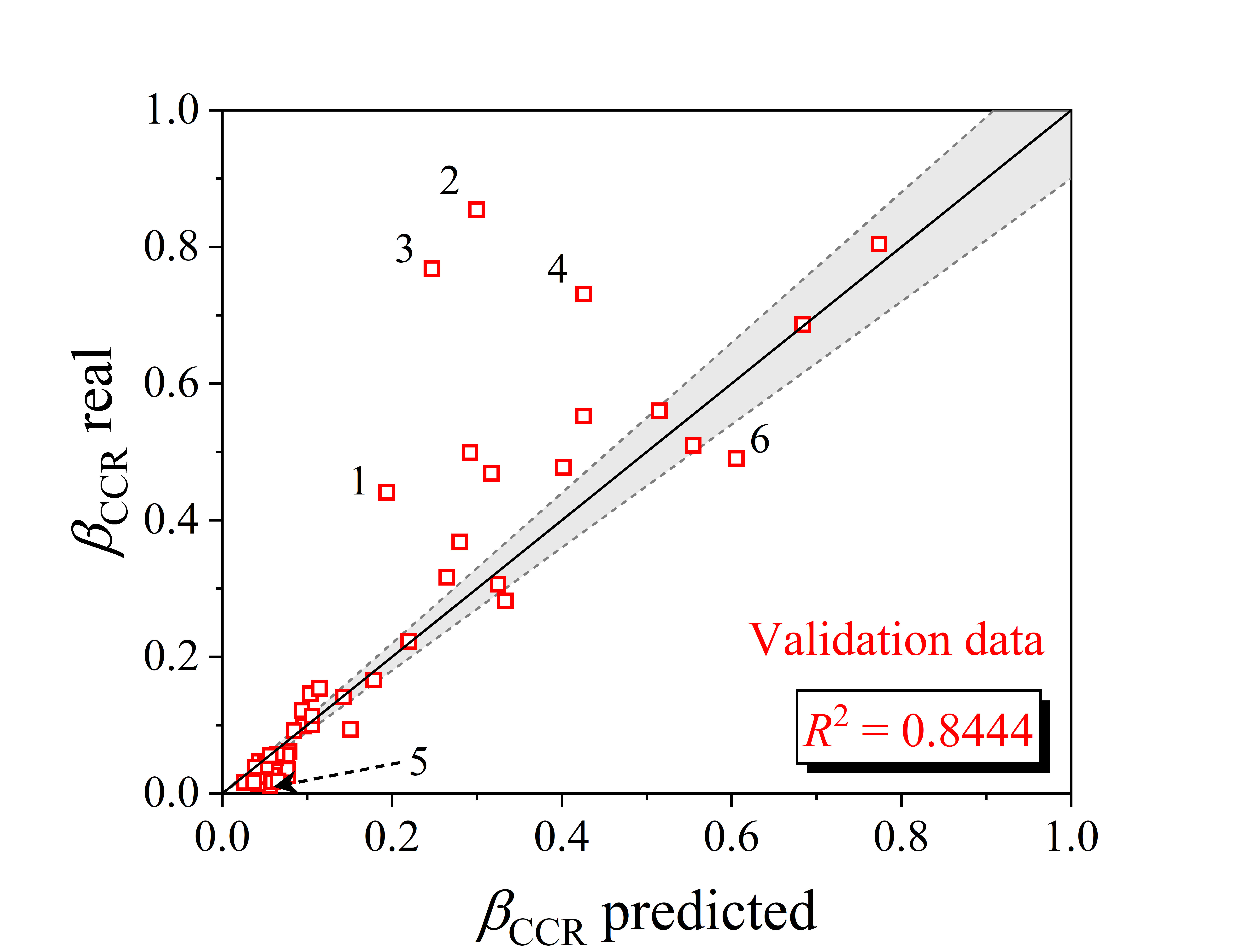}
    \caption{Parity plots for validation data for the RoLiE-Poly model parameters. RF algorithm was trained at $De = 1$ and $Wi = 50$. RF hyperparameters: $d_{\mathrm{max}} = 20$ and $N=200$. $n_p = 8^3$.}
    \label{fig:RPValidationFits}
\end{figure}

\begin{figure}[h!]
    \centering
    \includegraphics[trim={0 0 2.5cm 0},clip,width=0.325\textwidth]{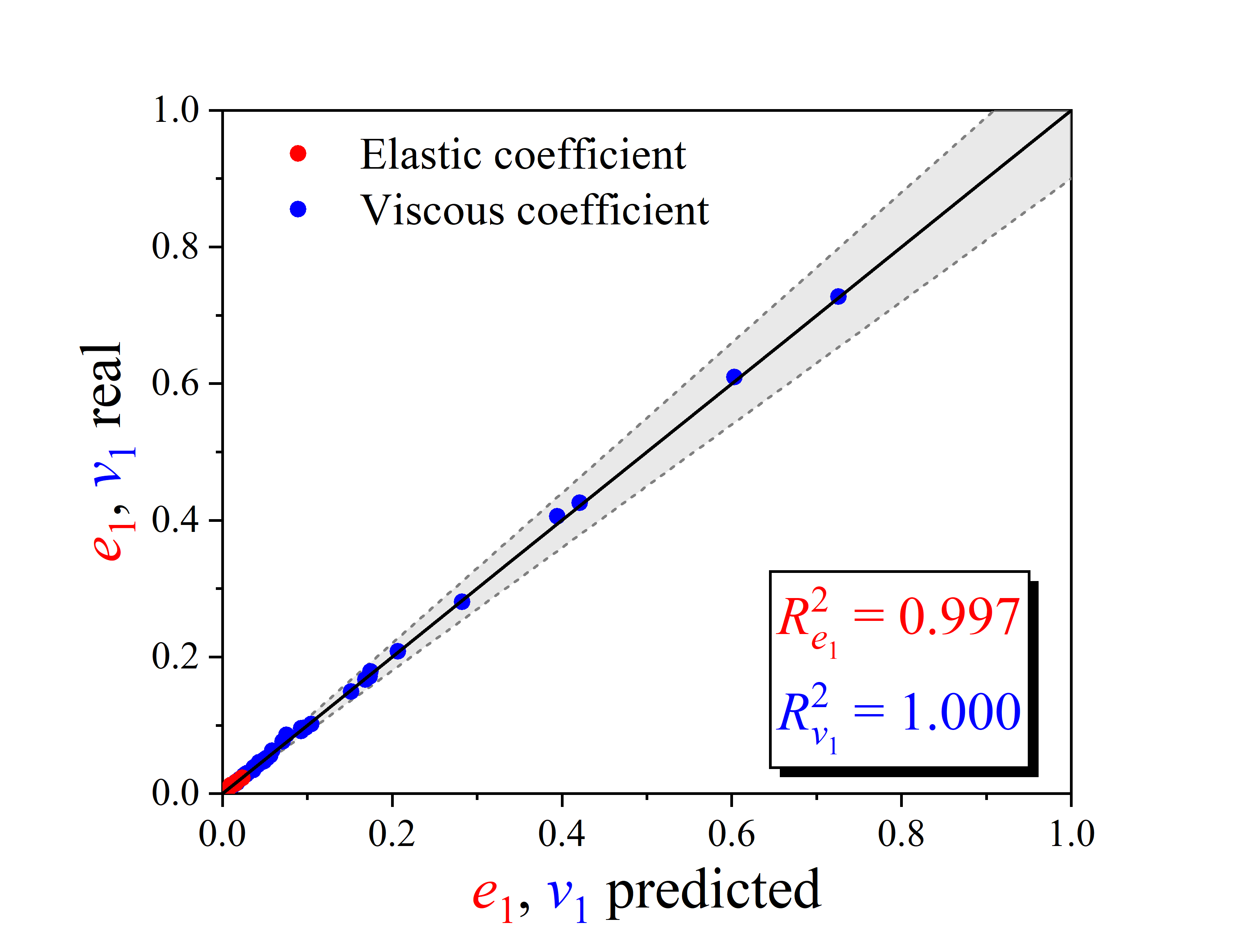}
    \includegraphics[trim={0 0 2.5cm 0},clip,width=0.325\textwidth]{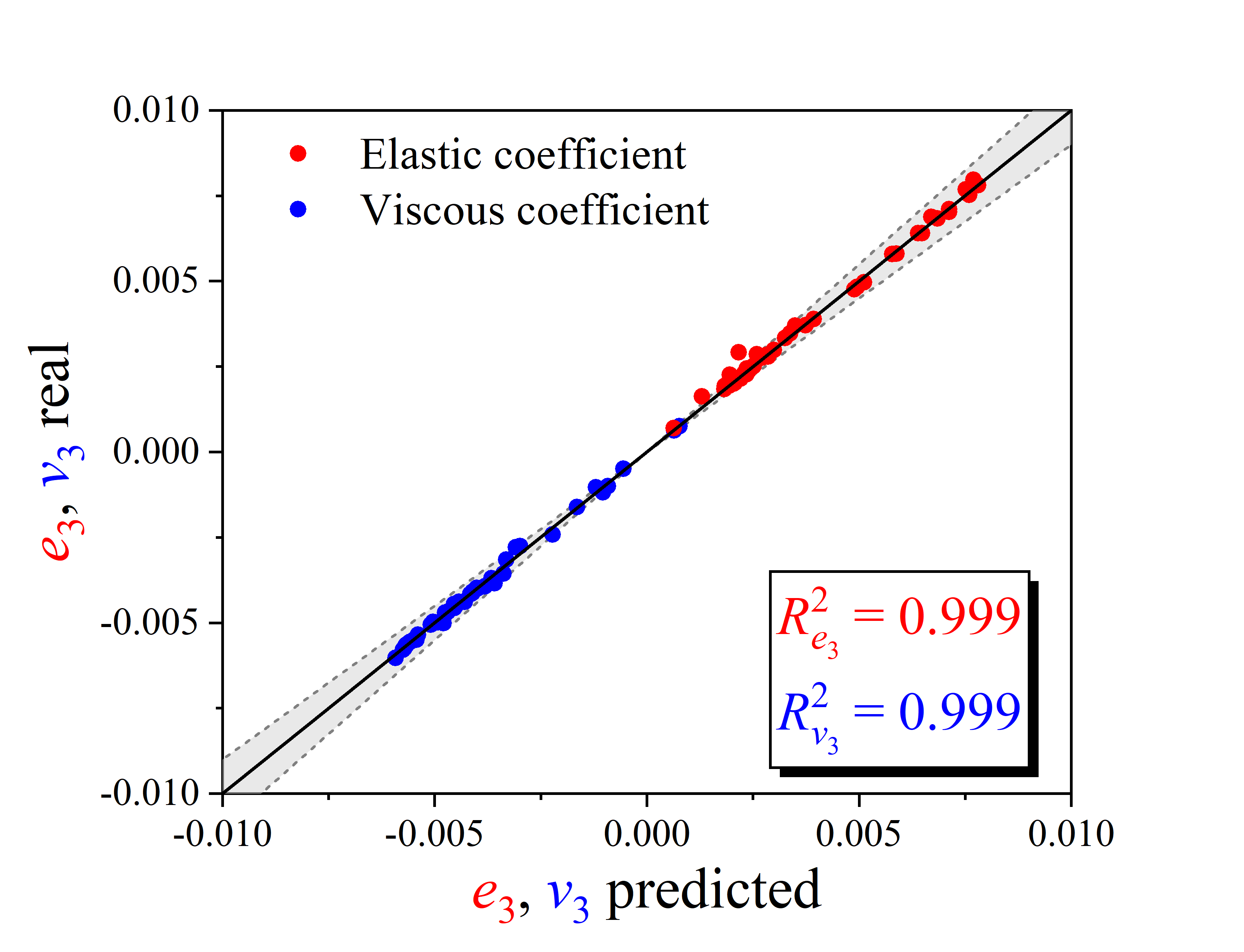}
    \includegraphics[trim={0 0 2.5cm 0},clip,width=0.325\textwidth]{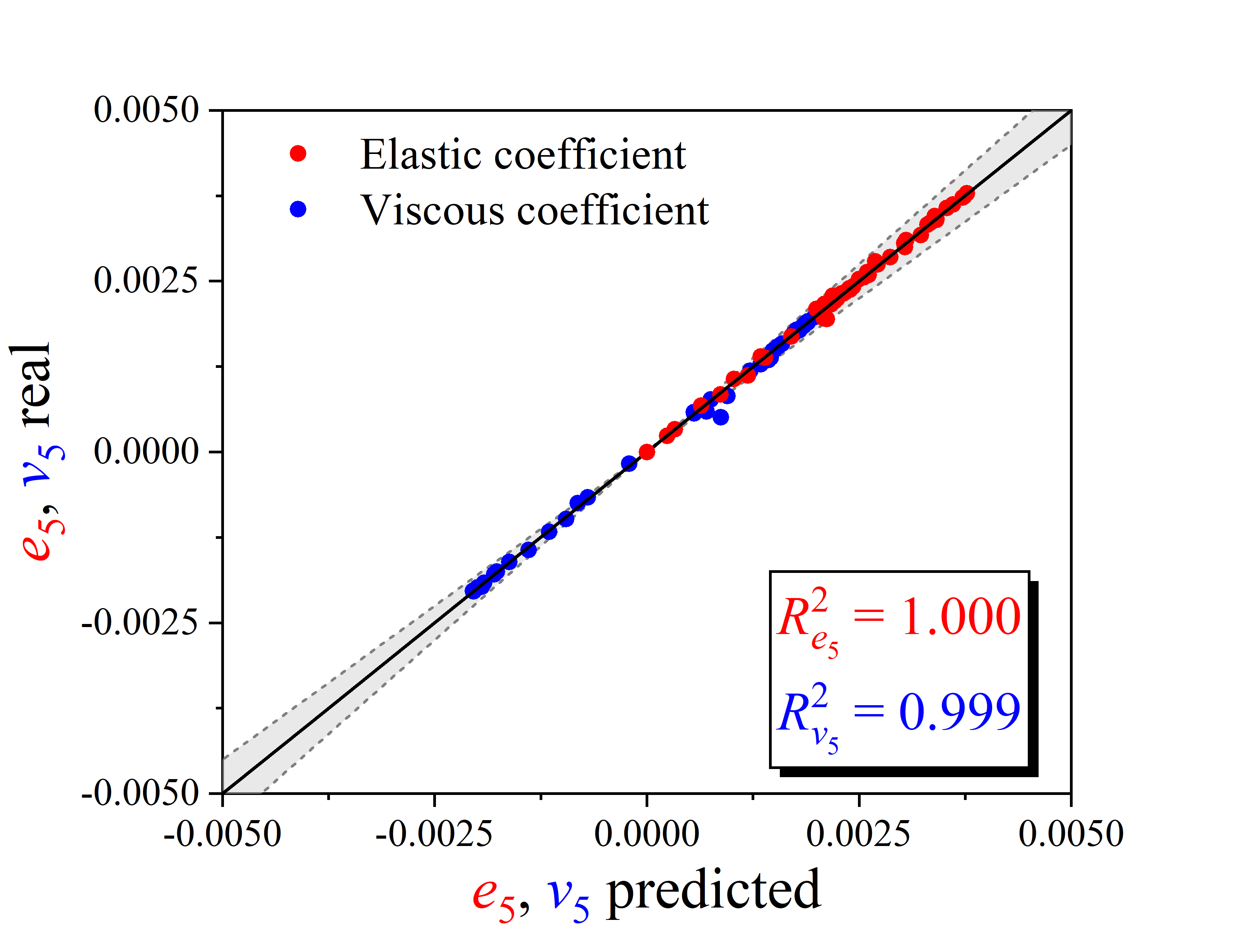}

    \includegraphics[trim={0 0 2.5cm 0},clip,width=0.325\textwidth]{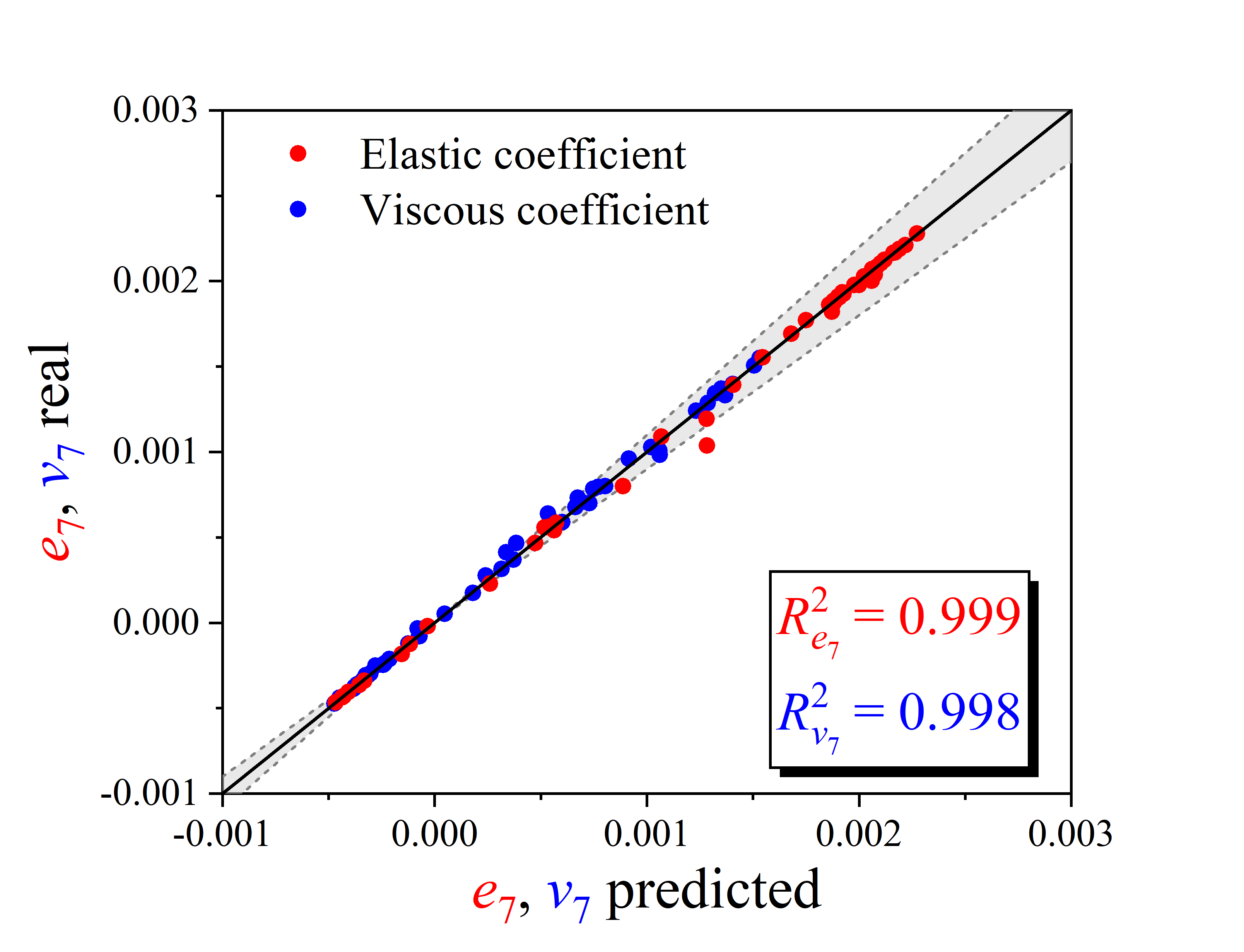}
    \includegraphics[trim={0 0 2.5cm 0},clip,width=0.325\textwidth]{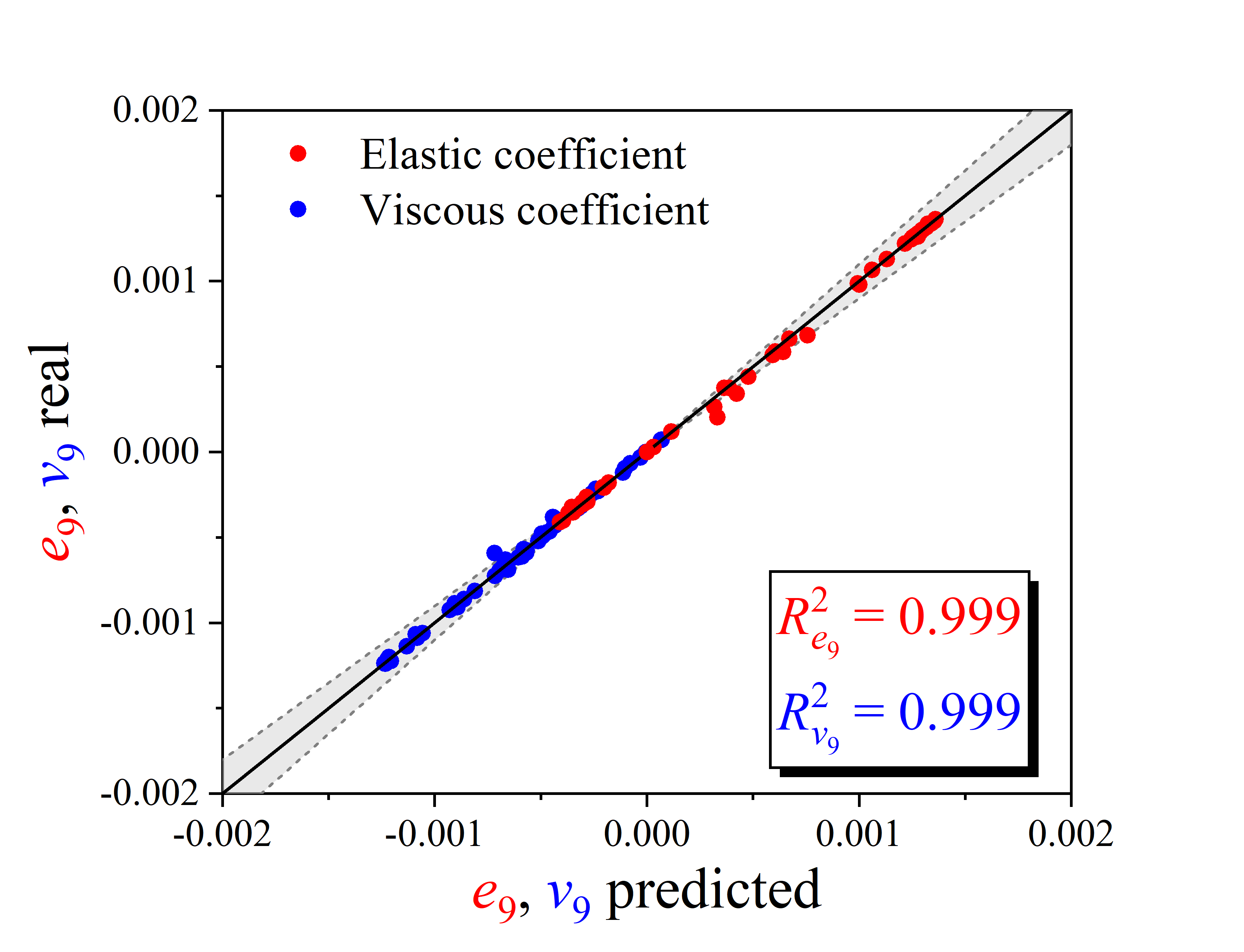}
    \caption{Parity plots for the first 5 Chebyshev coefficients for the RoLiE-Poly model under LAOS at $De = 1$ and $Wi = 50$ (i.e. the same conditions as that used for the training). Note how the prediction of the shear rheology is highly accurate even though the prediction of model parameters is poor in some cases.}
    \label{fig:RPchebparity}
\end{figure}

Figures \ref{fig:RPTrainingFits} and \ref{fig:RPValidationFits} show the model parameter parity plots for the training and validation data, respectively, for the RoLiE-Poly model. Here, we used $n_p = 8^3$. We also found the predictions were most accurate using $d_{\mathrm{max}} = 20$ and $N=200$ for the RF algorithm, which is a more complex structure than that which was used for the lPTT and ePTT models. Whilst the predictive accuracy for the training data seems reasonably high for both $z$ and $\beta_{\mathrm{CCR}}$, the accuracy of the predictions of the validation data is significantly reduced for $\beta_{\mathrm{CCR}}$. At this point, we must question whether this inaccuracy in terms of model parameter prediction actually translates into inaccuracy of the prediction of the rheological behaviour (which could suggest the RF algorithm is not adequate for this particular constitutive model). To this end, as before, we simulate the LAOS response at $De=1$ and $Wi=50$ using the real and RF predicted parameters in the validation data set, and plot the parity plots for the first 5 elastic and viscous Chebyshev coefficients in Figure \ref{fig:RPchebparity}. It is clear that, despite the fact the RF algorithm is not choosing the expected model parameters based on the inputted LAOS data, the model parameters it is predicting are still providing the correct rheological behavior under the oscillatory shear flow (at least at the same $De$ and $Wi$ as those used for the training). This essentially confirms that, for this particular flow and range of model parameters, there are multiple sets of model parameters which provide at least very similar LAOS behaviour (if not identical LAOS behaviour). Therefore, the RF algorithm is still performing as intended, and the results suggest there is a more fundamental problem regarding the identifiability of the constitutive model parameters from the LAOS data.

\begin{figure}[h!]
    \centering
    \includegraphics[width=0.325\textwidth]{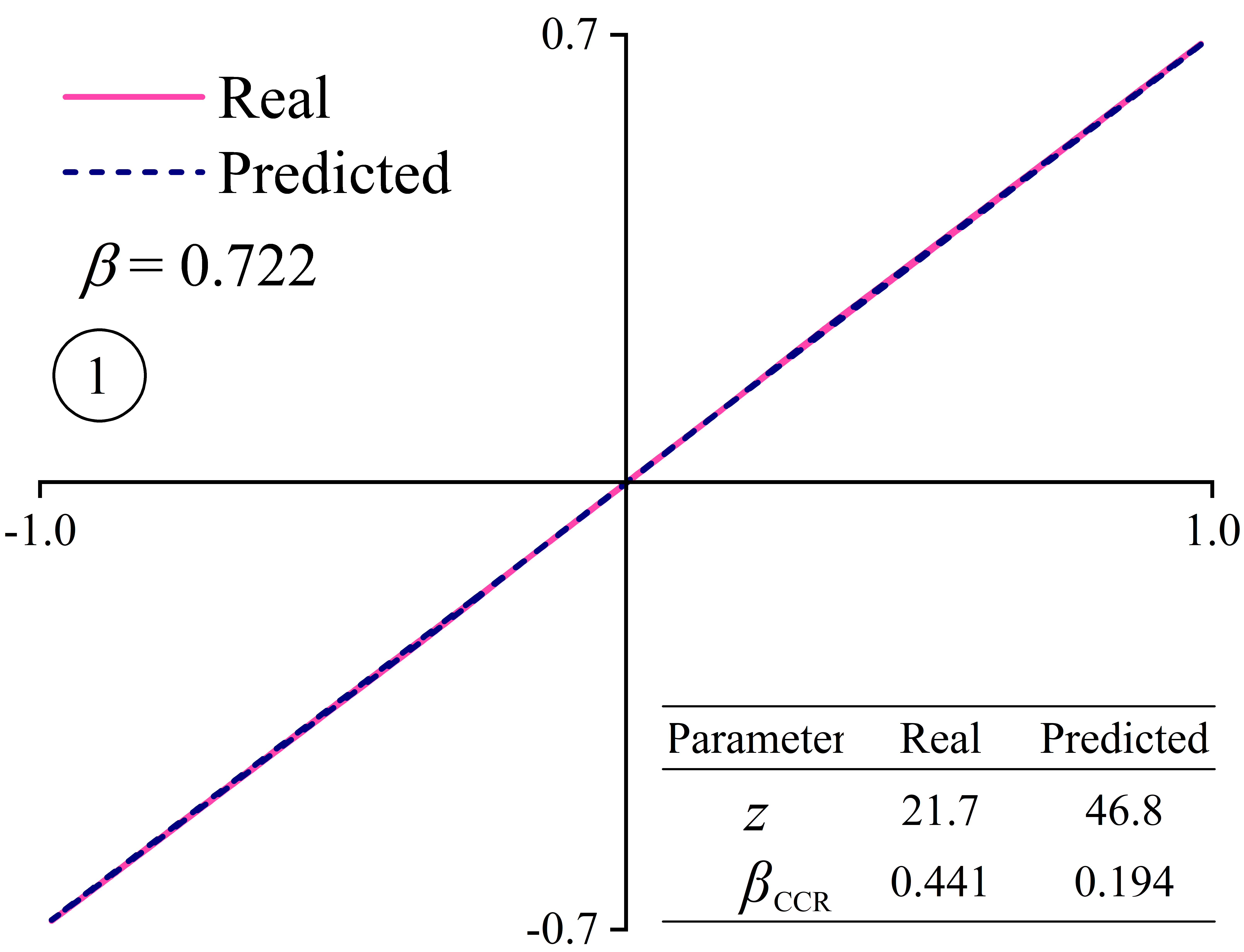}
    \includegraphics[width=0.325\textwidth]{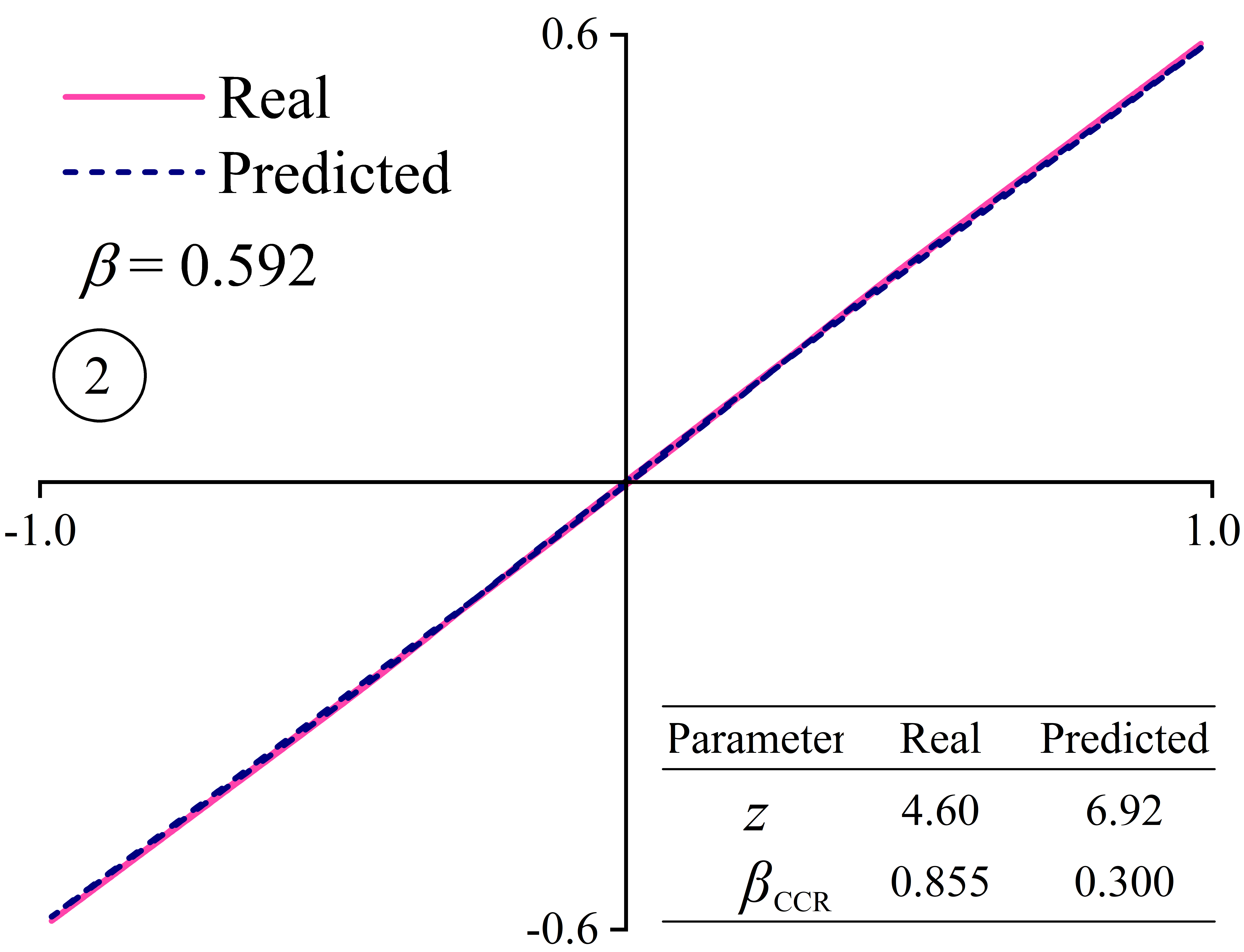}
    \includegraphics[width=0.325\textwidth]{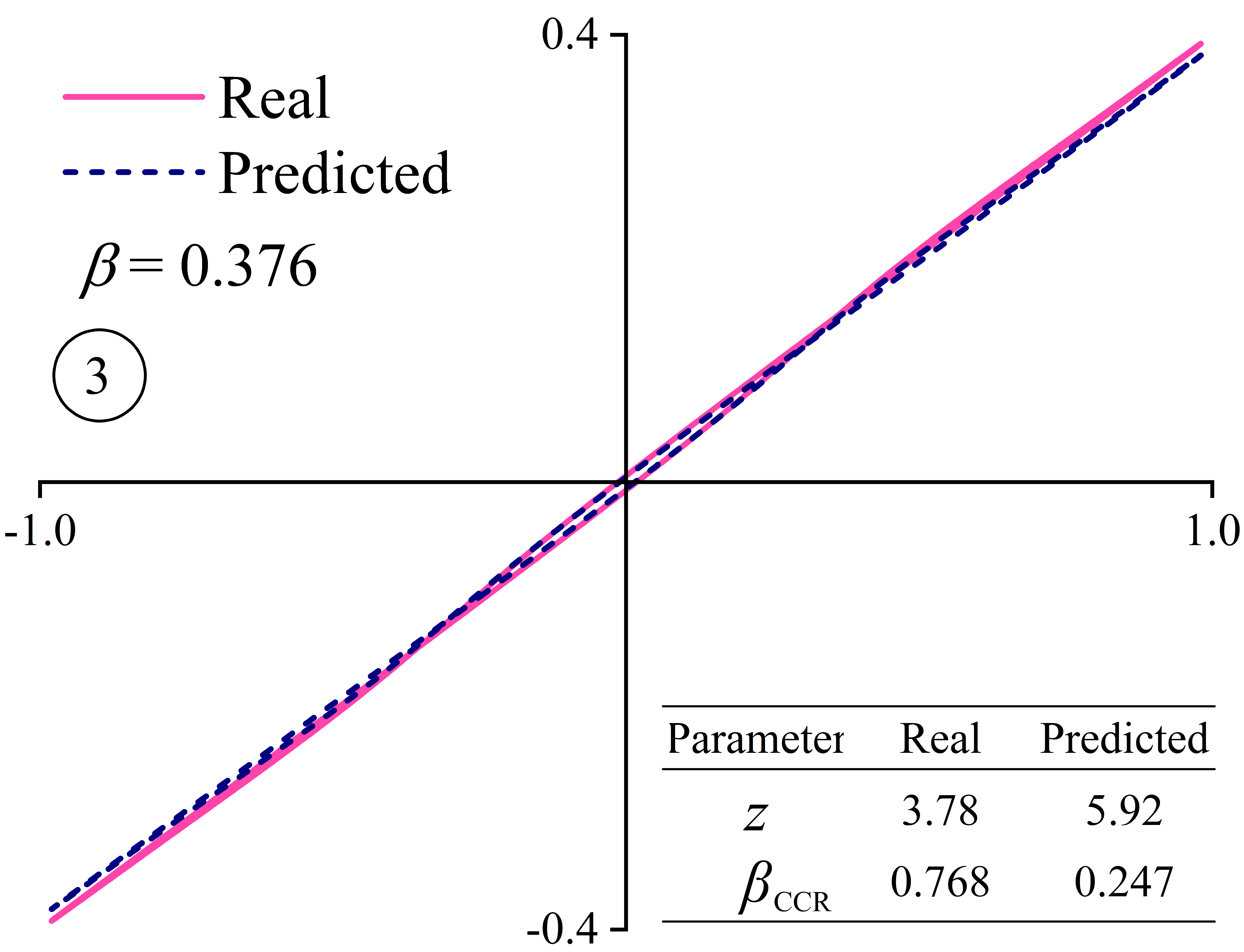}

    \includegraphics[width=0.325\textwidth]{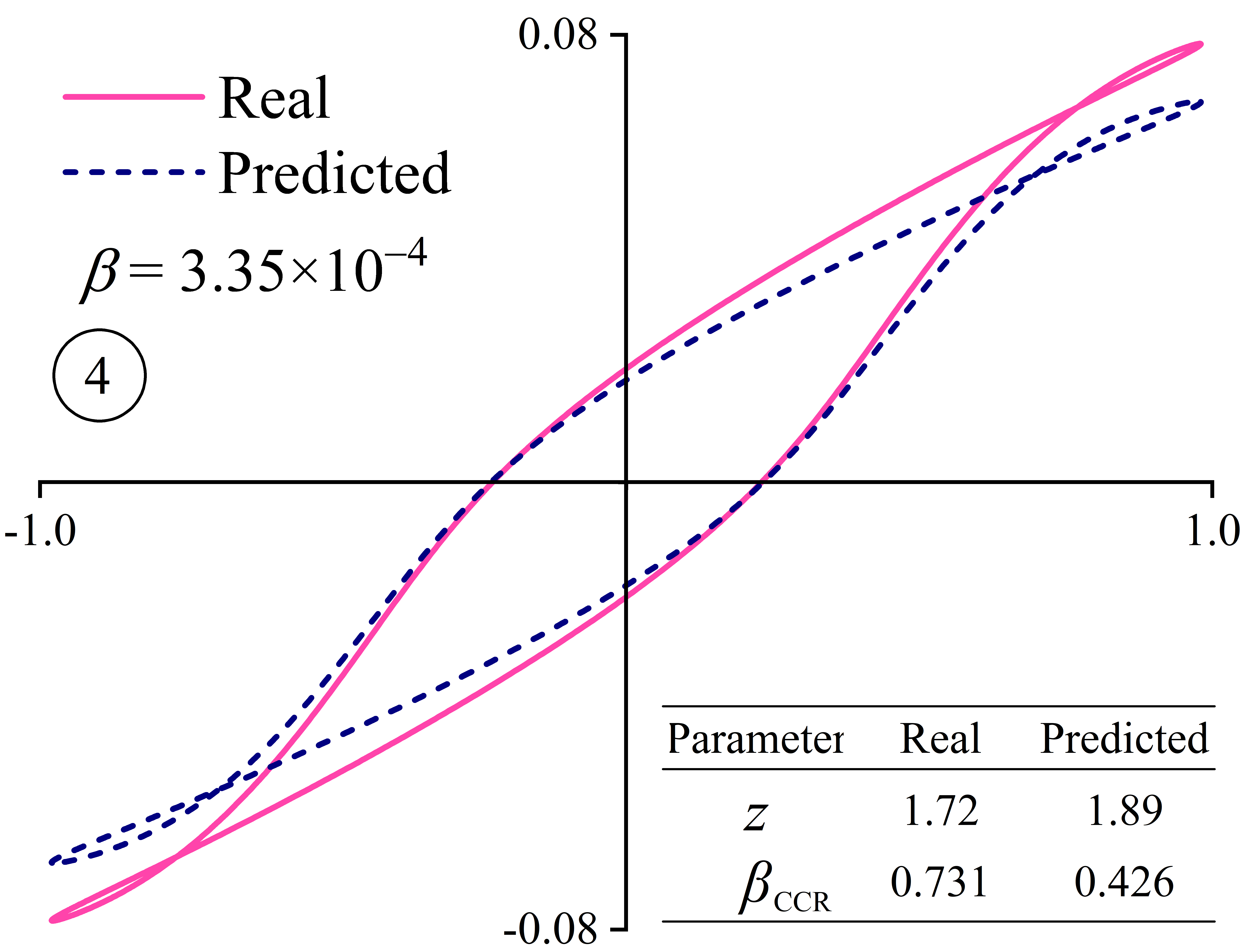}
    \includegraphics[width=0.325\textwidth]{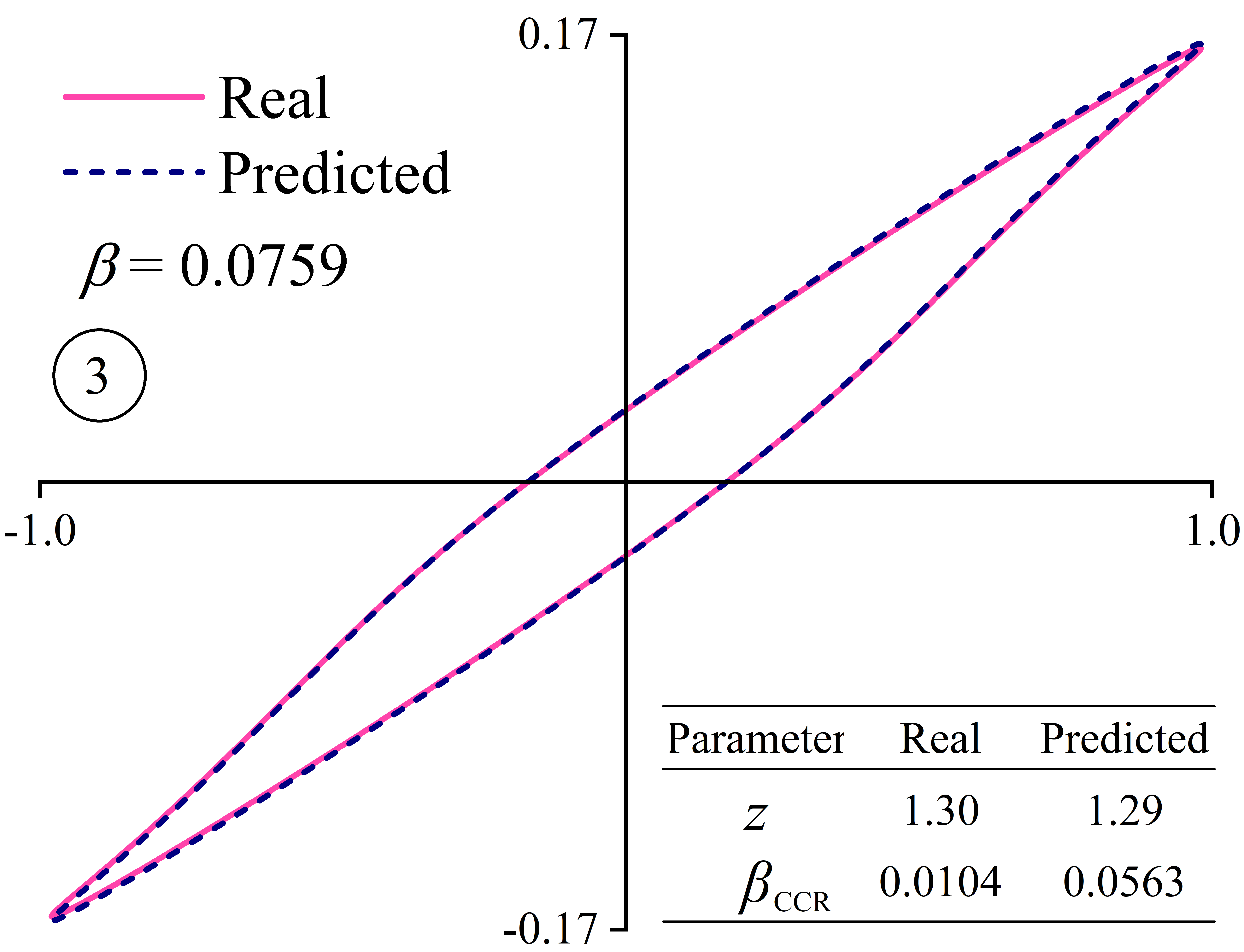}
    \includegraphics[width=0.325\textwidth]{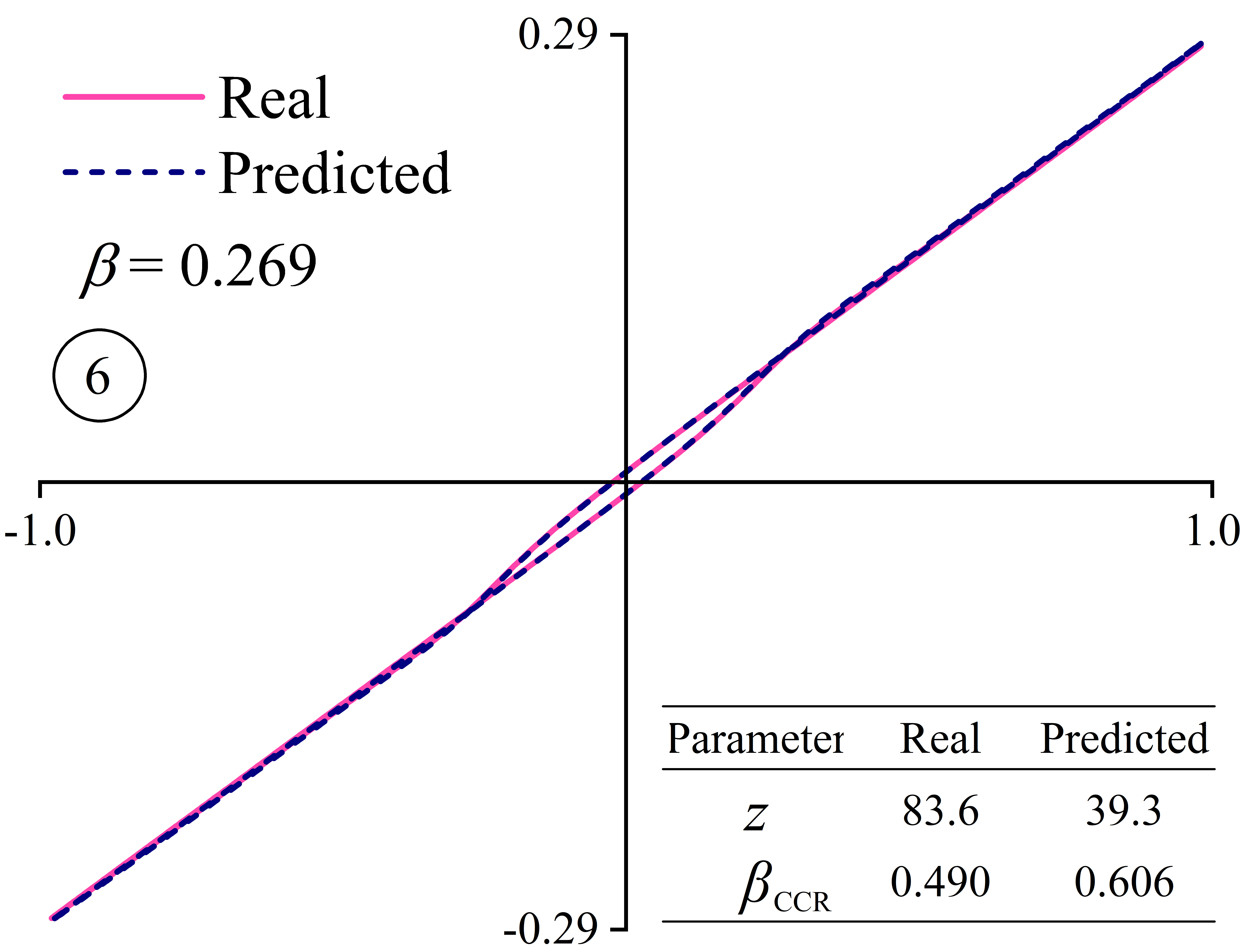}
    \caption{Examples of viscous Lissajous curves (stress \textit{vs} strain rate) for RoLiE-Poly model under LAOS at $De = 1$ and $Wi = 50$. The 6 plots correspond to the labelled data points with poor predictions in Figure \ref{fig:RPValidationFits}. This figure highlights clearly that more than one set of model parameters can give very similar shear rheology.}
    \label{fig:RPlissajousexamples}
\end{figure}

In Figure \ref{fig:RPValidationFits}(b), we label 6 data points for which the RF algorithm has predicted the value of $\beta_{\mathrm{CCR}}$ poorly from the inputted Chebyshev coefficients. In Figure \ref{fig:RPlissajousexamples}, we show the viscous Lissajous curves (real and predicted) for LAOS simulations for these 6 examples at $De = 1$ and $Wi = 50$. Again, the real value of $\beta$ has been used, rather than the predicted value. The value of $\beta$ is given in the plot for each example. Data points 1, 2, 3, and 6 have particularly large values of $\beta$ (close to unity), which means these LAOS responses (as is evident from the Lissajous curves) are close to being Newtonian. This likely explains why the RF predictions for $\beta_{\mathrm{CCR}}$ are poor. Essentially, since this flow is close to being Newtonian for these cases, the Chebyshev coefficient spectra are practically insensitive to the RoLiE-Poly model parameters. Note that in the Newtonian case $v_1 = 1$ and all other elastic and viscous Chebyshev coefficients are zero. Moreover, points 3 and 4 have particularly low values of $z$, meaning these responses are becoming close to the quasi-linear Oldroyd-B response. Again, as the response becomes closer to the Oldroyd-B response, there is a loss of sensitivity between LAOS behaviour and constitutive model parameters.

\begin{figure}[h!]
    \centering
    \includegraphics[trim={0 0 2.5cm 0},clip,width=0.325\textwidth]{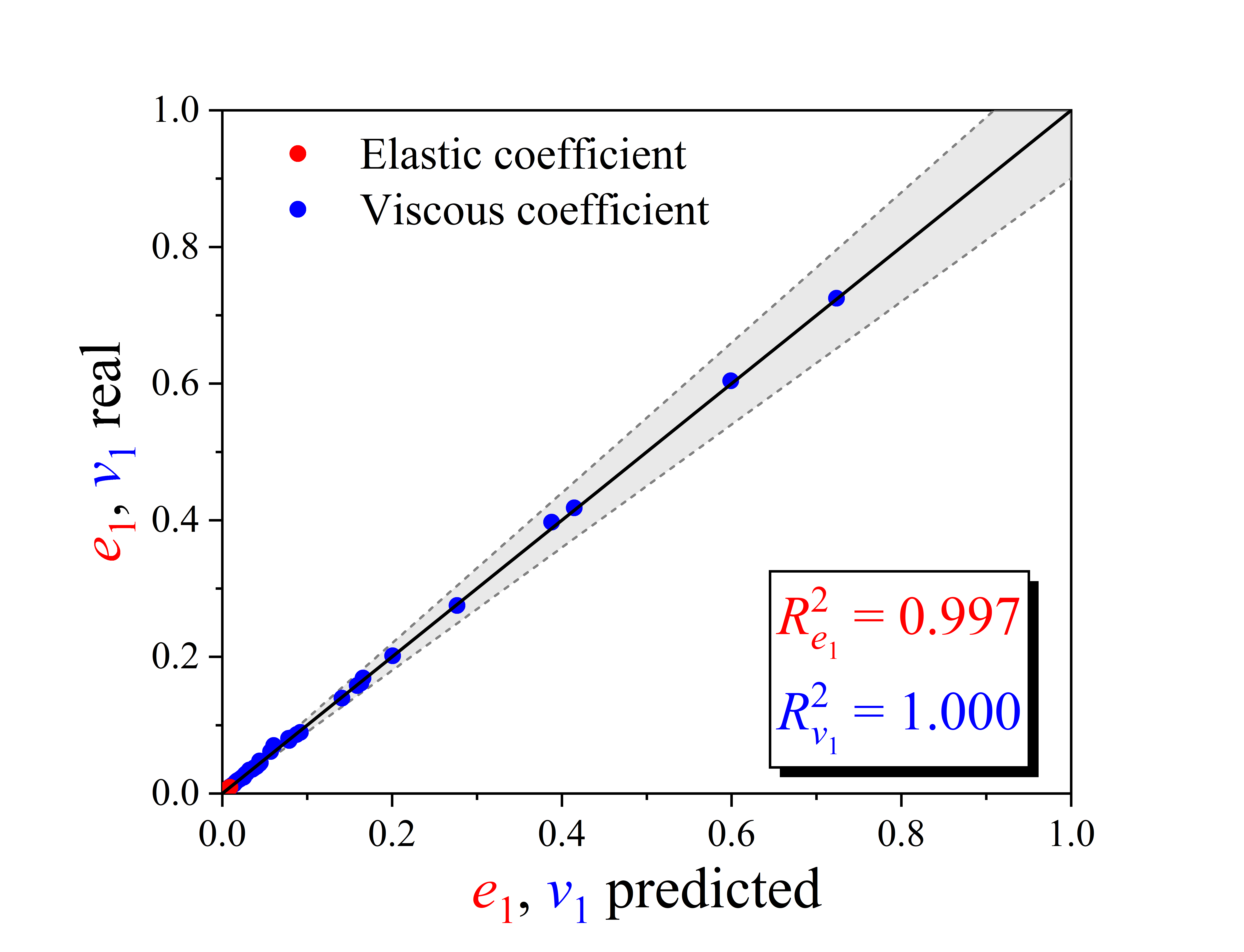}
    \includegraphics[trim={0 0 2.5cm 0},clip,width=0.325\textwidth]{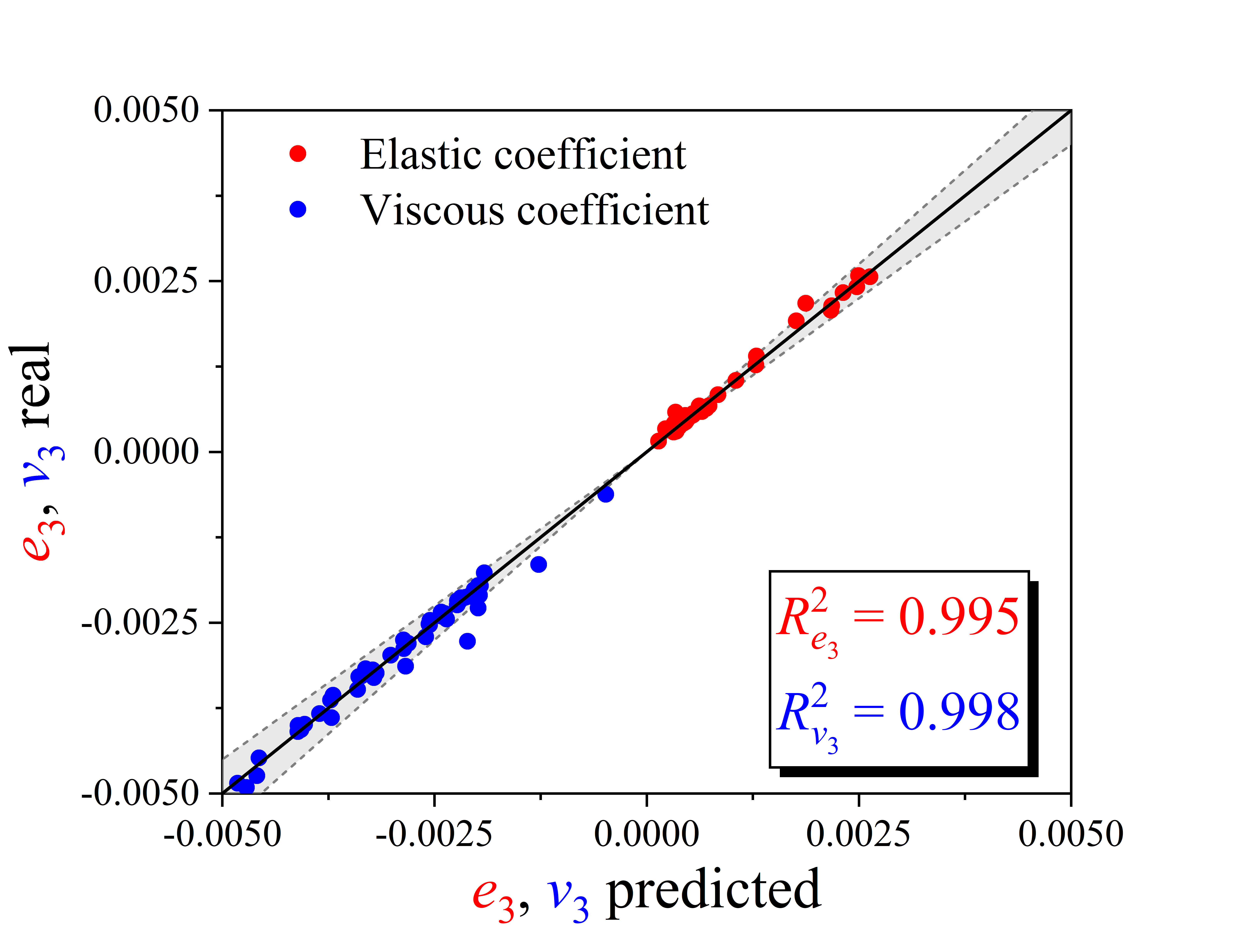}
    \includegraphics[trim={0 0 2.5cm 0},clip,width=0.325\textwidth]{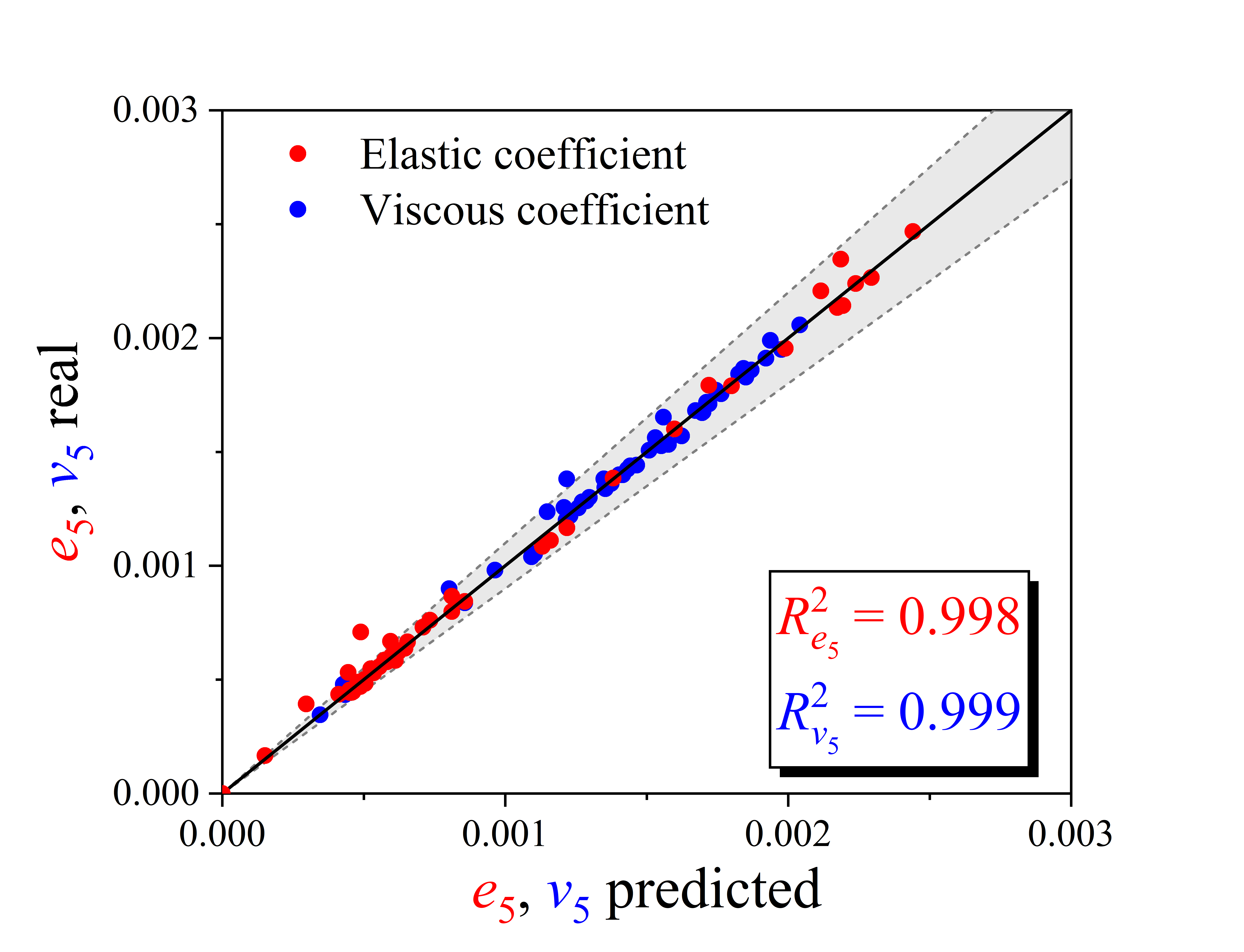}

    \includegraphics[trim={0 0 2.5cm 0},clip,width=0.325\textwidth]{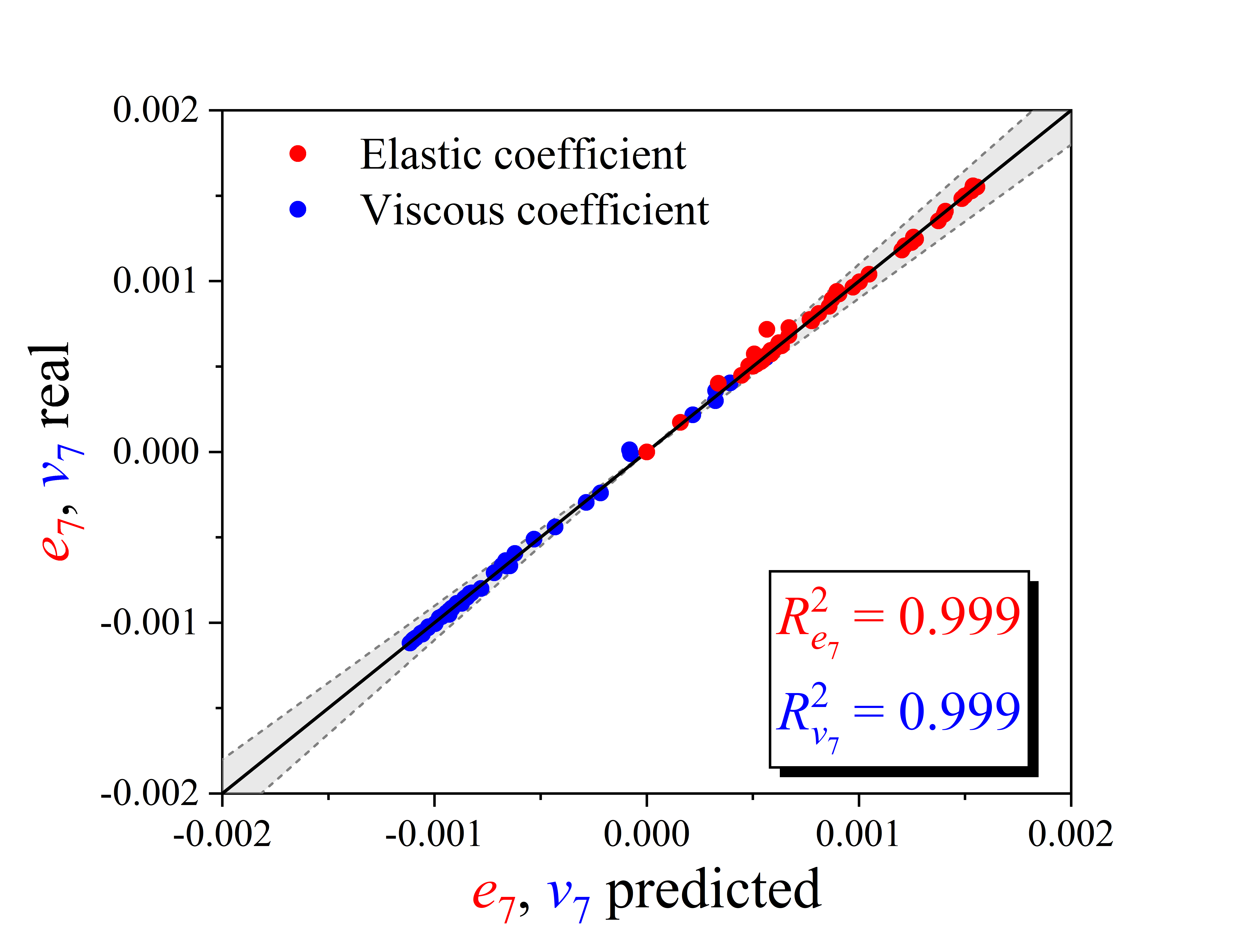}
    \includegraphics[trim={0 0 2.5cm 0},clip,width=0.325\textwidth]{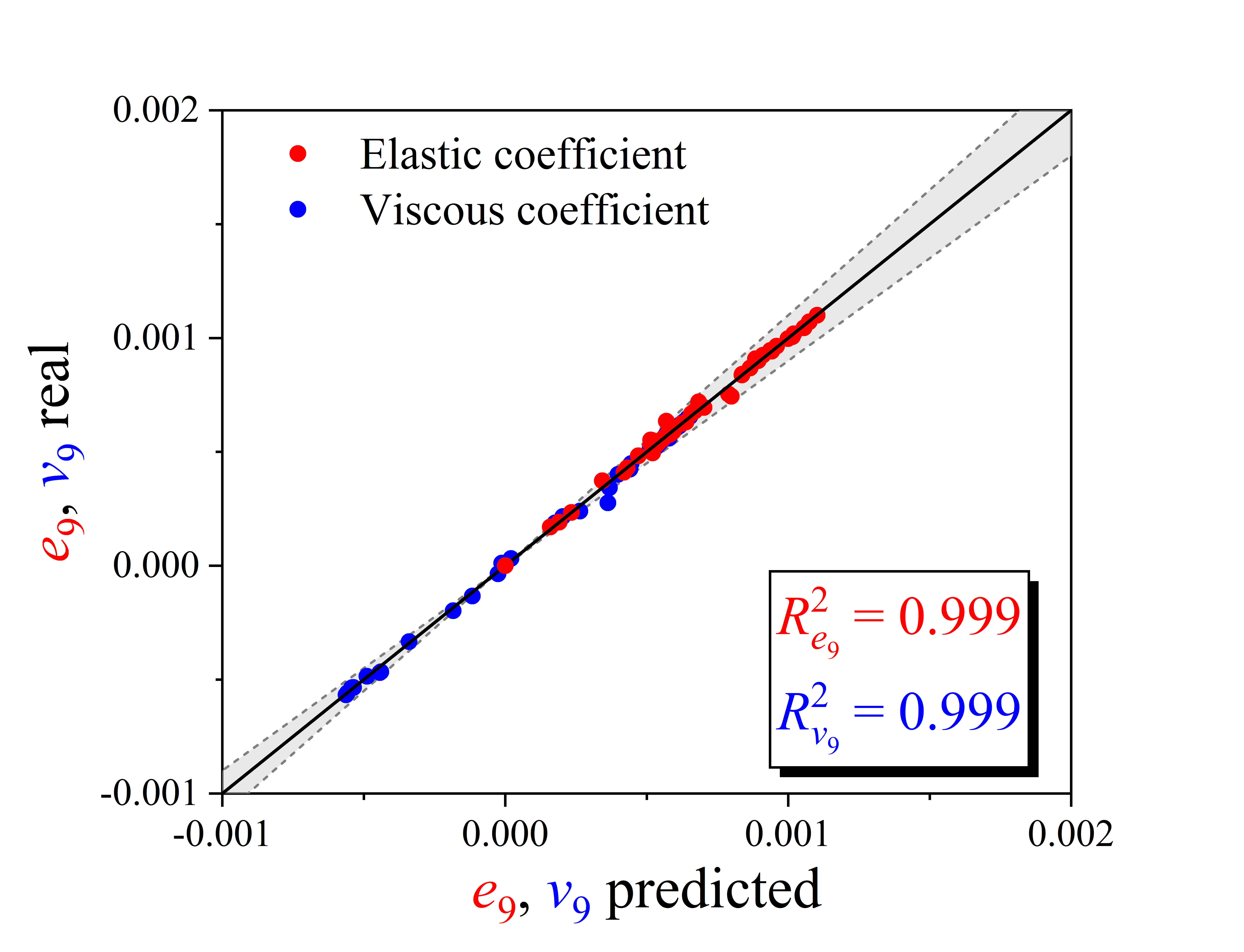}
    \caption{Parity plots for the first 5 Chebyshev coefficients for RoLiE-Poly model under LAOS at $De = 0.5$ and $Wi = 100$. The parameter predictions were made for LAOS data at $De = 1$ and $Wi = 50$ (see Figure \ref{fig:RPValidationFits}). This figure highlights that the shear rheology is still predicted remarkably well even in a flow which was not used in the training, despite the seemingly poor parameter predictions in some cases.}
    \label{fig:RPchebparity_De_0_5}
\end{figure}

Next, we check how accurately the predicted model parameters recover the rheological behaviour in a LAOS flow which does not correspond to the training condition. Figure \ref{fig:RPchebparity_De_0_5} shows, for $De = 0.5$ and $Wi = 100$ the parity plots for the first 5 Chebyshev cofficients using the validation data displayed in Figure \ref{fig:RPValidationFits}, where the training was undertaken at $De = 1$ and $Wi = 50$. Again, the LAOS behaviour given by the RF parameters almost perfectly matches the real LAOS behaviour, despite the fact the predictions of $\beta_{\mathrm{CCR}}$ are fairly inaccurate. This highlights that the RF algorithm, despite not predicting the "correct" parameters, is predicting model parameter sets which provide the correct rheological behaviour even in flows which were not specifically used for the training. This also means we could not increase the accuracy of the parameter predictions by including this flow for $De = 0.5$ and $Wi = 100$ into the ML training, since the same problem regarding non-uniqueness of the solution exists.

\begin{figure}[h!]
    \centering
    \includegraphics[trim={0 0 2.5cm 0},clip,width=0.325\textwidth]{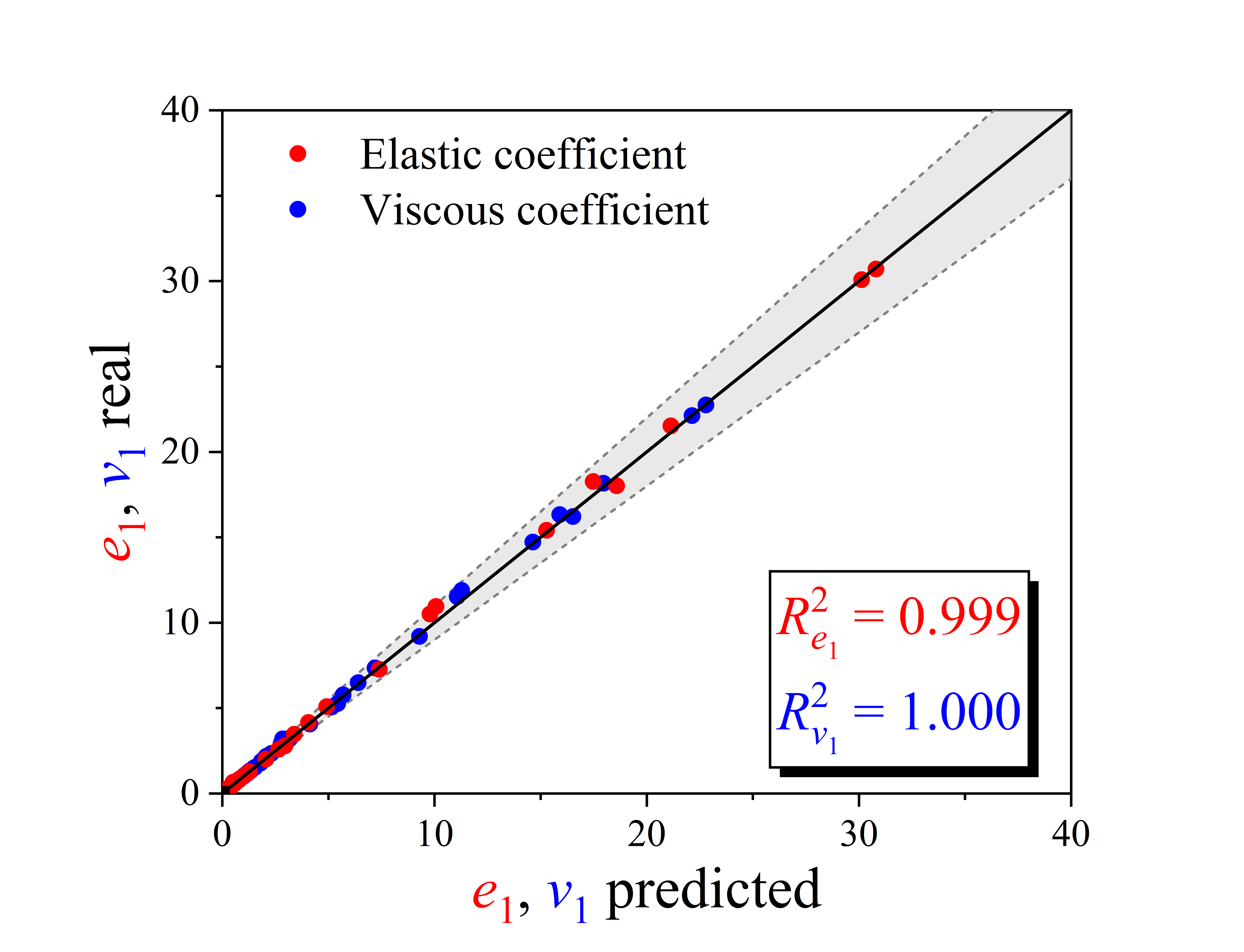}
    \includegraphics[trim={0 0 2.5cm 0},clip,width=0.325\textwidth]{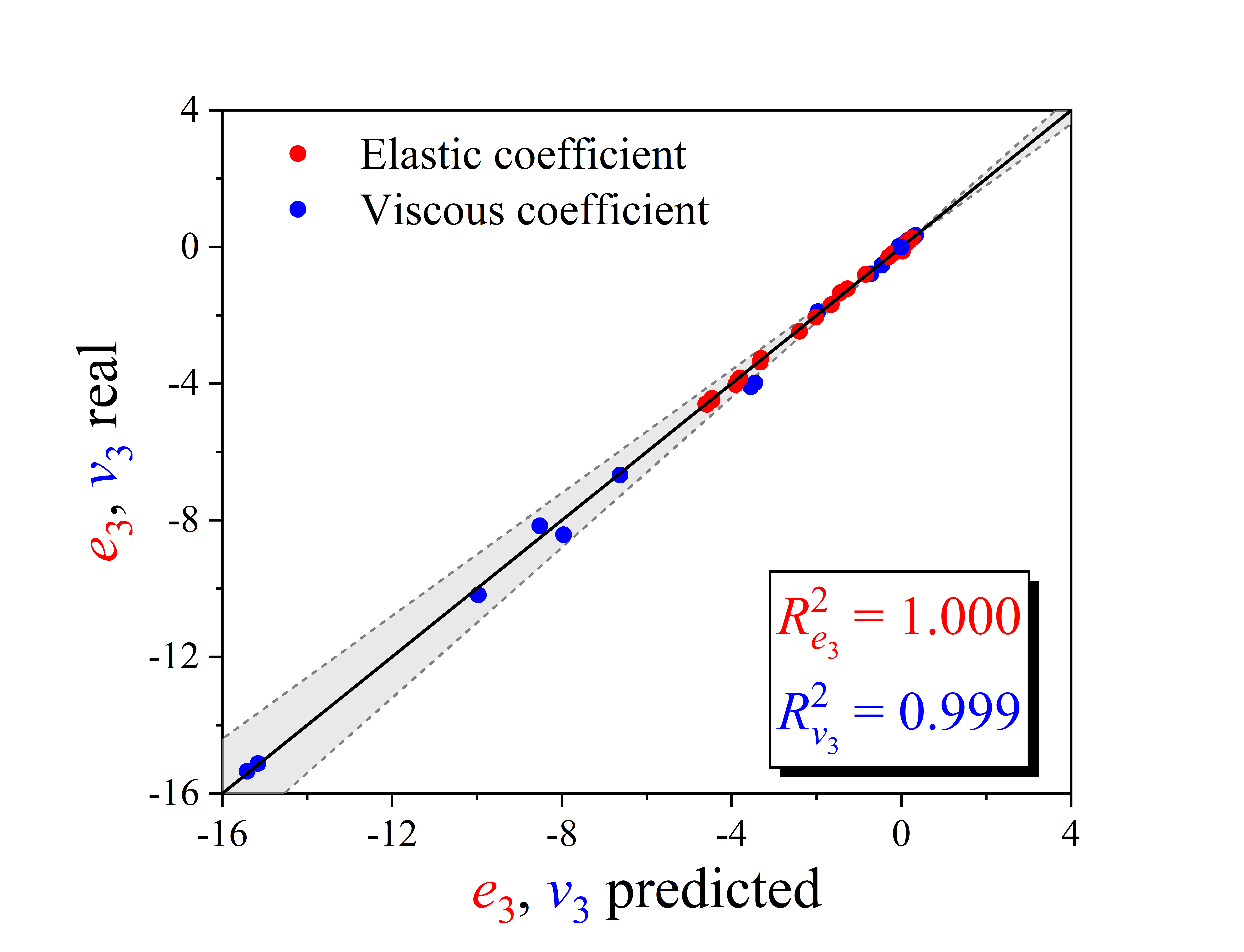}
    \includegraphics[trim={0 0 2.5cm 0},clip,width=0.325\textwidth]{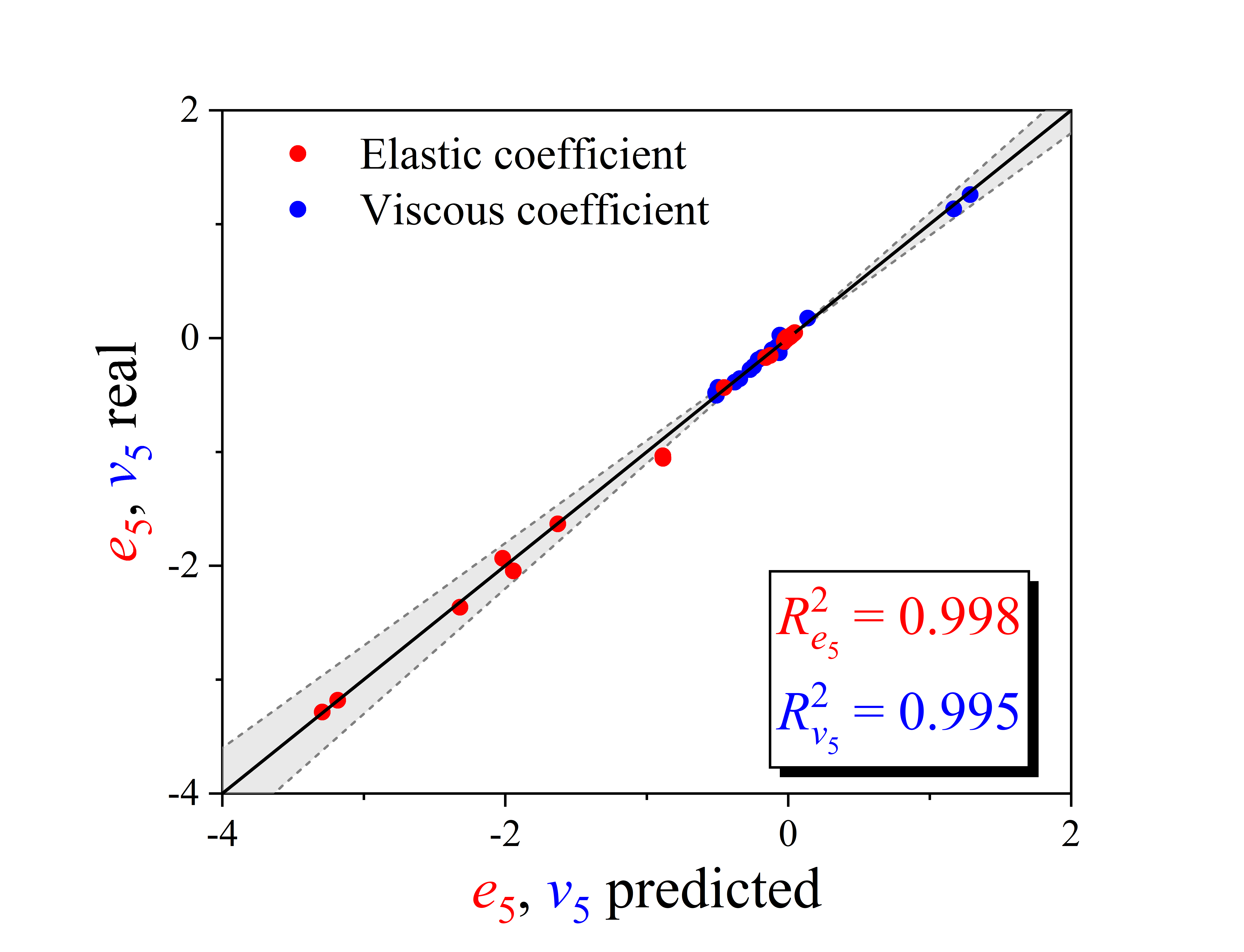}

    \includegraphics[trim={0 0 2.5cm 0},clip,width=0.325\textwidth]{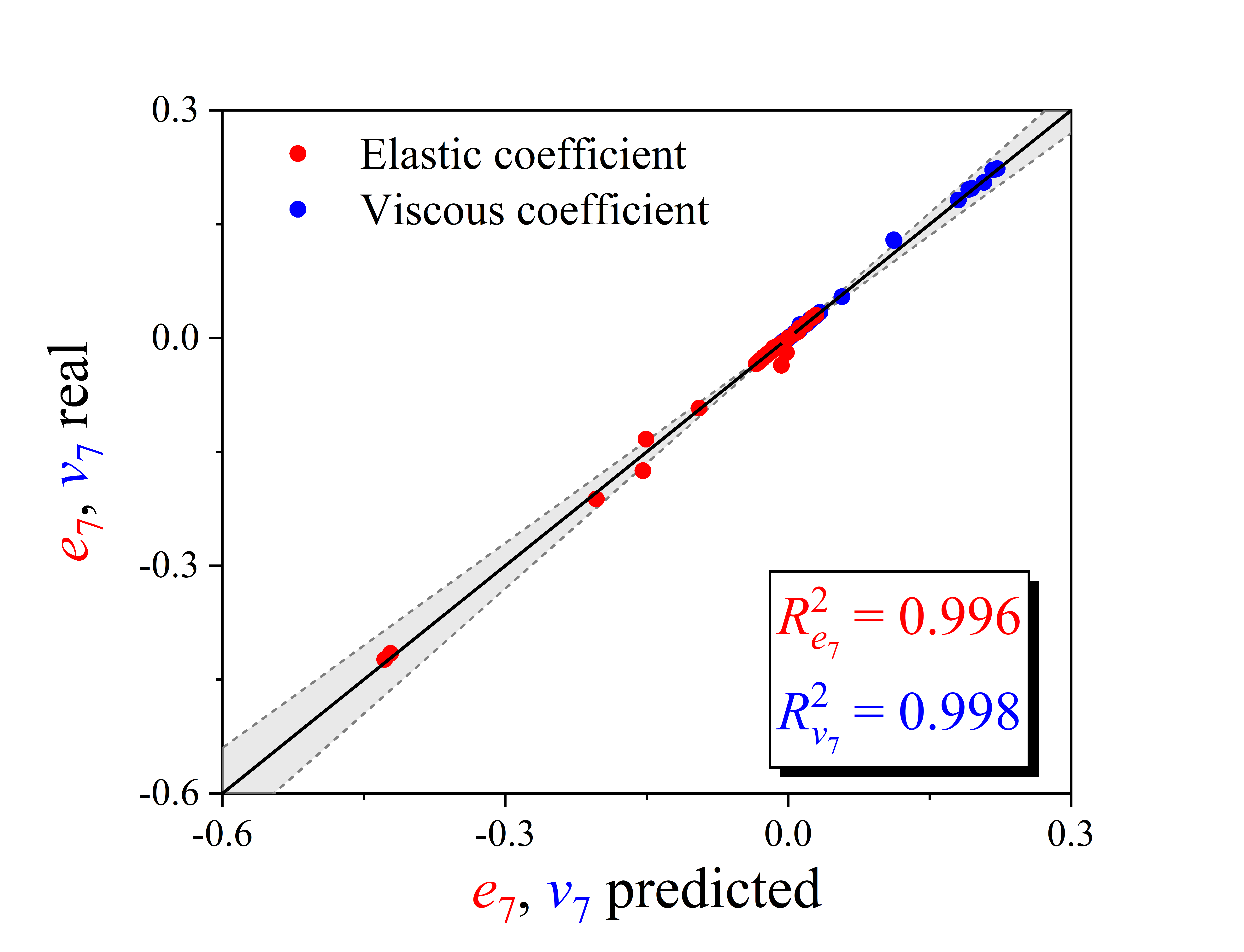}
    \includegraphics[trim={0 0 2.5cm 0},clip,width=0.325\textwidth]{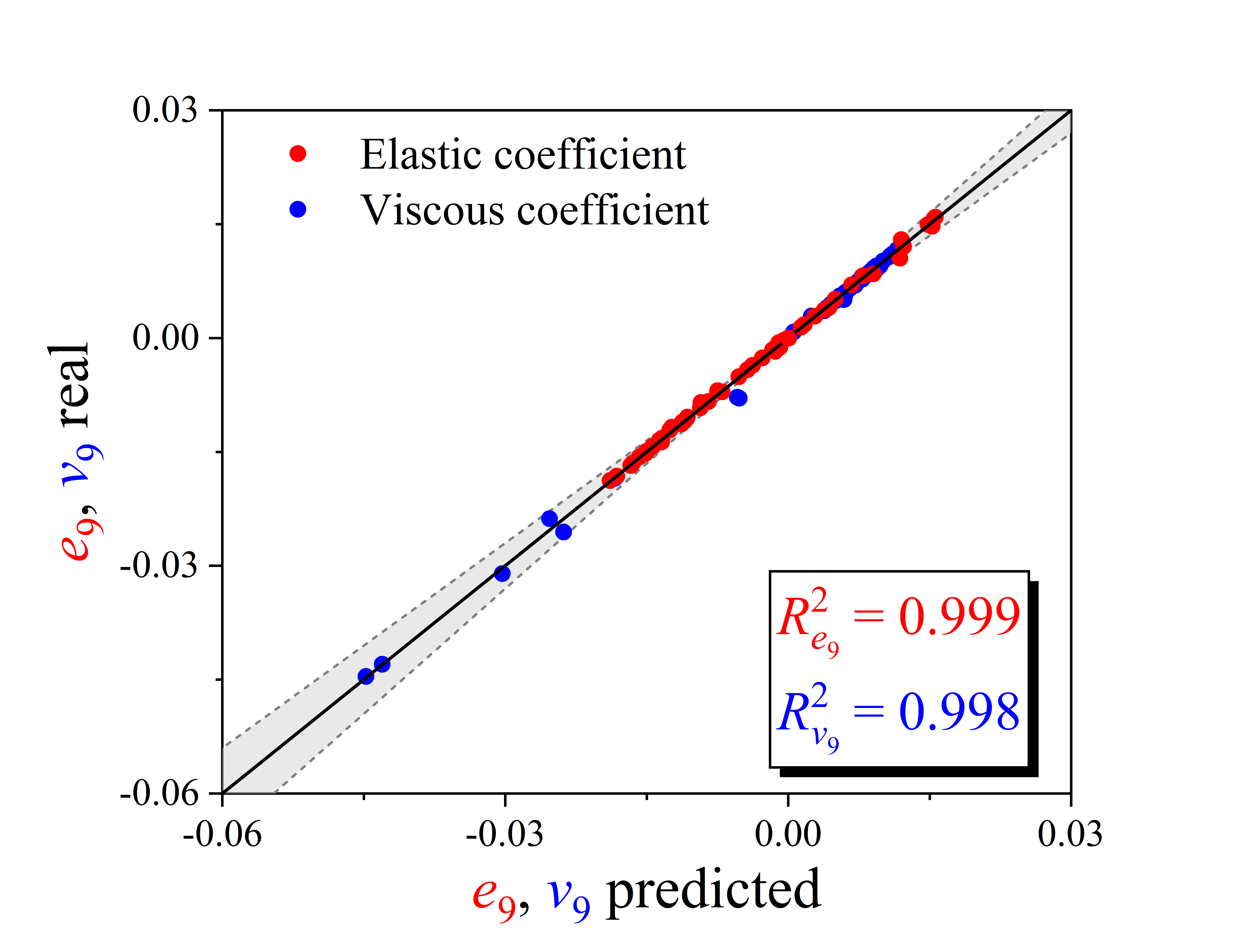}
    \caption{Parity plots for the first 5 Chebyshev coefficients for RoLiE-Poly model under LAOE at $De = 1$ and $Wi = 3$. Parameter predictions were made using LAOS data at $De = 1$ and $Wi = 50$ (see Figure \ref{fig:RPValidationFits}). This figure highlights that the rheology is still predicted with almost perfect accuracy even though some of the model parameters were not predicted well.}
    \label{fig:RP_LAOE_ChebParity}
\end{figure}

Figure \ref{fig:RP_LAOE_ChebParity} shows the parity plots for the first 5 Chebyshev coefficients for an oscillatory extensional flow at $De = 1$ and $Wi = 3$. It is clear that, even in this extensional flow, the RoLiE-Poly model parameters predicted by the RF algorithm trained using LAOS data at $De = 1$ and $Wi = 50$ (see Figure \ref{fig:RPValidationFits}) still work very well at predicting the rheological behavior.

%\paragraph{$De=1$ and $Wi=6$}

\begin{figure}[h!]
    \centering
    \includegraphics[width=0.325\textwidth]{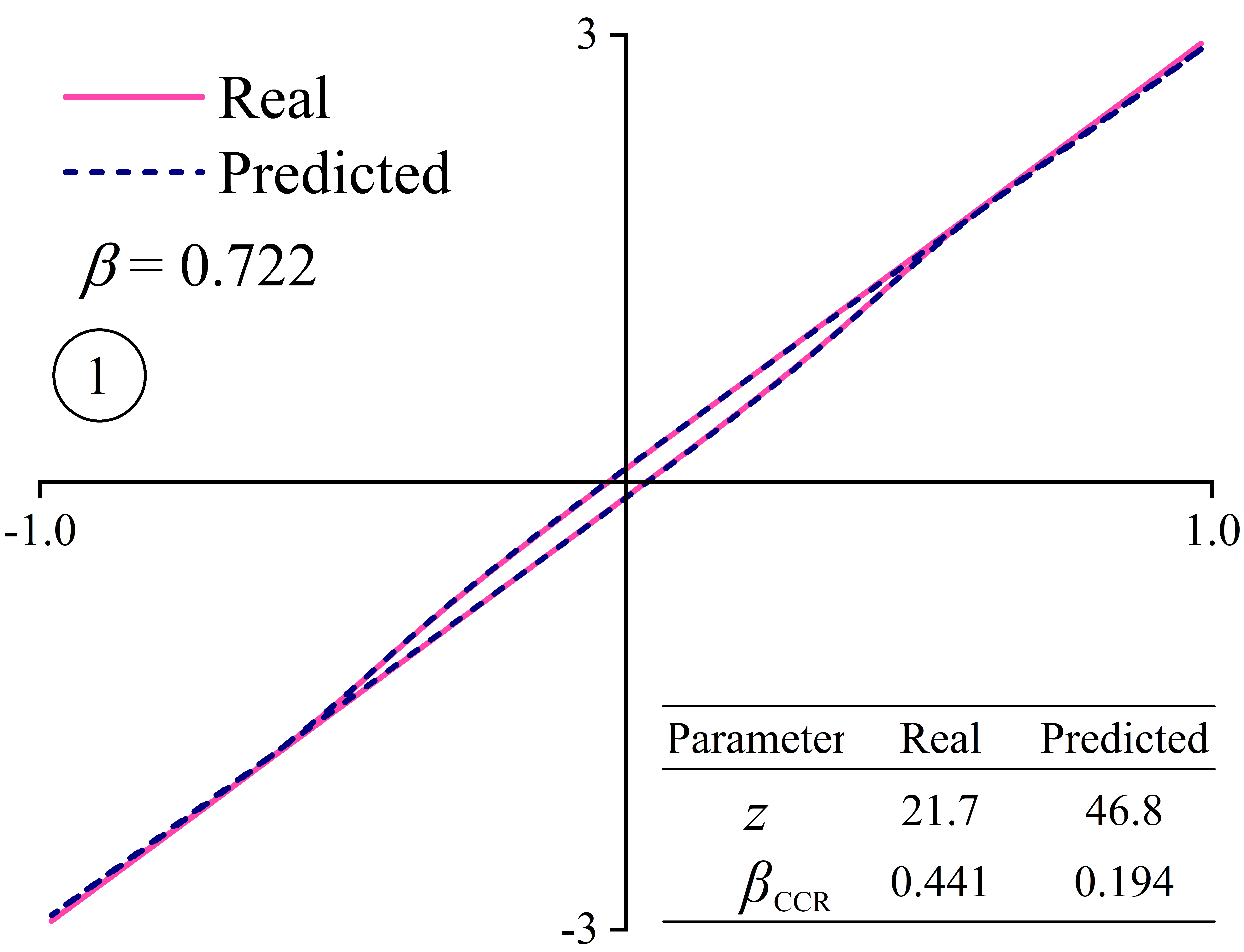}
    \includegraphics[width=0.325\textwidth]{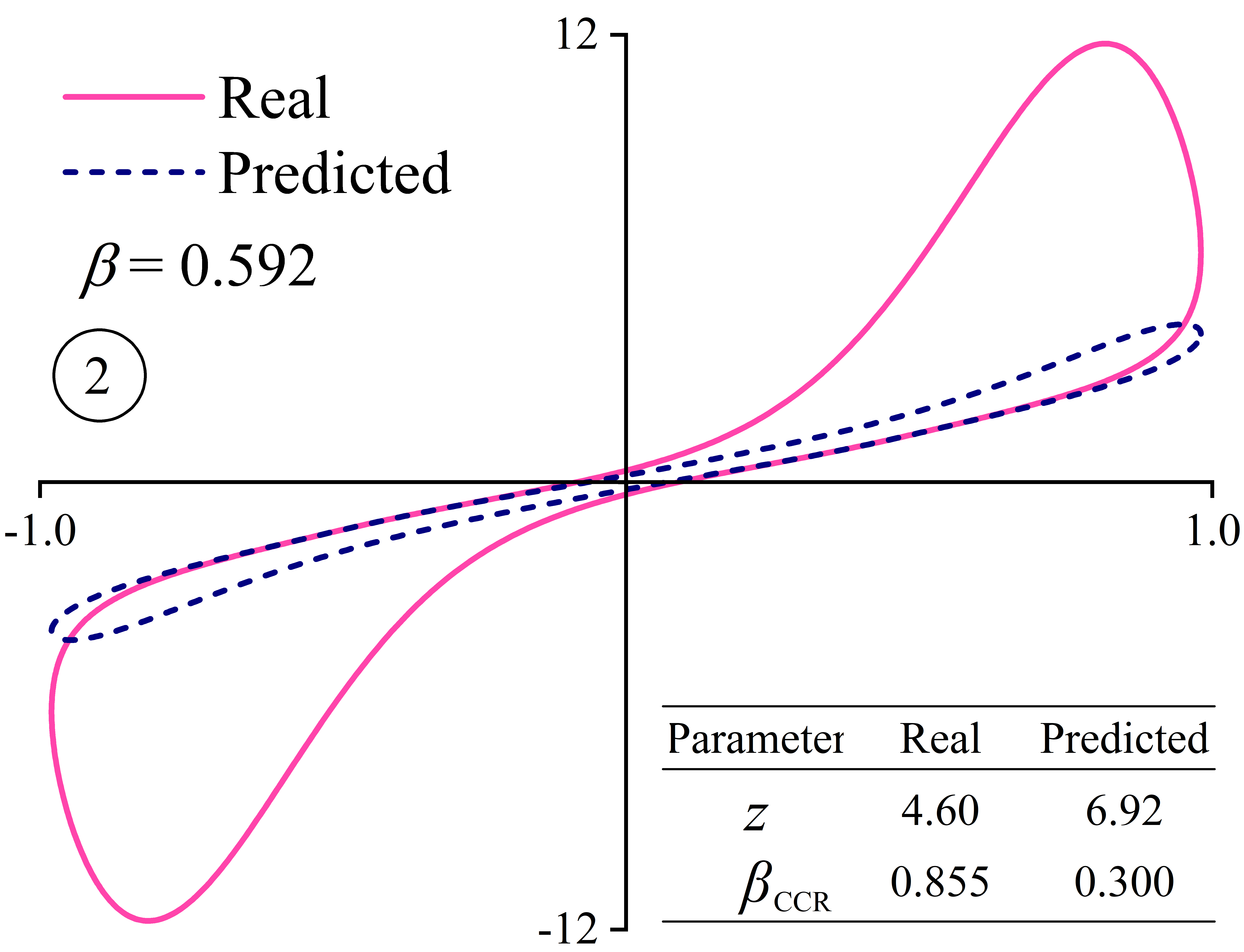}
    \includegraphics[width=0.325\textwidth]{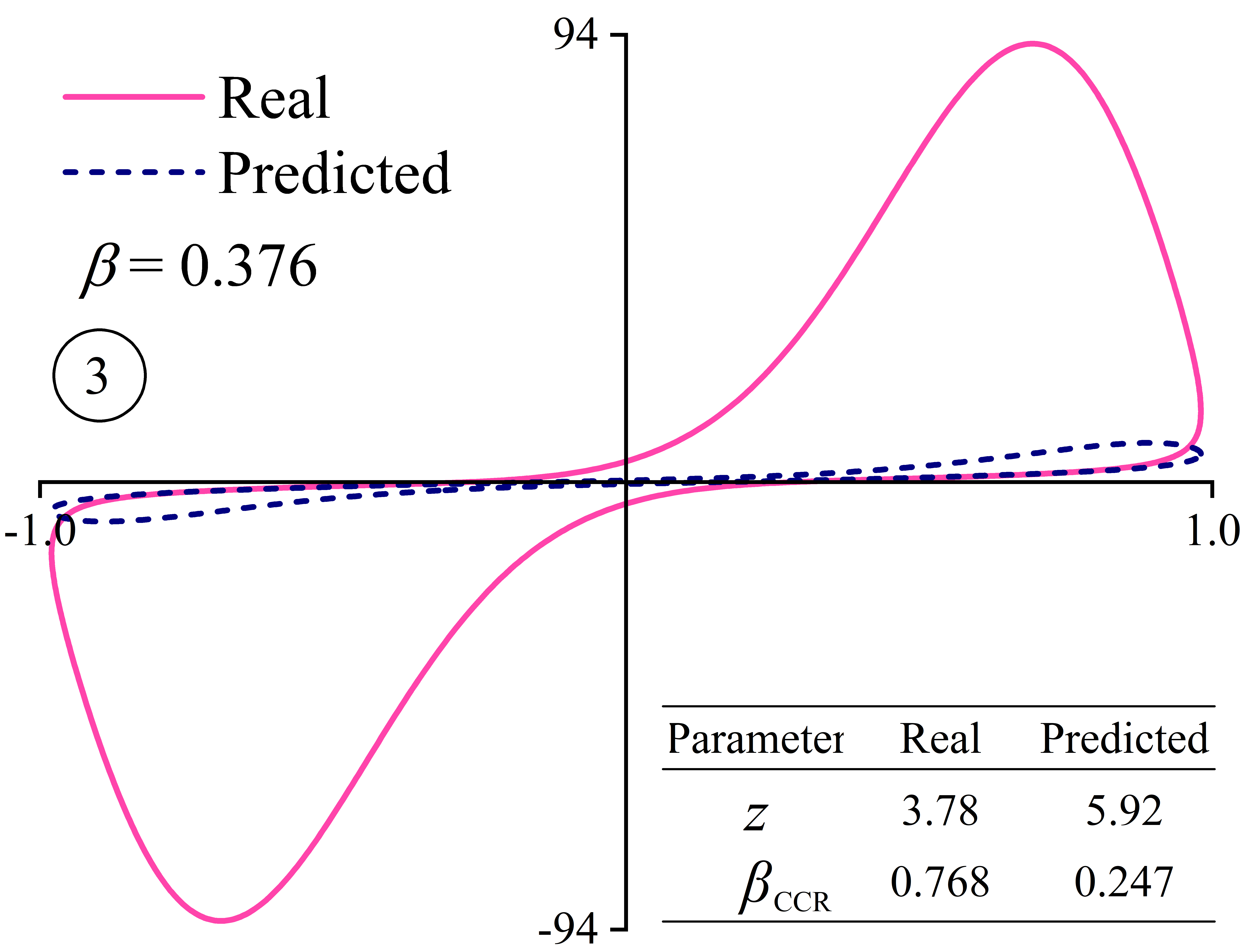}

    \includegraphics[width=0.325\textwidth]{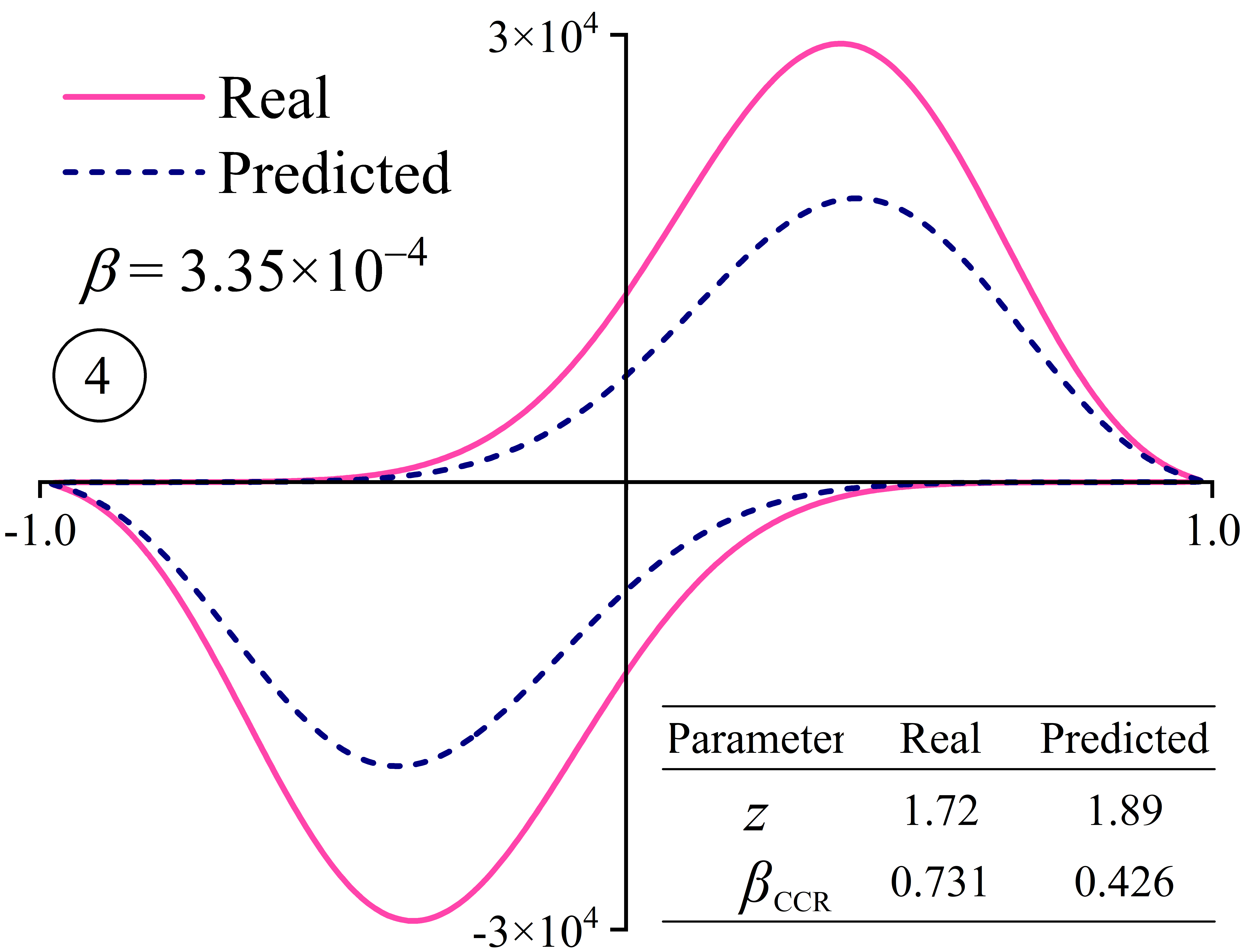}
    \includegraphics[width=0.325\textwidth]{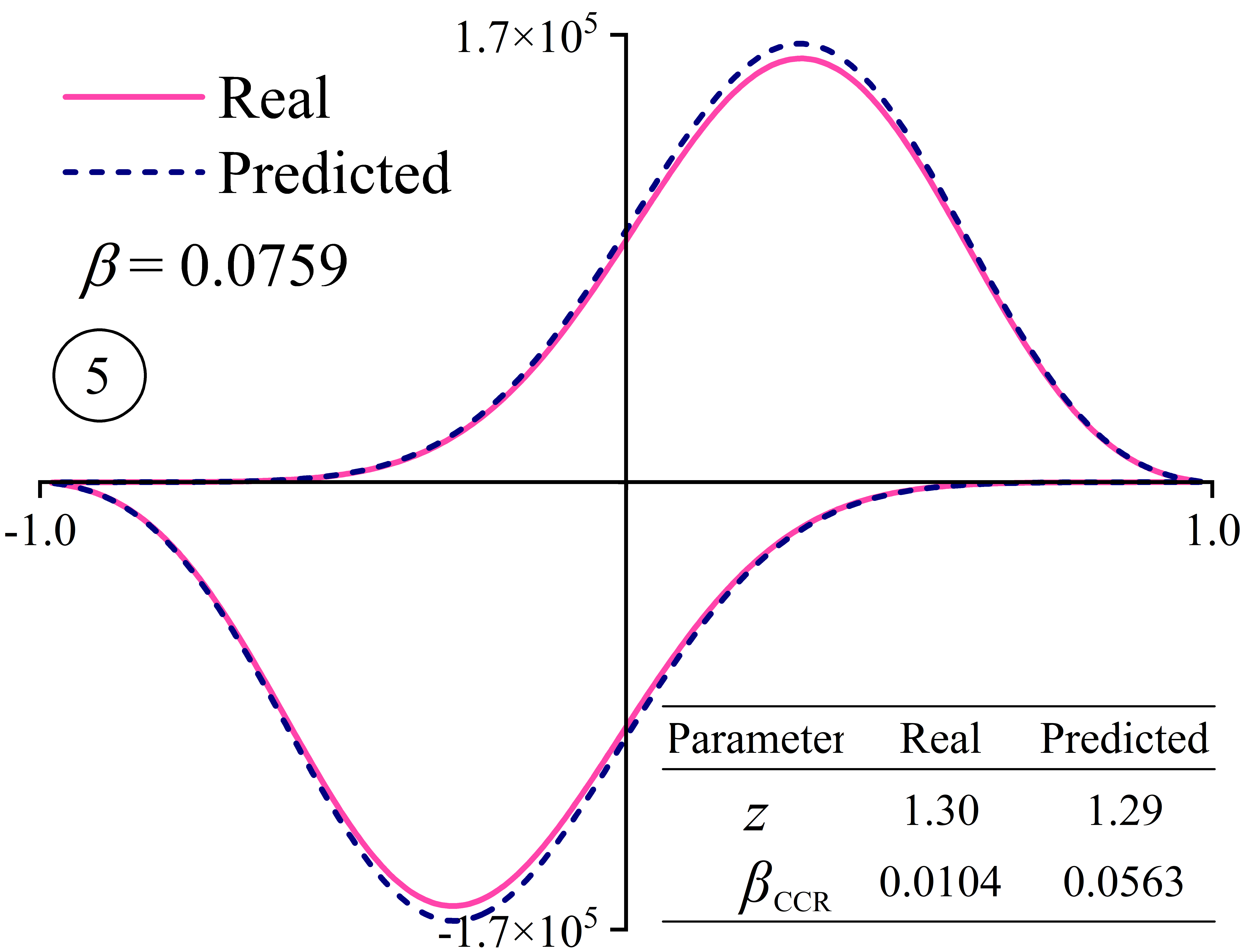}
    \includegraphics[width=0.325\textwidth]{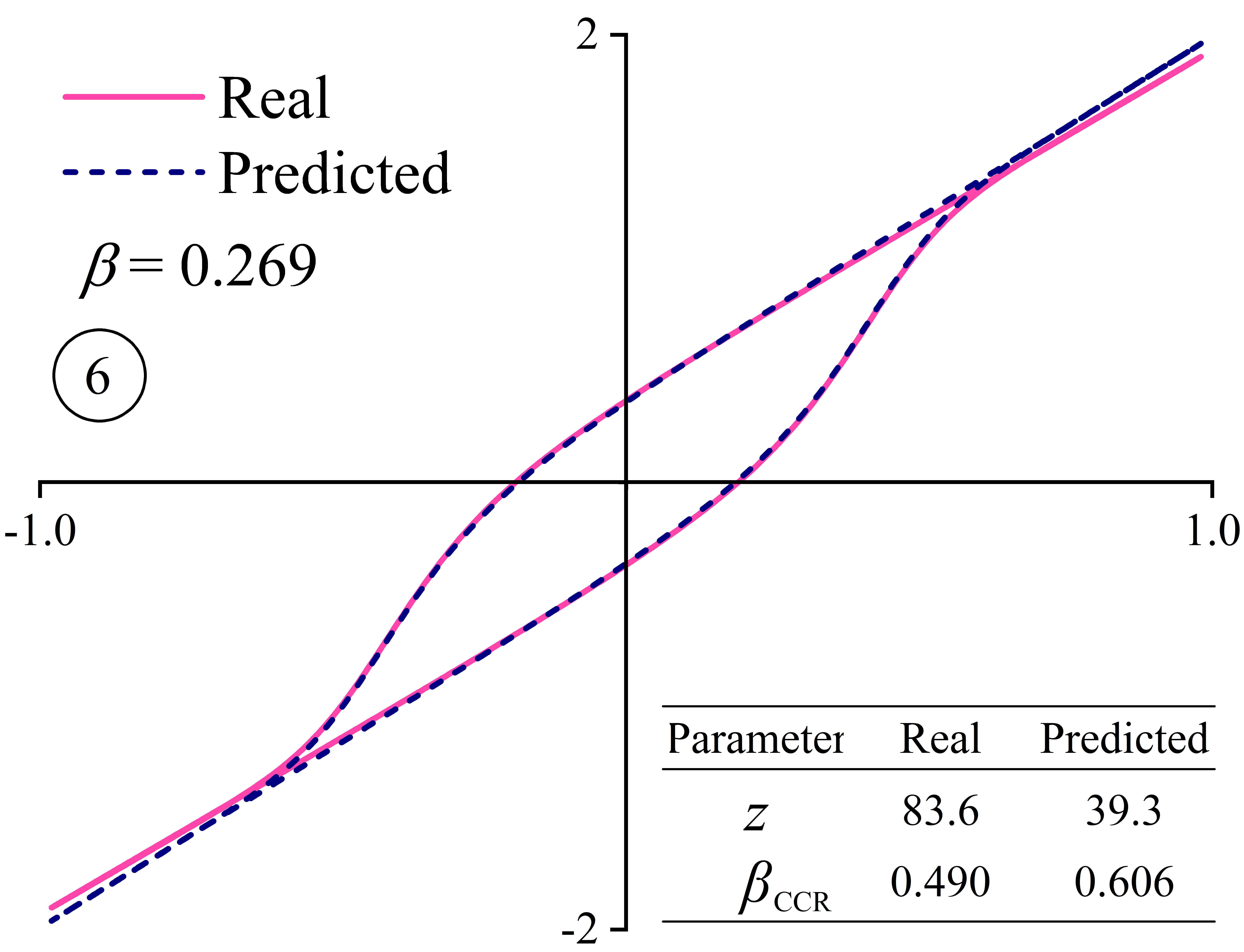}
    \caption{Examples of viscous Lissajous curves for RoLiE-Poly model under LAOE at $De = 1$ and $Wi = 6$. See Figure \ref{fig:RPValidationFits} for the parity plots for the RF parameter predictions. Training was undertaken using LAOS at $Wi = 1$ and $Wi = 50$. The numbers 1-6 in the plots correspond to the data points with poor predictions labelled in Figure \ref{fig:RPValidationFits}. This figure highlights that the extensional rheological behaviour is now poorly captured for some of the data points with poor parameter predictions.}
    \label{fig:RP_LAOE_Lissajous}
\end{figure}

Finally, we now check if the predicted parameters in Figure \ref{fig:RPValidationFits} (based on training data for LAOS at $De = 1$ and $Wi = 50$) work at predicting homogeneous oscillatory extensional flow at $De = 1$ and $Wi = 6$. The accumulated strain during the oscillation will be higher in this case, and the non-Newtonian behaviour will be expected to be more pronounced. In Figure \ref{fig:RP_LAOE_Lissajous}, we show the 6 examples (the 6 data points labelled in Figure \ref{fig:RPValidationFits}) of the viscous Lissajous curves for the oscillatory extensional flow at $De = 1$ and $Wi = 6$. For points 1, 5, and 6, the oscillatory extensional rheology is still predicted remarkably well considering the model parameters are not predicted particularly well. For points 2, 3, and 4, it is now clear that the poor parameter predictions do translate to poor predictions of the rheological behaviour in this flow. This highlights two interesting points. Firstly, the RF algorithm, despite not predicting the correct parameters in some cases, can still provide parameter predictions provide the correct rheological behavior even in different flow types and deformation rates. Secondly, to improve the accuracy of the RF predictions, extensional flow data with strong enough velocity gradients should likely be used in conjunction with, or potentially instead of, shear flow data for training. This may help resolve some of the practical identifiability issues observed, and will be explored in future work. However, unless the user wishes to simulate such a flow using the parameterised model, our results show that the RF algorithm will provide an adequate set of model parameters (even if it is one set of a number of sets).

\section{Conclusion}

We have investigated the use of Random Forest (RF) regression for the prediction of viscoelastic constitutive model parameters from LAOS data. We use the RF regression method to essentially solve the inverse problem for parameterisation of viscoelastic constitutive models. For a given constitutive model, we solve the model under oscillatory shear flow at a given value of $De$ and $Wi$ for a wide range of model parameters. The RF regression is then used to create a black-box type function mapping from Chebyshev spectra to model parameters. This tool can then be used to rapidly parameterise viscoelastic models. Previously, one would need to solve the ODEs for each constitutive model numerically in each iteration of an optimisation-based fitting procedure, which currently hinders the widespread use of digital modelling for viscoelastic fluid flows.

We investigated the linear and exponential PTT models, as well as the RoLiE-Poly model. For both versions of the PTT model tested, the RF algorithm was able to predict the model parameters using the input Chebyshev coefficients for a LAOS simulation with high levels of accuracy. The accuracy of the predictions was, naturally, better when the amount of training data was increased. It was found that using a max depth of 10 and 100 estimators for the RF algorithm gave the best model performance. The model performance was found to decrease slightly when the max depth for the algorithm was decreased. However, reducing the number of estimators did not significantly affect the predictive performance, at least within the range investigated.

The RF algorithm was less capable of predicting the model parameters for the RoLiE-Poly model, at least with the hyper-parameter sets we investigated. However, despite not being able to predict all of the data points accurately, the model parameters predicted still represented the rheological behaviour with high levels of accuracy. This was shown to be caused by the fact that multiple sets of model parameters can yield very similar LAOS behaviours. It was found that the shear rheological behaviour was predicted accurately at the conditions used for the training ($De = 1$ and $Wi = 50$), as well as at different conditions ($De = 0.5$ and $Wi = 100$), despite the difference between the real and predicted parameter values. We also tested the parameter predictions in a homogeneous oscillatory extensional flow. At $De = 1$ and $Wi = 3$, the model parameters predicted by the RF algorithm still captured the extensional rheological behaviour well. At $De = 1$ and $Wi = 6$, however,  the extensional rheological was not captured well for some of the data points where the parameter predictions were inaccurate. This suggests combining LAOS and LAOE data (and training at high enough values of $Wi$) might help to improve predictive performance.

In this current study, we have limited ourselves to homogeneous flows (i.e.  spatially uniform velocity gradients). To implement this methodology in real-world applications where many materials will exhibit shear banding during the rheolometric testing, the RF model should be trained using simulated rheological data where the flow is allowed to shear-band. This will involve solving the constitutive models in both space and time, and obtaining the waveform for the stress at the top boundary. However, overall, we have shown that the RF algorithm is capable of learning the complex relationship between non-Newtonian rheology and constitutive model parameters, and therefore has strong potential to aid in modelling and simulation of complex fluid flows, which is of particular use across a wide range of sectors and industries. Particularly, the methodology that we have developed and proposed in this investigation could be used to strong effect in industry to drive a change from wasteful trial-and-error experimentation towards digital modelling for robust and sustainable process scale-up and optimisation. Moreover, since our methodology can allow for rapid parameterisation of constitutive models, one can quickly and accurately model the rheological behaviour of new products made with novel and sustainable raw ingredients.

\section*{Acknowledgments}

The authors acknowledge the financial support of the Center in Advanced Fluid Engineering for Digital Manufacturing, UK (CAFE4DM) project (Grant No. EP/R00482X/1). We would like to acknowledge the British Society of Rheology, who funded a summer research internship for Michail Vousvoukis. We would also like to acknowledge the University of Manchester's School of Engineering, who funded a summer research internship for Ahmed Alalwyat.

\appendix

\section{ODEs for PTT/ePTT model under LAOS}\label{eq:PTTODES}

Noting that $A_{13} = A_{23} = 0$ in simple shear flow, and that $\boldsymbol{A}$ is a symmetric second-order tensor, Equation \eqref{eq:pttdimensionless} can be written out in full as

\begin{multline}
\frac{\dif}{\dif \tilde{t}} \begin{bmatrix}
A_{11} & A_{12} & 0\\
A_{12} & A_{22} & 0 \\
0 & 0 & A_{33}
\end{bmatrix} = 
\frac{Wi}{De} \left( \begin{bmatrix}
2A_{12}\mathrm{cos}(\tilde{t}) & A_{22}A_{12}\mathrm{cos}(\tilde{t}) & 0\\
A_{22}A_{12}\mathrm{cos}(\tilde{t}) & 0 & 0 \\
0 & 0 & 0
\end{bmatrix} \right) \\
-\frac{1}{2} \frac{Wi}{De} \, \zeta \left( \begin{bmatrix}
2A_{12}\mathrm{cos}(\tilde{t}) & (A_{11}+A_{22})\mathrm{cos}(\tilde{t}) & 0\\
(A_{11}+A_{22})\mathrm{cos}(\tilde{t}) & 2A_{12}\mathrm{cos}(\tilde{t}) & 0 \\
0 & 0 & 0
\end{bmatrix} \right) \\
- \frac{1}{De} D(A) \left(\begin{bmatrix}
A_{11}-1 & A_{12} & 0\\
A_{12} & A_{22}-1 & 0 \\
0 & 0 & A_{33}-1
\end{bmatrix}\right),
\end{multline}

\noindent which gives the following system of Equations

\begin{subequations}
\begin{gather}
\frac{\dif {A}_{11}}{\dif \tilde{t}} = \left(2 - \zeta \right) \left(\frac{Wi}{De}\right)  A_{12} \, \mathrm{cos}(\tilde{t})- \frac{1}{De}D(A)(A_{11} - 1),
\\[12pt]
\frac{\dif {A}_{12}}{\dif \tilde{t}} = \left( \left(1 - \frac{\zeta}{2} \right) \left(\frac{Wi}{De}\right) A_{22} - \frac{A_{11}\zeta}{2} \right)  \, \mathrm{cos}(\tilde{t})- \frac{1}{De}D(A)A_{12},
\\[12pt]
\frac{\dif {A}_{22}}{\dif \tilde{t}} = - \zeta \left(\frac{Wi}{De}\right) \, A_{12} \, \mathrm{cos}(\tilde{t}) - \frac{1}{De}D(A)(A_{22}-1),
\\[12pt]
\frac{\dif {A}_{33}}{\dif \tilde{t}} = - \frac{1}{De}D(A)(A_{33}-1),
\end{gather}
\label{eq:pttoscillatoryshear}
\end{subequations}

\noindent where the functional form of $D(A)$ depends on the specific PTT model used (i.e. PTT or ePTT). See Equation \eqref{eq:destructionrates} for the forms of $D(A)$ for the PTT and ePTT models. Note that regardless of the form of $D(A)$, $\dif A_{33}/\dif \Tilde{t} = 0$ at all times, and so, for initial conditions of $\boldsymbol{A}=\boldsymbol{I}$, $A_{33} = 1$ at all times.

\section{ODEs for RoLiE-Poly model under LAOS and LAOE}\label{eq:RPODES}

To solve the RoLiE-Poly model under LAOS, Equation \eqref{eq:roliepolymodel} can be written out in full as

\begin{multline}
\frac{\dif}{\dif \tilde{t}} \begin{bmatrix}
A_{11} & A_{12} & 0\\
A_{12} & A_{22} & 0 \\
0 & 0 & A_{33}
\end{bmatrix} = 
\frac{Wi}{De}\begin{bmatrix}
2A_{12}\mathrm{cos}(\tilde{t}) & A_{22}A_{12}\mathrm{cos}(\tilde{t}) & 0\\
A_{22}A_{12}\mathrm{cos}(\tilde{t}) & 0 & 0 \\
0 & 0 & 0
\end{bmatrix} \\
- \frac{1}{De}\begin{bmatrix}
A_{11}-1 & A_{12} & 0\\
A_{12} & A_{22}-1 & 0 \\
0 & 0 & A_{33}-1
\end{bmatrix} \\
-
2z\left(1-\sqrt{\frac{3}{A_{11}+A_{22}+A_{33}}}\right) \\
\left(
\begin{bmatrix}
A_{11} & A_{12} & 0\\
A_{12} & A_{22} & 0 \\
0 & 0 & A_{33}
\end{bmatrix}
+
\beta_{\mathrm{CCR}} \left(\frac{A_{11}+A_{22}+A_{33}}{3}\right)^{-0.5}
\begin{bmatrix}
A_{11}-1 & A_{12} & 0\\
A_{12} & A_{22}-1 & 0 \\
0 & 0 & A_{33}-1
\end{bmatrix} \right),
\end{multline}

\noindent which gives the following system of Equations

\begin{subequations}
\begin{gather}
\begin{split}
\frac{\dif {A}_{11}}{\dif \tilde{t}} = 2\left(\frac{Wi}{De}\right) A_{12} \, \mathrm{cos}(\tilde{t})- \frac{1}{De}(A_{11} - 1) \\ -2z\left(1-\sqrt{\frac{3}{A_{11}+A_{22}+A_{33}}}\right)\left( A_{11} + \beta_{\mathrm{CCR}} \left(\frac{A_{11}+A_{22}+A_{33}}{3}\right)^{-0.5} \left( A_{11}-1 \right) \right),
\end{split}
\\[12pt]
\begin{split}
\frac{\dif {A}_{12}}{\dif \tilde{t}} = \left(\frac{Wi}{De}\right) A_{22} \, \mathrm{cos}(\tilde{t})- \frac{1}{De}(A_{12}) \\ -2z\left(1-\sqrt{\frac{3}{A_{11}+A_{22}+A_{33}}}\right)\left( A_{12} + \beta_{\mathrm{CCR}} \left(\frac{A_{11}+A_{22}+A_{33}}{3}\right)^{-0.5} \left( A_{12} \right) \right),
\end{split}
\\[12pt]
\begin{split}
\frac{\dif {A}_{22}}{\dif \tilde{t}} = - \frac{1}{De}(A_{22} - 1) \\ -2z\left(1-\sqrt{\frac{3}{A_{11}+A_{22}+A_{33}}}\right)\left( A_{22} + \beta_{\mathrm{CCR}} \left(\frac{A_{11}+A_{22}+A_{33}}{3}\right)^{-0.5} \left( A_{22}-1 \right) \right),
\end{split}
\\[12pt]
\begin{split}
\frac{\dif {A}_{33}}{\dif \tilde{t}} = - \frac{1}{De}(A_{33} - 1) \\ -2z\left(1-\sqrt{\frac{3}{A_{11}+A_{22}+A_{33}}}\right)\left( A_{33} + \beta_{\mathrm{CCR}} \left(\frac{A_{11}+A_{22}+A_{33}}{3}\right)^{-0.5} \left( A_{33}-1 \right) \right).
\end{split}
\end{gather}
\label{eq:RPoscillatoryshear}
\end{subequations}

The RoLiE-Poly model under LAOE can be written out in full as

\begin{multline}
\frac{\dif}{\dif \tilde{t}} \begin{bmatrix}
A_{11} & 0\\
0 & A_{22}
\end{bmatrix} = 
\frac{Wi}{De}\begin{bmatrix}
-2A_{11}\mathrm{cos}(\tilde{t}) & 0\\
 & 2A_{22}\mathrm{cos}(\tilde{t})\\
\end{bmatrix} \\
- \frac{1}{De}\begin{bmatrix}
A_{11}-1 & 0\\
0 & A_{22}-1
\end{bmatrix} \\
-
2z\left(1-\sqrt{\frac{3}{A_{11}+A_{22}}}\right) \\
\left(
\begin{bmatrix}
A_{11} & 0\\
0 & A_{22}
\end{bmatrix}
+
\beta_{\mathrm{CCR}} \left(\frac{A_{11}+A_{22}}{3}\right)^{-0.5}
\begin{bmatrix}
A_{11}-1 & 0\\
0 & A_{22}-1
\end{bmatrix} \right),
\end{multline}

\noindent which yields the following system of equations

\begin{subequations}
\begin{gather}
\begin{split}
\frac{\dif {A}_{11}}{\dif \tilde{t}} = -2\left(\frac{Wi}{De}\right) A_{11} \, \mathrm{cos}(\tilde{t})- \frac{1}{De}(A_{11} - 1) \\ -2z\left(1-\sqrt{\frac{3}{A_{11}+A_{22}}}\right)\left( A_{11} + \beta_{\mathrm{CCR}} \left(\frac{A_{11}+A_{22}}{3}\right)^{-0.5} \left( A_{11}-1 \right) \right),
\end{split}
\\[12pt]
\begin{split}
\frac{\dif {A}_{22}}{\dif \tilde{t}} = 2\left(\frac{Wi}{De}\right) A_{22} \, \mathrm{cos}(\tilde{t})- \frac{1}{De}(A_{22} - 1) \\ -2z\left(1-\sqrt{\frac{3}{A_{11}+A_{22}}}\right)\left( A_{22} + \beta_{\mathrm{CCR}} \left(\frac{A_{11}+A_{22}}{3}\right)^{-0.5} \left( A_{22}-1 \right) \right),
\end{split}
\end{gather}
\label{eq:RPoscillatoryextensional}
\end{subequations}

\bibliographystyle{unsrtnat}
\bibliography{bib}

\end{document}